\documentclass[%
 reprint,
superscriptaddress,
nofootinbib,
 amsmath,amssymb,bbm,
 aps,
pra,
]{revtex4-2}

\usepackage{graphicx}
\usepackage{dcolumn}
\usepackage{bm}
\usepackage{braket}
\usepackage{array,ragged2e}
\usepackage{tabularx}
\usepackage{multirow}
\usepackage{enumitem}
\usepackage[caption=false]{subfig}
\usepackage{amsmath}
\usepackage{xcolor}
\usepackage{hyperref}
\hypersetup{
    colorlinks,
    linkcolor={orange!50!black},
    citecolor={blue!50!black},
    urlcolor={blue!80!black}
}
\usepackage{booktabs}

\renewcommand{\vec}{\boldsymbol}

\begin{document}

\title{Better than classical? The subtle art of benchmarking quantum machine learning models}

\author{Joseph Bowles}
 \email{joseph@xanadu.ai}
\affiliation{Xanadu, Toronto, ON, M5G 2C8, Canada
}
\author{Shahnawaz Ahmed}
 \email{shahnawaz.ahmed95@gmail.com}
\affiliation{Xanadu, Toronto, ON, M5G 2C8, Canada
}
\affiliation{Chalmers University of Technology
}
\author{Maria Schuld}
 \email{maria@xanadu.ai}
\affiliation{Xanadu, Toronto, ON, M5G 2C8, Canada
}
\date{\today}

\begin{abstract}
Benchmarking models via classical simulations is one of the main ways to judge ideas in quantum machine learning before noise-free hardware is available. However, the huge impact of the experimental design on the results, the small scales within reach today, as well as narratives influenced by the commercialisation of quantum technologies make it difficult to gain robust insights. To facilitate better decision-making we develop an open-source package based on the PennyLane software framework and use it to conduct a large-scale study that systematically tests 12 popular quantum machine learning models on 6 binary classification tasks used to create 160 individual datasets. We find that overall, out-of-the-box classical machine learning models outperform the quantum classifiers. Moreover, removing entanglement from a quantum model often results in as good or better performance, suggesting that ``quantumness'' may not be the crucial ingredient for the small learning tasks considered here. Our benchmarks also unlock investigations beyond simplistic leaderboard comparisons, and we identify five important questions for quantum model design that follow from our results.
\end{abstract}


\maketitle

Much has been written about the ``potential'' of quantum machine learning, a discipline that asks how quantum computers fundamentally change what we can learn from data \citep{biamonte2017quantum, schuld2021machine}. While we have no means of running quantum algorithms on noise-free hardware yet, there are only a limited number of tools available to assess this potential. Besides proving advantages for artificial problem settings on paper, certain ideas -- most prominently, variational models designed for near-term quantum technologies -- can be tested in classical simulations using small datasets. Such benchmarks have in fact become a standard practice in the quantum machine learning literature and are found in almost every paper. 

A taste for the results derived from small-scale benchmarks can be obtained through a simple literature review exercise. Out of $55$ relevant papers published on the preprint server \textit{arXiv}\footnote{It is standard practice in the field of quantum computing to publish articles on the arXiv server, and we therefore expect it to provide representative samples of the literature. The $55$ papers are a subset of $73$ papers returned by the keyword search, and were selected by manually reading the abstracts and discarding papers that related to quantum-inspired methods, other fields than quantum machine learning, or did not explicitly mention a method outperforming another.} until December 2023 that contain the terms ``quantum machine learning'' and ``outperform'' in title or abstract, one finds that about $40\%$ claim that a quantum model outperforms a classical model, while about $50\%$ claim that some improvement to a quantum machine learning method outperforms the original one (such as optimisers \citep{ito2023santaqlaus, wiedmann2023empirical}, pre-training strategies \citep{kashif2023alleviating} or symmetry-aware ansatze \citep{le2023symmetry, west2023reflection}). Only $3$ papers or $4\%$ find that a quantum model does \textit{not} outperform a classical one \citep{bordoni2023long, schreiber2023classical, piatkowski2022towards}; two of these are quick to mention that this is ``due to the noise level in the available quantum hardware'' \citep{bordoni2023long} or that ``the proposed methods are ready for [...] quantum computing devices which have the potential to outperform classical systems in the near future'' \citep{piatkowski2022towards}. Only one paper \citep{schreiber2023classical} draws critical conclusions from their empirical results. If we assume that this literature review is representative\footnote{Other search terms than ``outperform*'', such as ``better than'' or ``benchmark*'', were tried to detect a possible selection bias caused by the keyword search, but showed similar patterns.}, then the overwhelming signal is that quantum machine learning algorithm design is progressing rapidly on all fronts, with ample evidence from small-scale experiments that it is already beating classical machine learning in generic domains. But can we trust this picture when judging the potential of current ideas? 

\begin{figure}[t!]
    \centering
    \includegraphics[width=0.4\textwidth]{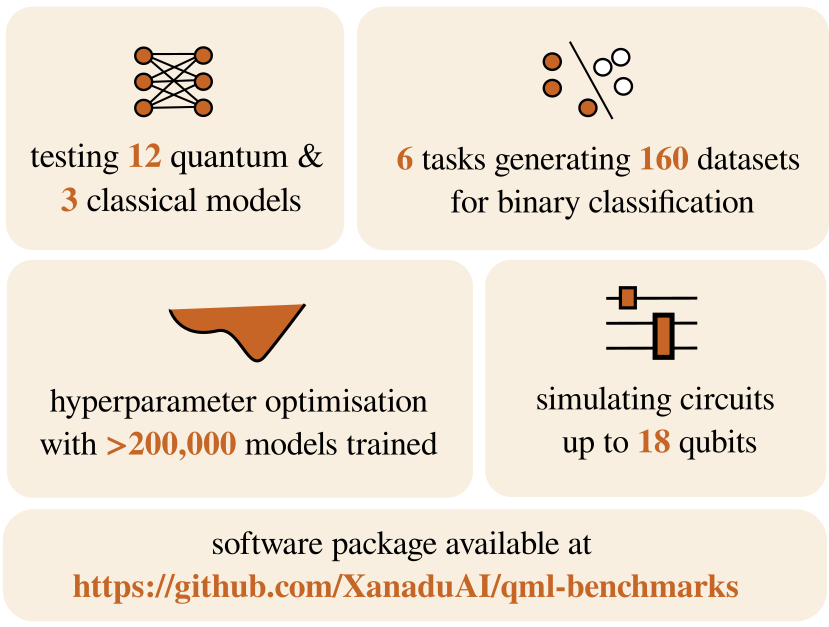}
    \caption{The scope of the benchmark study at a glance.}
    \label{fig:facts}
\end{figure}

In search for an evidence-based evaluation of proposals in near-term quantum machine learning algorithm design, this paper conducts a large-scale benchmark study that systematically tests popular quantum machine learning models on classification tasks. The code, built on the \textit{PennyLane} software framework \cite{bergholm2018pennylane}, is made available under \url{https://github.com/XanaduAI/qml-benchmarks}, and the datasets can also be found under  \url{https://pennylane.ai/datasets/}. We designed the study with a strong emphasis on scientific best practices that aim at reducing a positivity (or negativity) bias. To achieve this, we selected $12$ prototypical, influential quantum machine learning models ranging from so-called \textit{quantum neural networks} \citep{farhi2018classification, circuitcentric} to \textit{convolutional quantum neural networks} \citep{cong2019quantum, quanvolutional} and \textit{quantum kernel methods} \citep{schuld2019quantum, havlivcek2019supervised} through a randomised procedure and implemented them as faithfully as possible.

The $6$ data generation procedures for the classification tasks are chosen based on principles such as structural diversity, comparability, control of important variables, and theoretical relevance. They generate $160$ individual datasets grouped into $10$ benchmarks, where each benchmark varies either the dimension or a parameter controlling the difficulty of the learning problem that the models have to solve. Note that not all models are tested on all datasets and benchmarks; some, like translation-invariant tasks, were designed for specific models, while others required too many computational resources for the extensive hyperparameter optimisation performed in this paper.\footnote{We ran our simulation on the \textit{Digital Research Alliance of Canada}'s Cedar supercomputer and limited runtimes to at most $24$ hours for one $5$-fold cross-validation (i.e., training the model five times) using a 10-core cluster with 40GB RAM. The largest size of a circuit simulated in the study has 18 qubits, a number that we are determined to push further going forward. } 

We compare the quantum models to likewise prototypical and influential classical machine learning models like Support Vector Machines and simple Neural Networks (i.e., \textit{not} ``state-of-the-art'', but rather ``out-of-the-box'' models for small datasets). The goal is to find signals across the experiments that are consistent and can give us clues about which research questions are worth investigating further.

Our results show that -- contrary to the picture emerging from the literature sample above -- the prototypical classical baselines perform systematically better than the prototypical quantum models on the small-scale datasets chosen here. Furthermore, for most models there are no significant drops in performance if we use comparable quantum models that do not use entanglement and are classically simulable at scale, suggesting that ``quantumness'' may not be a defining factor. 

With small variations, the overall rankings between models are surprisingly consistent throughout the different benchmarks and allow some more nuanced observations. For example, hybrid quantum-classical models -- such as models that use quantum circuits inside neural networks or a support vector machine -- perform similarly to their purely classical ``hosts'', suggesting that the quantum components play a similar role to the classical ones they replace. While a layer-wise, trainable encoding with trainable input feature scaling (``data-reuploading'' \citep{datareuploading}) shows some promising behaviour, models based on so-called ``amplitude encoding'' \citep{circuitcentric, treetensor, weinet} struggle with the classification tasks, even if given copies of inputs. Interestingly, almost all quantum models perform particularly badly on the benchmarks we deemed simplest, a linearly separable dataset. 

Although the quantum models we tested failed to provide compelling results, we are not necessarily advocating for a research program that attempts to optimise them for the datasets in this work. Rather, the poor performance relative to baseline classical models across a range of tasks should bolster a hard-to-swallow fact about current quantum machine learning research: namely, that the inductive bias of near-term quantum models, the added benefit of ``quantumness'' as well as the problems for which both are useful, are still poorly understood \cite{schreiber2023classical, kubler2021inductive, larocca2022group, bowles2023contextuality}. 

We finally note that while independent benchmark ``meta-studies'' like ours are still rare, an increasing number of papers aim at systematically studying aspects of quantum model design. For example, \citet{kashif2023alleviating} look at the role of parameter initialisation for trainability, and \citep{kiwit2023application} provide a software framework to test generative models. \citet{Moussa2023} investigate the role of hyperparameters on the generalisation performance, and some of their findings are confirmed by our results. 

The remainder of the paper will discuss common benchmarking pitfalls  (Section~\ref{sec:background}), the model (Section~\ref{sec:models}) and data  (Section~\ref{sec:data}) selection method used in this study, and insights from hyperparameter optimisation (Section~\ref{sec:pipeline}). Our main results are presented in Section~\ref{sec:results} and Section~\ref{sec:questions} discusses questions that follow for the design of quantum models. The conclusion (Section~\ref{sec:conclusion}) will reflect on important lessons learnt in this study, which can hopefully contribute to more robust and critical benchmarking practices for quantum machine learning.

\section{What can we learn from benchmarks?}\label{sec:background}

Eagerly following its parent discipline of classical machine learning, benchmarks in the quantum machine learning literature are commonly motivated by statements like ``to demonstrate the performance of our quantum model we test it on the XYZ dataset''. But how much can benchmarks give us an insight into the quality of a model or idea? Before entering into the technical details we want to discuss this question critically as a cautionary tale that informs our own benchmark study.

\subsection{The need for scientific rigour}\label{sec:rigour}

In classical machine learning, increasing doubts about the singular reliance on performance metrics have emerged in recent years  \citep{dehghani2021benchmark, sculley2018winner, ethayarajh2020utility}\footnote{Calls for more methodological rigour also led to the inception of datasets \& benchmarking tracks at leading conferences like NeurIPS.}. Concerns are supported by a range of introspective studies that show how benchmarking results are subject to high variance when seemingly small experimental design decisions are changed.

\begin{figure}
    \centering
    \includegraphics[width=0.23\textwidth]{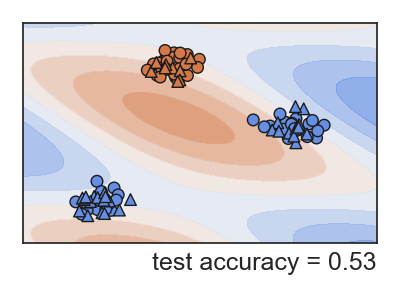}
    \includegraphics[width=0.23\textwidth]{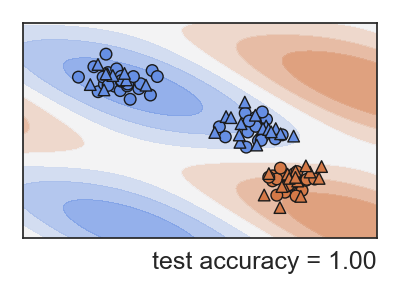}
    \caption{Illustrative example showing the effect of slight variations in a dataset on model performance. The same quantum model is trained on two different datasets to predict two classes, red and blue. The decision regions are displayed as the shaded areas. Depending on a small variation of the classification task the same model can perform poorly (left) or have a perfect test score (right). We used a "vanilla" quantum neural network model with two layers of \hyperref[glos:angleemb]{\emph{angle embedding}} interspersed with CNOT entanglers, and three layers of a trainable variational circuit, followed by a $Z$-measurement on the first qubit. The classifier is trained on the points with round markers and tested on the points marked as triangles.}
    \label{fig:pitfalls}
\end{figure}

Firstly, the dataset selection has a significant impact on the comparative performance of one model over others. This is epitomised by \textit{no-free-lunch theorems} \citep{wolpert1996lack} suggesting that for a large enough set of problems, the average performance of any model is the same, and we can only hope for a good performance on a relevant subset  (automatically paying the price of a bad performance on another subset).\footnote{This raises once more the question whether existence proofs for quantum advantages in learning brings us any closer to useful quantum machine learning.} An illustrative, even if highly simplified, example is shown in Figure~\ref{fig:pitfalls}, where a minor rearrangement of the data in a classification task causes an -- admittedly adversarially hand-crafted -- quantum model to switch from being nearly unable to learn, to gaining perfect accuracy. 

In more serious examples, studies like \citet{dehghani2021benchmark} in their aptly named paper ``The benchmark lottery'', show for a range of hotly contested benchmarks that significantly different leaderboard rankings are obtained when excluding a few datasets from benchmarking suites or making changes to the way that scores are aggregated. 
\citet{northcutt2021pervasive} find that correcting labeling errors in the ubiquitous ImageNet validation set likewise changes the ranking of models, and~\citet{recht2018cifar, recht2019imagenet} demonstrate that using a different test-train split in popular image recognition datasets decreases the accuracies of models by as much as 14\%\footnote{Although, reassuringly, the ranking between models is largely unaffected by this intervention.}. While ``every benchmark makes a statement about what it perceives to be important'' \citep{dehghani2021benchmark}, it can be shown that \textit{which} benchmarks are deemed relevant is largely influenced by trends in research \citep{koch2021reduced, paullada2021data, dehghani2021benchmark, dotan2019value}, and therefore by the social makeup of a scientific community rather than methodological considerations. 

Secondly, even if the data is selected carefully, variations in the remainder of the study design can hugely influence the results of benchmarking, leading to a host of contradicting results when judging the performance of a new method. For example, large-scale comparative studies suggest that small adjustments to transformer architectures advertised in the literature have no measurable effect on their performance \citep{narang2021transformer}, and that contrary to a common belief, standard convolutional neural networks still beat transformers when trained on a comparable amount of data \citep{smith2023convnets}.
Similarly, extensive hyperparameter optimisation can give baseline models of generative adversarial networks \cite{lucic2018gans}, deep reinforcement learning  \citep{henderson2018deep} and Bayesian deep neural nets \citep{riquelme2018deep} the performance of more advanced proposals. These findings question the large effort invested into optimising model design. And not only are there variations in the model and data, but the choice of scoring metric like accuracy, F-score and Area-Under-Curve make different assumptions on what is deemed important and may correlate poorly with each other \citep{flach2019performance}. Lastly, \citet{dodge2019show} demonstrate how integrating computational costs into performance considerations for Natural Language Processing models highlight regimes where a method as basic as logistic regression is superior to deep learning. 

\begin{figure}
    \centering
    \includegraphics[width=0.49\textwidth]{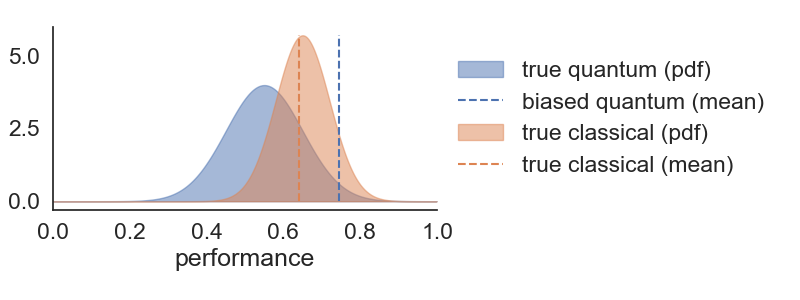}
    \caption{Numerical illustration of the thought experiment on a positivity bias. Assume that the ``true'' performance of classical and quantum models is distributed normally (blue and red curves) with the mean for classical model performance higher than in the quantum case. The dashed lines report numerical calculations of the mean of model performance if $100$ researchers report only on the top-performing candidate out of $20$ quantum models, but do not select the best classical model in a similar manner. The bias from discarding the $19$ worst-performing quantum models reverses the observed average performance with respect to the true one.}
    \label{fig:positivity-bias}
\end{figure}

A third danger of benchmarks that judge the ``power'' of a model over another is a systematic positivity bias, since the goal of designing models that are better than existing ones creates an incentive to publish positive results. This can be illustrated in a simple thought experiment (see Figure~\ref{fig:positivity-bias}): Consider $100$ researchers, each investigating $20$ different types of quantum models before settling on a promising design and publishing benchmarks of the best one against a classical model of their choice. Let us assume there is some kind of ground truth that the performance of quantum and classical machine learning models is normally distributed, and on average classical models perform better and more consistently (say, with a mean of $0.65$ and standard deviation of $0.07$) than quantum ones (mean $0.55$, standard deviation $0.1$). But the bias created from discarding $19$ models means that the researchers will overall find that quantum models are better than classical models; when running a simple simulation we find that the published mean performance of the quantum models is $0.74$ vs. $0.65$ of classical ones: the scales are flipped! Since leaderboard-driven research actively searches for good models, a positivity bias of this nature is not a question of ethical misconduct, but built into the research methodology.

Together, these arguments make a convincing case that ``perfomance'' or ``power'' of a model cannot be viewed as a property that can be ``demonstrated'' by an empirical study, but as a highly context-dependent signal that has to be artfully coaxed out of a system with many interacting parts -- not unlike the challenge of an experimental physicist trying to make sense of a complex physical system. 

Luckily, we have a tried-and-tested weapon for this challenging task: scientific rigour. For example, benchmarks can become a powerful tool when asking questions of a clearly defined scope. ``Does my quantum model beat classical machine learning?'' is impossible to answer by a numerical experiment, while ``Does the performance of a model on a specific task increase as measured by some metric, if we change the data, or training conditions?'' might be. A carefully selected task for such a question, for example an artificial dataset allowing systematic control of the properties of interest, allow much clearer interpretations of the results. Testing when and how a model can be made to perform poorly will give other researchers clues about mechanisms at play, rather than advertising one method over another. And lastly, searching for patterns that can be reproduced across widely different settings will constitute a signal that allows the field to take objective decisions on which ideas show promise. In this study we aim to be explicitly mindful of these pitfalls in the process of model and data selection, experimental design and interpretation of the results.

\subsection{Should we use MNIST for quantum machine learning?}\label{sec:mnist}

In order to illustrate some specific problems in \textit{quantum} machine learning research with its peculiar challenges, let us get back to the question of data selection once more. Unless a quantum researcher focuses on their favourite domain, such as physics itself, the choice of dataset tends to be directly adopted from the classical machine learning literature. But without error-free hardware to run algorithms on, quantum machine learning research finds itself in a very different situation: data is either borrowed from machine learning some decades in the past, or needs to be downscaled and simplified in order to suit the scale of simulation capabilities or noisy small-scale hardware. We want to demonstrate some issues arising from this predicament with a discussion of the (in)famous MNIST handwritten digits datasets which is used widely in quantum machine learning studies. Note that somewhat contradictory, we will still add MNIST as a task here in order to make our results easier to compare with so many other studies.

The  MNIST dataset consists of 60.000 $28\times 28$ images of handwritten digits \citep{lecun1998mnist} and played a crucial historical role in the benchmarking of classical machine learning. While largely superseded by data like ImageNet \citep{deng2009imagenet} and CIFAIR \citep{krizhevsky2009learning}, it is still occasionally used as a sanity check of new models, as this quote by Geoffrey Hinton, one of the pioneers of deep learning, demonstrates:

\begin{quote}
    MNIST has been well studied and the performance of simple neural networks trained with backpropagation is well known. This makes MNIST very convenient for testing out new learning algorithms to see if they actually work. \citep{hinton2022forward}
\end{quote}

Hinton goes on to summarise the baselines for a reasonable success on MNIST: 

\begin{quote}
Sensibly-engineered convolutional neural nets with a few hidden layers typically get about 0.6\% test error. In the ``permutation-invariant'' version of the task, the neural net is not given any information about the spatial layout of the pixels [and] feed-forward neural networks with a few fully connected hidden layers of Rectified Linear Units (ReLUs) typically get about 1.4\% test error. This can be reduced to around 1.1\% test error using a variety of regularizers [...] \citep{hinton2022forward}
\end{quote}

While this may suggests MNIST as a useful benchmark for innovation in quantum machine learning, there is a serious caveat to consider: The typical input size that can be handled with simulations (not to mention hardware) is of the order of tens of features, but downscaling the 784 original pixels by that much voids our extensive knowledge about the baseline to beat. 
    
For example, a representative analysis of $15$ randomly selected papers in quantum machine learning using MNIST benchmarks\footnote{These were sampled out of $46$ papers that we identified to use MNIST for supervised learning benchmarks listed in the arxiv `quant-ph' category between January 2018 and June 2023. The papers were in turn selected from a total of over 100 papers in the `quant-ph' category on the arXiv in the same period that mention MNIST in title or abstract.} shows that more than half use a pre-processing strategy like resolution reduction or PCA to lower the number of input features, while most others investigate hybrid models where the dimensionality reduction is implicitly achieved by a classical layer such as a convolutional or tensor network.\footnote{A notable exception are papers based on so-called ``quanvolutional'' architectures, where the first layer uses a small ``filter'' or ``kernel'' composed of a quantum circuit \citep{quanvolutional}.} Preprocessing the data changes the hardness and nature of the learning problem and the meaningfulness of using MNIST has to be reassessed. For example, the 2-dimensional PCA-reduced MNIST 3-5 classification problem consists of about $10,000$ datapoints that lie in overlapping blobs (see Figure~\ref{fig:data} further below), a task that is hardly meaningful. Furthermore, as \citet{bausch2020recurrent} remarks, reducing images to a $4 \times 4$ resolution leads to $70\%$ of test images also being present during training. And of course,  using powerful classical layers to read in the data makes it hard to judge the impact of the quantum model.

There are other issues besides pre-processing. For example, only four of the papers in our representative sample of 15 use the original MNIST multi-class problem, while all others distinguish between two of the $10$ digits -- which tends to make the problem easier. Distinguishing between $0-1$, for instance, can be achieved by a linear classifier to almost $100\%$ accuracy. Possibly blinded by the size of the original dataset, not all authors seem to be aware that their quantum model scores highly on an almost trivial task.

It is no surprise that the overall results -- which in the sample of $15$ papers range from accuracies of $70-99.6\%$ are mixed and hard to interpret with respect to the known baseline. Collectively, they certainly do not give convincing evidence for quantum models to systematically reach the test accuracies of $98.6\%$ or even $99.4\%$ put forward by Hinton with respect to the original MNIST task -- at least not if unaided by powerful classical models.

Overall, the arguments summarised here put into question the practice of adopting datasets from classical machine learning blindly when their benefits are lost. Alternatives, such as 1-d MNIST \citep{greydanus2020scaling} designed to scale down deep learning experiments, may be a lot more suitable.

With the methodological concerns sufficiently emphasised, we now proceed to introducing the technical details and decisions taken in this work.

\section{Models to be benchmarked}\label{sec:models}
    \subsection{Selection methodology and bias} 
    The goal of our paper selection procedure was to identify a diverse set of influential ideas in quantum machine learning that are implementable on current-day standard simulators. Importantly, we are not claiming to capture the ``best'' or most advanced algorithmic designs, but rather those that are the foundation of models proposed today. 
    
\begin{figure}
        \centering
        \includegraphics[width=0.40\textwidth]{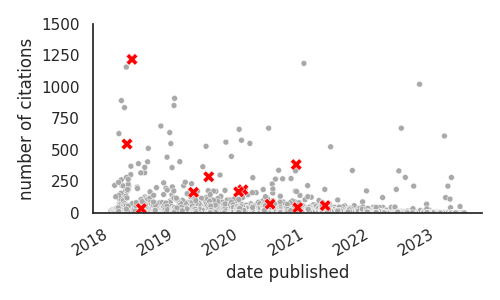}
        \caption{Publishing date versus citations of the eleven selected papers (red crosses) from an initial set of 3500 papers drawn from the ArXiv API (gray dots). Outliers with over $1500$ citations are not shown. Selecting papers with $30$ or more citations introduces a bias towards less recent work. }
        \label{fig:paper-selection}
    \end{figure}

    After attempting other strategies that failed to produce a representative sample we settled on the following step-by-step selection method:

    \begin{enumerate}
        \item \textbf{Thematic match:} We first collected all papers accessible via the arXiv API in the period 1 January 2018 to 1 June 2023 in the ``quant-ph'' category with keywords ``classif*'', ``learn*'', ``supervised'', ``MNIST''. By this we intended to find the overwhelming share of papers that are part of the current quantum machine learning discourse. [$3500$ papers]
        \item \textbf{Influential:} From the previous set we sub-selected papers with $30$ or more Google Scholar citations on the cutoff date 2 June 2023. This was intended to bias the selection towards ideas that stood the test of time and are used widely in quantum machine learning research. It inevitably also biases towards older papers (see Figure~\ref{fig:paper-selection}). [$561$ papers]
        \item \textbf{Suitable for benchmarks:} From reading the abstract, we sub-selected papers that propose a new quantum model for classification on conventional classical data which can be numerically tested by a standard qubit-based simulator. This ensured that the models can be meaningfully compared via a focused set of benchmarks, and that we limit ourselves to original proposals. [$29$ papers]
        \item \textbf{Implementable:} To achieve a reasonable workload, we sampled a random subset of $15$ papers from the $29$ candidates. After reading the full body of these papers we had to exclude another four that we could not reasonably implement, either because the data encoding was too expensive \citep{farhi2018classification, zhao2019building}, or because the paper did not specify important parts of our classification pipeline like training \citep{tacchino2019artificial} or an embedding strategy for classical datasets \citep{cong2019quantum}. [$11$ papers]
    \end{enumerate}

    From this final selection of 11 papers \citep{datareuploading, dressed, iqp, circuitcentric, treetensor, qboltzmann, projectedquantumkernel, kitchensinks, metriclearner, quanvolutional, weinet} we implemented 12 quantum machine learning models. \citet{iqp} compared two distinct architectures, both of which we implemented. 

    \subsection{Summary of the models}

    The 12 models we implemented can be split into three families: \textit{quantum neural networks}, \textit{quantum kernel methods} and \textit{quantum convolutional neural networks}. Here we explain the principle of each family and give a brief overview of the central idea of each model. Technical terms commonly known within the quantum machine learning community, but not necessarily clear to every reader, are highlighted in orange italic font and explained in a glossary at the end of this paper (see Appendix~\ref{app:glossary}). More detailed descriptions of the models can be found in Appendix~\ref{app:models}, and Table~\ref{tab:model-overview} provides the basic facts at a glance. We use camel case to refer to models, and the names correspond to the classes used in the software package.

    \subsubsection{Quantum neural network models}
        So-called ``quantum neural networks'' are -- somewhat misleadingly named -- simply variational quantum circuits that encode data and are trainable by gradient-descent methods. Variational quantum circuits have been extensively studied in the quantum machine learning literature and take the form
          \begin{align}\label{PQC}
            f(\vec{\theta},\vec{x}) = \text{tr}[O(\vec{x},\vec{\theta})\rho(\vec{x},\vec{\theta})],
        \end{align}
        where $\rho$ is a density matrix, $O$ is an observable, and both may depend on an input data point $\vec{x}$ and trainable parameters $\vec{\theta}$. For our purposes, a quantum neural network model is one that combines one or more such variational quantum circuits $\{f_{1},...,f_{L}\}$ to classify data into one of two labels $y=\pm1$ through a class prediction function $f_{\text{pred}}$:
        \begin{align}
            y = f_{\text{pred}}(f_{1}(\vec{\theta}_1,\vec{x}), \cdots , f_{L}(\vec{\theta}_L,\vec{x})). 
        \end{align}
        In the simplest case, where the model uses a single quantum function ($L =1$) and an observable $O(\vec{x},\vec{\theta})$ with eigenvalues $\pm1$, $f_{\text{pred}}$ is typically the sign function, so that the sign of the measured observable indicates the predicted class. 
        
        To train these models, one defines a differentiable cost function  
        \begin{align}
           \mathcal{L}(\vec{\theta},\mathbf{X},\vec{y}) = \frac{1}{N}\sum_i \ell(\vec{\theta},\vec{x}_i,y_i), 
        \end{align}
        where the loss  $\ell(\vec{\theta},\vec{x}_i,y_i)$ measures the model performance for a specific training data point $\vec{x}_i$ whose true label $y_i$ we know, and $\mathbf{X},\vec{y}$ summarise all training inputs and labels into a matrix/vector. Like neural networks, the loss can be numerically minimised in an end-to-end fashion via stochastic gradient descent by application of the chain rule, where so-called parameter-shift rules \citep{mitarai2018quantum, schuld2019evaluating} allow us to evaluate gradients of quantum circuit parameters on hardware. 
        
        We study six quantum neural network classifiers:
        \begin{itemize}
            \item 
            {\texttt{CircuitCentricClassifier}} \citep{circuitcentric}: 
            A generic quantum neural network model in which copies of \hyperref[glos:ampemb]{\emph{amplitude embedded}} data are followed by a trainable unitary and measurement of a single-qubit observable. 
            \item 
            {\texttt{DataReuploadingClassifier}} \citep{datareuploading}: 
            A model in which the data is scaled by trainable classical parameters and fed to a quantum circuit via layers of trainable \hyperref[glos:angleemb]{\emph{angle embedding}}. One single-qubit rotation thereby takes three input features at once. The aim of training is to maximise the fidelities of the output qubits to the desired class state (either $\ket{0}$ or $\ket{1}$).
            \item  
            {\texttt{DressedQuantumCircuitClassifier}} \citep{dressed}:
            A model that preprocesses the data using a classical neural network, which is  fed into a generic quantum circuit via \hyperref[glos:angleemb]{\emph{angle embedding}}. The expectation values at the output are then post-processed by another classical neural network for prediction. Both the classical neural networks and the quantum circuit are trainable. 
            \item 
            {\texttt{IQPVariationalClassifier}} \citep{iqp}: A model that encodes the input features via an \hyperref[glos:angleemb]{\emph{angle embedding}} using a circuit structure inspired from \hyperref[glos:iqp]{\emph{Instantaneous Quantum Polynomial}} (IQP) circuits, which are known to be hard to simulate classically. \citep{bremner2011classical}. 
            \item 
            {\texttt{QuantumBoltzmannMachine}} \citep{qboltzmann}: A model inspired from classical Boltzmann machines that trains the Hamiltonian of a multi-qubit \hyperref[glos:gibbsstate]{\emph{Gibbs state}} and measures an observable on a subset of its qubits for prediction. 
            \item 
            {\texttt{QuantumMetricLearner}}  \citep{metriclearner}:
            A model that optimises a trainable embedding of the data to increase the distance of states with different labels. Prediction relies on evaluating state overlaps  of a new embedded input with the embedded training data.  
            \item  
            {\texttt{TreeTensorClassifier}}  \citep{treetensor}: 
            A model that uses \hyperref[glos:ampemb]{\emph{amplitude embedding}} followed by a trainable unitary with a tree-like structure that is designed to avoid \hyperref[glos:vanishinggrads]{\emph{vanishing gradients}}. 
        \end{itemize}

        The choice of loss function for training varies among models\footnote{The choice of square and linear loss in some models is curious, since it can be argued that a more natural choice for a classification loss is the cross entropy loss, as it corresponds to the maximum likelihood estimator of the data \cite{hastie2009elements}. A square loss, instead, is known to be more naturally suited to regression problems with continuous labels.
        }. Two models use a \hyperref[glos:crossentropy]{\emph{cross entropy loss}}, three use a \hyperref[glos:squareloss]{\emph{square loss}}, one uses a \hyperref[glos:linearloss]{\emph{linear loss}}, and one uses a loss based on distances between embedded quantum states (see Table~\ref{tab:model-overview}).

        \subsubsection{Quantum kernel methods}
        Kernel methods \cite{steinwart2008support,hofmann2008kernel} form a well known family of machine learning model that take the form 
        \begin{align}
            f(\vec{x}) = \sum_{i}\alpha_i k(\vec{x}_i,\vec{x}),
        \end{align}
        where $\alpha_i$ are real trainable parameters and $k$ is a kernel function: a positive definite function that measures the similarity between data points. Since the values $\alpha_i$ typically take the same sign as $y_i$, these models have the flavour of a weighted nearest neighbour classifier in which the distance to neighbours is mediated by the kernel function. A fundamental result in the theory of kernel methods states that any such model is equivalent to a linear classifier in a potentially infinite dimensional complex feature space $\ket{\phi(\vec{x})}$ defined via the inner product
        \begin{align}
            k(\vec{x},\vec{x}')  = \langle\phi(\vec{x})\vert\phi(\vec{x'})\rangle,
        \end{align}
        and a rich mathematical theory of these methods has been developed as a result. 
        To train such models, one typically seeks a \hyperref[glos:maxmargin]{\emph{maximum margin classifier}} of the data, which can be shown to be equivalent to solving a simple convex optimization problem in the parameters $\alpha_i$ \cite{steinwart2008support}.
        
        A quantum kernel method is one in which the kernel function is evaluated with the aid of a quantum computer. A common strategy is to define an embedding $\rho(\vec{x})$ of the classical data into quantum states, and use the function 
        \begin{align}\label{qkernelfunction}
            k(\vec{x}_i,\vec{x}_j) = \text{tr}[\rho(\vec{x}_i)\rho(\vec{x}_j)], 
        \end{align}
        which is a kernel by virtue of being an inner product, and can be evaluated by methods that evaluate state overlaps. In principle, the kernel function and any of the quantum circuits needed to evaluate it may also depend on trainable parameters, and in some models these are optimized as part of a wider training pipeline. For a deeper discussion on the connection between kernel methods and quantum models, we point the reader to \citep{schuld2021supervised}. 
        
        We implemented three quantum kernel methods:
        \begin{itemize}
            \item  
            {\texttt{IQPKernelClassifier}}  \citep{iqp}: A model that uses a quantum kernel of the form of Eq.~\eqref{qkernelfunction}, where the embedding is the same IQP-inspired angle embedding used in \texttt{IQPVariationalClassifier}. 
            \item 
            {\texttt{ProjectedQuantumKernelClassifier}} \citep{projectedquantumkernel}: 
            A model that attempts to avoid problems related to the exponential size of Hilbert space by projecting the embedded state to a smaller space defined via its reduced density matrices, and using a \hyperref[glos:rbfkernel]{\emph{Gaussian kernel}} in that space. The initial quantum embedding corresponds to a trotterized evolution given by a 1D Heisenberg Hamiltonian.
            \item 
            {\texttt{QuantumKitchenSinks}} \citep{kitchensinks}: A model in which input data is first transformed by random linear maps and then by a number of fixed quantum circuits via \hyperref[glos:angleemb]{\emph{angle embedding}}. Output bit-string samples of the quantum circuits are concatenated to form a feature vector that is fed to a linear classifier for training and prediction. 
        \end{itemize}

        We loosely include \texttt{QuantumKitchenSinks} in the above list since, as in quantum kernel methods, the linear classifier finds an optimal hyperplane in a feature mapped space given by the quantum model. However, note that the implementation does use a SVM as in the other two models.
        
        \subsubsection{Quantum convolutional neural network models}
        Our third family consists of models can be seen as analogues of convolutional neural networks \citep{cnn1, cnn2}: a class of classical neural network model designed for computer vision related tasks which exploit a form of translation symmetry between layers called ``equivariance'' \citep{weiler2023EquivariantAndCoordinateIndependentCNNs}. The literature features a large number of quantum convolutional models, undoubtedly due to the enormous success of classical convolutional models and the general tendency to `quantumify' any model that gains sufficient fame in the classical literature. We do not attempt to capture these models in a strict mathematical definition, but rather identify them by the fact that they are examples of quantum neural network models that---like classical convolutional models---also exploit a form of translation symmetry \citep{cohen2019general}, and are therefore designed for data that respects such symmetries. 
        
        We study two such models: 
        
        \begin{itemize}
            \item 
            {\texttt{QuanvolutionalNeuralNetwork} } \citep{quanvolutional}:
            This model is equivalent to a classical convolutional neural network in which the convolutional filter that defines the first layer is evaluated by a random quantum circuit that encodes the input data via angle encoding.
            \item  
            {\texttt{WeiNet}} \citep{weinet}: Uses a quantum version of a convolutional layer based on \hyperref[glos:ampemb]{\emph{amplitude embedding}} and linear combination of unitaries, with the goal of having fewer trainable parameters than a classical convolutional neural network. 
        \end{itemize}

    {\tiny
    \begin{table*}[]
    \def\arraystretch{2}  
        \centering
        \begin{tabular}{>{\RaggedRight\arraybackslash}p{2.5cm}  >{\RaggedRight\arraybackslash}p{2.5cm} >{\RaggedRight\arraybackslash}p{2.5cm} >{\RaggedRight\arraybackslash}p{3.9cm} >{\RaggedRight\arraybackslash}p{3.3cm} >{\RaggedRight\arraybackslash}p{1.8cm} }
             \textbf{Model} &  \textbf{Embedding} &
             \textbf{Measurement} &\textbf{Hyperparameters} & \textbf{Classical processing} & \textbf{Loss}  \\ \hline \hline
             \multicolumn{6}{l}{\textbf{Quantum Neural Networks}}
             \\ \hline
             \textit{Circuit Centric Classifier}  & copies of amplitude embedding & single-qubit Pauli $Z$ & - {\emph{learning\_rate}}\newline - {\emph{n\_input\_copies}}\newline - {\emph{n\_layers}} & trainable bias added to output of circuit & square \\
             \textit{Data Reuploading Classifier} & layers of trainable angle embedding & multi-qubit Pauli Z & - {\emph{learning\_rate}}\newline - {\emph{n\_layers}} \newline - {\emph{observable\_type}}  & input features and output fidelities multiplied by trainable weights & square \\
             \textit{Dressed Quantum Circuit Classifier} & layers of angle embedding & multi-qubit Pauli Z & - {\emph{learning\_rate}}\newline - {\emph{n\_layers}}  & input and output features processed by trainable classical neural network & cross entropy (softmax) \\ 
             \textit{IQP Variational Classifier} & layers of IQP-inspired angle embedding & two-qubit ZZ & - {\emph{learning\_rate}}\newline - {\emph{n\_layers}} \newline - {\emph{repeats}}  & input extended by product of features &  linear \\ 
             \textit{Quantum Boltzmann Machine} & angle embedding & multi-qubit Pauli Z &  - {\emph{learning\_rate}}\newline - {\emph{temperature}} \newline - {\emph{visible\_qubits}} & input features multiplied by trainable weights & cross entropy \\
             \textit{Quantum Metric Learner} & layers of QAOA-inspired angle embedding & pairwise state overlaps & - {\emph{learning\_rate}}\newline - {\emph{n\_layers}} & None & distance between embedded classes \\
             \textit{Tree Tensor Classifier} & amplitude embedding & single-qubit Pauli $Z$ & - {\emph{learning\_rate}} & trainable bias added to output of circuit & square \\  
             \hline 
             \multicolumn{6}{l}{\textbf{Quantum Kernel Methods}}
             \\ \hline
             \textit{IQP Kernel Classifier} & layers of angle embedding & pairwise state overlaps & - {\emph{repeats}} \newline - {\emph{C}} (SVM regularisation) & quantum kernel used in SVM & hinge \\ 
             \textit{Projected Quantum Kernel} & layers of Hamiltonian-inspired angle embedding & X, Y, Z on all qubits & - {\emph{trotter\_steps}} \newline - {\emph{C}} (SVM regularisation)  \newline - {\emph{t}} (evolution time) \newline - {\emph{gamma\_factor}} (RBF bandwidth) & quantum kernel used in SVM & hinge  \\
             \textit{Quantum Kitchen Sinks} & angle embedding & computational basis samples & - {\emph{n\_episodes}} \newline - {\emph{n\_qfearures}} & quantum features used in logistic regression & cross entropy \\ \hline 
             \multicolumn{6}{l}{\textbf{Quantum Convolutional Neural Networks}}
             \\ \hline
             \textit{Quanvolutional Neural Network} & angle embedding & computational basis samples & - {\emph{learning\_rate}} \newline - {\emph{n\_qchannels}} \newline - {\emph{qkernel\_shape}} \newline - {\emph{kernel\_shape}} & classical convolutional neural network & cross entropy (sigmoid) \\
             \textit{Wei Net} & amplitude embedding & single- and double-qubit Z & - {\emph{learning\_rate}}\newline - {\emph{filter\_type}} & single layer neural network applied to the circuit output values & cross entropy (sigmoid)
        \end{tabular}
        \caption{Overview of models used in the benchmarks. For definitions of the terms, consult the glossary in Appendix~\ref{app:glossary}. More details on the models are found in Appendix~\ref{app:models}.}
        \label{tab:model-overview}
    \end{table*}
    }

    \subsubsection{Classical models}\label{sec:classical}
    In addition to the above quantum models, we use a set of standard classical models, 
    which we define as algorithms that are classically simulable at scale (even if they might be quantum-inspired).

    Typical strategies for selecting a baseline to compare quantum models with try to match architectural components, like the number of parameters or layers in the model. However, the role that these components play and the effect they have on the computational resources differ vastly between architectures, and we do not believe that this comparison is meaningful in our context -- much like one does not enforce the same number of parameters when comparing kernel methods with neural networks. 
    
    Instead, we employ two selection criteria for the classical competitors. On the one hand we use a standard feed-forward neural network, a support vector classifier with Gaussian kernel, and a convolutional neural network model as natural equivalents to the three quantum model families defined above. The first two of these were implemented using scikit-learn's \texttt{MLPClassifier} and \texttt{SVC} classes, and the third that we call \texttt{ConvolutionalNeuralNetwork} was implemented using \textit{Flax} \citep{flax}. Similarly to the quantum model selection, these are out-of-the-box versions of models that represent popular ideas in machine learning research, and that are widely used by practitioners for small-scale datasets. They are not intended to be state-of-the-art models.

    We also conduct experiments with models that are classically simulable but inspired by quantum models. For example, the \texttt{SeparableVariationalClassifier} represents a standard quantum neural network with layer-wise, trainable angle embedding but uses no entangling gates or non-product measurements. The \texttt{SeparableKernelClassifier} is a support vector machine with a quantum kernel that embeds data using (non-trainable) layers of non-entangling angle embedding.  The quantum circuits of these models can be simulated by circuits consisting of a single qubit only.

    \subsection{Implementation}\label{sec:implementation}

    We develop a software pipeline built from \textit{PennyLane} \citep{bergholm2018pennylane}, \textit{JAX} \citep{jax2018github}, \textit{optax} \citep{deepmind2020jax} and \textit{scikit-learn} \citep{scikit-learn} as our software framework. While PennyLane's differentiable state-vector simulators in combination with JAX's just-in-time compilation tools allow us to run trainable quantum circuits, scikit-learn provides a simple API for model training and evaluation, as well as a wealth of machine learning functionalities like data pre-processing, cross-validation and performance metrics. It also offers a broad range of standard classical machine learning models to compare with. The code can be found in the repository \href{https://github.com/XanaduAI/qml-benchmarks}{https://github.com/XanaduAI/qml-benchmarks}.

    When implementing the models from the selected papers we followed these principles:
    \begin{enumerate}
        \item \textbf{Faithful implementation:} We carefully deduced the model design and training procedure from the paper. If specific details needed for implementation were missing in the paper, they were defined based on what appeared natural given our own judgement (see also Appendix~\ref{app:models}). 
        \item \textbf{Convergence criteria:} All quantum neural network models and quantum convolutional models used the same convergence criterion, which was based on tracking the variance of the loss values over time and stopping when the mean loss value did not change significantly over 200 parameter updates\footnote{As with all variational models, it can be difficult to know whether a model has converged or if the model is stuck on a particularly flat part of the optimisation landscape, and test accuracies can both improve or worsen with more training. The choice to decide on convergence over 200 updates is therefore to some extent arbitrary.} (see Appendix~\ref{app:convergence} for details). Loss histories were routinely inspected by eye to ensure the criterion was working as desired. Quantum kernel methods followed the default convergence criterion of the specific scikit-learn model class (\texttt{SVC}, \texttt{Logistic Regression}) that they employ. Pure scikit-learn baseline models (\texttt{MLPClassifier}, \texttt{SVC}) used their default convergence criterion except for the \emph{max\_iter} parameter of \texttt{MLPClassifier} which was increased to 3000. 
        If a training run did not converge during grid search, which happened very rarely, that particular result was ignored in the hyperparameter optimisation procedure. 
        \item \textbf{Batches in SGD:} Since the batch size for models using gradient descent training plays an important role in the runtime and memory resources, we did not optimize this hyperparameter with respect to the performance of the model, but with respect to simulation resources. Note that in an unrelated study, \citet{Moussa2023} found that the batch size did not have a significant impact on model performance. All models use a batch size of 32, except for the computationally expensive \texttt{QuantumMetricLearner} that uses a batch size of 16. 
        \item \textbf{Data preprocessing:} Not all models define a data preprocessing strategy, even though most data embeddings are sensitive to the range on which the input data is confined. If no preprocessing was specified, we  pre-scaled to natural intervals; for example, if angle embedding is used all features were scaled to lie in $[-\frac{\pi}{2},\frac{\pi}{2}]$ (also consistent with findings in \citep{Moussa2023}). For models that use amplitude embedding, we set the first $d$ values of the state vector equal to $\vec{x}$, pad the remaining values with a constant $1/2^n$ (with $n$ the number of qubits), then normalize all values as required.  
    \end{enumerate}

\subsection{
Difficulties of scaling to higher qubit numbers}
As quantum software gets better, simulations in the two-digit qubit numbers have become a standard for benchmark studies. However, quantum machine learning benchmarks pose particular challenges to numerical computations, as a circuit is not run once, but hundreds or thousands of times to compute one test accuracy during hyperparameter optimisation using the complex workflows quantum models are constructed of. Although hyperparameter search was parallelized such that each 5-fold cross validation run (i.e. training a model 5 times for fixed a hyperparameter setting) was sent to a cluster of 10 CPUs with a maximum compute time of 24 hours, we nevertheless ran into runtime limits for even modest qubit numbers. 

There were different causes for models to be computationally demanding. For example, \texttt{ProjectedQuantumKernel} has a large hyperparameter grid which consists of 108 distinct hyperparameter settings. Since we perform 5-fold cross validation, we therefore need to train 540 models for every dataset we consider, and training each model involves evaluating hundreds of quantum circuits. For other models the number of circuits required poses the biggest issue. For example, to train on $N=250$ datapoints we need to evaluate a quadratic number of circuits -- around $30,000$ -- for \texttt{IQPKernelClassifier}. \texttt{QuantumMetricLearner}, instead, requires multiple circuits for prediction, as it computes the distance of a new data sample to training samples. The \texttt{QuanvolutionalNeuralNetwork} model involves performing a convolution over the input image, and needs to evaluate a number of circuits that scales with the number of pixels, which in case of a $16\times 16$ pixel grid amounts to many millions of circuit evaluations. Another costly example is the \texttt{QuantumBoltzmannMachine} which is based on simulating a density matrix that requires quadratically more memory than state vector simulation. For large qubit numbers, some quantum neural network models were also very slow to reach convergence. This was particularly the case for \texttt{DressedQuantumCircuit}, which for a 15 qubit circuit failed to converge after 24 hours of training on some of the datasets. As a result of these challenges, computing a quantum model's test accuracy on a single dataset can already take over a day on a cluster for single-digit qubit numbers, and reaching of the order of $20$ qubits becomes unfeasible for the scope of our study.

There are many mitigation strategies that can speed up simulations, such as a more judicious choice of parameter initialization, more resources on the cluster, better runtime-versus-memory trade-off choices, snapshotting jobs or by turning to GPU simulation such as available in the \textit{PennyLane Lightning} suite. The variety of computational bottlenecks however requires different solutions to be found for different models, and we therefore have to defer more extensive code optimisation to future work on the benchmark software package.   
                       
\section{Datasets}\label{sec:data}

\begin{figure*}
    \centering
    \includegraphics[width=\textwidth]{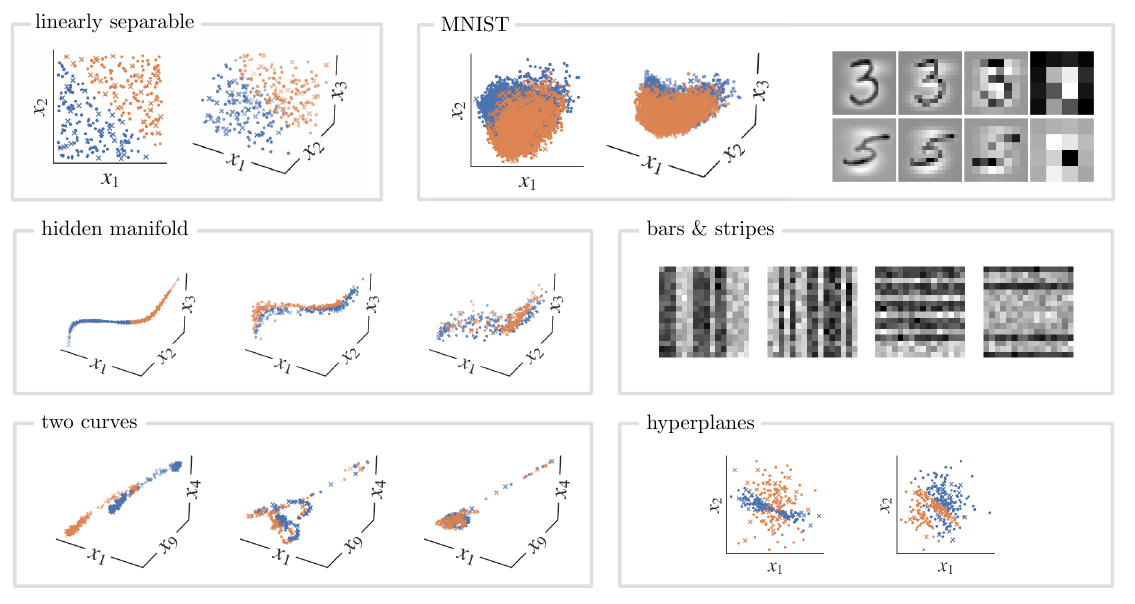}
    \caption{Illustrative examples of datasets created by the different data generation procedures. For the scatter plots, the two classes are shown in blue and orange, and training points are shown in round vs.\ test points in an `x' shape. The linearly separable pannel shows data for the 
    \textsc{linearly separable} benchmark in 2 and 3 dimensions. The left two plots for the MNIST data correspond to 2d and 3d \textsc{mnist pca} data, and the rightmost image shows examples from the \textsc{mnist-cg} dataset for 32 x 32, 16 x 16, 8 x 8 and 4 x 4 pixel grids. The hidden manifold examples correspond to a $1$d (left) and $2$d (center) and $3$d (right) manifold embedded into $3$ dimensions. The bars and stripes panel shows examples from the \textsc{bars \& stripes} dataset for a 16 x 16 pixel grid. The examples from the \textsc{two curves diff} benchmark show a degree of $2, 10, 20$ for the Fourier series, embedding the curves into $10$ dimensions (of which three are plotted). The hyperplanes pannel shows data from the \textsc{hyperplanes diff} benchmark, where there are two (left) and five (right) hyperplanes used to decide the class labels.}
    \label{fig:data}
\end{figure*}

    Choosing meaningful datasets for general benchmarking studies is difficult, and, as discussed in Section~\ref{sec:background}, can have a huge impact on the findings. For example, should we use datasets that suit the inductive bias of the quantum models, since these would be likely future applications? Shall we use small datasets that were relevant for machine learning in the 1990s? Shall we use popular current-day benchmarking tasks and reduce them to manageable scales? Should we focus on data in the form of quantum states \citep{schatzki2021entangled, perrier2022qdataset}? While we do not claim to provide satisfying answers to these questions -- an endeavour that is worth a multi-year research programme and will unlikely find a single answer -- we want to make transparent the rationale behind choosing the $6$ different flavours of data that we employ in this study, and what we expect them to measure. 
    
    We followed three overarching principles: Firstly, we predominantly use artificially generated data which allows us to understand and vary the properties and size of the datasets. This may limit conclusions with respect to ``real-world'' data, but is an essential ability in the early research stage that quantum machine learning is in. Secondly, we aim at maximising the diversity of the datasets by using fundamentally different functional relationships and procedures as generators -- in the hope to increase the chance that consistent trends found in this study may be found in other data as well. Thirdly, in the last three out of six data generation procedures introduced below we follow the ``manifold hypothesis'' \citep{bengio2013representation, narayanan2010sample, pope2021intrinsic}, which states that typical data in modern machine learning effectively lives on low-dimensional manifolds.

    With this in mind we define $6$ data generation procedures, with which we generate data for $10$ benchmarks (in the following named in capital letters). Each benchmark consists in turn of several datasets that differ by varying parameters in the data generation procedure (in most cases the input dimension). Overall, the benchmarks consist of $160$ individual datasets. While the $6$ data generation procedures and their associated benchmarks are summarised in the following list and illustrated in Figure~\ref{fig:data}, the precise generation settings can be found in Appendix~\ref{app:data}.
    \begin{enumerate}
        \item \textit{Linearly separable}. This data generation procedure consists of linearly separable data and serves as the ``fruit-fly'' example of learning: it is easy to understand and has well-defined performance baselines, since it is known to be solvable by a perceptron model -- a neural network with one layer and output ``neuron'' -- since the early days of artificial intelligence research \citep{rosenblatt1958perceptron, minsky2017perceptrons}\footnote{Note that a subtlety here is the size of the margin between the classes in increasing dimensions, which has an influence on how easy it is to generalise from the training data. }. The datasets are generated by sampling inputs uniformly from a $d$-dimensional hypercube and dividing them into two classes by the hyperplane orthogonal to the $(1, ...., 1)^T$ vector (including a small data-free margin). The benchmark that we will refer to as \textsc{linearly separable} consists of $19$ datasets that vary in the dimension $d=2,...,20$ of the input space.    
        \item \emph{Bars and stripes.} As a second ``fruit-fly'' task, but this time tailor-made for the convolutional models, we create images of noisy bars and stripes on 2-dimensional pixel grids. These datasets are among the simplest examples of translation invariant data and can thus be used as a sanity check of convolutional models. The data generation procedure involves sampling a number of images with values $\pm 1$,  corresponding to either bars or stripes, and adding independent Gaussian noise to each pixel value with standard deviation $0.5$. The \textsc{bars \& stripes} benchmark consists of four datasets where we vary the image size between $4\times 4$, $8\times 8$, $16\times 16$ and $32\times 32$. 
        \item \textit{Downscaled MNIST}. While we cautioned against the use of downsized MNIST datasets in Section~\ref{sec:mnist}, we want to report on this ubiquitous dataset here for the sake of comparability with other studies. We define three benchmarks: For the quantum neural network models we use Principal Component Analysis (PCA) to reduce the dimensions to $d=2,...,20$ (\textsc{mnist pca}). For quantum kernel methods, which need to simulate up to $N(N-1)/2$ quantum circuits during training if $N$ is the number of training samples, $250$ training and test points are subsampled from the \textsc{mnist pca} datasets (\textsc{mnist pca}$^-$). For the CNN architectures we reduce the resolution of the images by ``coarse-graining'' or extending the images to size $4\times4$, $8\times 8$, $16\times 16$, and $32 \times 32$ in order to keep the spatial pattern of the data intact (\textsc{MNIST-CG}). The three benchmarks consist of $42$ datasets in total.
        \item \textit{Hidden manifold model}. \citet{goldt2020modeling} introduced this data generation procedure as a means to probe the effect of structure in the data, such as the size of a hidden manifold conjectured to control the difficulty of the problem, on learning. In particular, it allows the analytical computation of average generalisation errors, a property that could be of interest in quantum machine learning beyond this study. We generate inputs on a low-dimensional manifold and label them by a simple neural network initialised at random. The inputs are then projected to the final $d$-dimensional space. We generate two benchmarks this way: \textsc{hidden manifold} varies only the dimension $d=2,...,20$ and keeps the dimension of the manifold at $m=6$, while \textsc{hidden manifold diff} keeps the dimension constant at $d=10$ and varies the dimensionality $m$ of the manifold between $m=2,...20$; in total we produce $38$ datasets.
        \item \textit{Two curves}. This data generation procedure is inspired by a theoretical study \citep{buchanan2021deep} that proves how the performance of neural networks depends on the curvature and distance of two $1$-dimensional curves embedded into a higher-dimensional space. Here we implement their proposal by using low-degree Fourier series to embed two sets of data sampled from a 1-d interval -- one for each class -- as curves into $d$ dimensions while adding some Gaussian noise. The curves are embedded using identical functional relationships, except from an offset applied to one of them, which controls their distance. We generate two benchmarks with in total $38$ datasets. \textsc{two curves} fixes the degree to $D = 5$, offset to $\Delta = 0.1$ and varies the dimension $d=2,..,20$. \textsc{two curves diff} fixes the dimension $d=10$ and varies the degree $D$ of the polynomial $D=2,...,20$ while adapting the offset $\Delta=\frac{1}{2D}$ between the two curves. 
        \item \textit{Hyperplanes and parity}. Finally, we devise a data generation procedure that fixes several hyperplanes in a $k$-dimensional space and labels randomly sampled points consulting the parity of perceptron classifiers that have these hyperplanes as decision boundaries. In other words, a label tells us whether a point lies on the ``positive'' side of an even number of hyperplanes. The motivation for this data generation procedure is to add a labeling strategy that requires information about the ``global'' structure of the problem, i.e. the position of all hyperplanes. A single benchmark, \textsc{hyperplanes diff}, fixes the dimension of the data to $d=10$ and varies the number $k$ of hyperplanes $k=2,...,20$ defined on a $3$-dimensional submanifold.
    \end{enumerate}

    \begin{figure}
        \centering
        \includegraphics[width=0.4\textwidth]{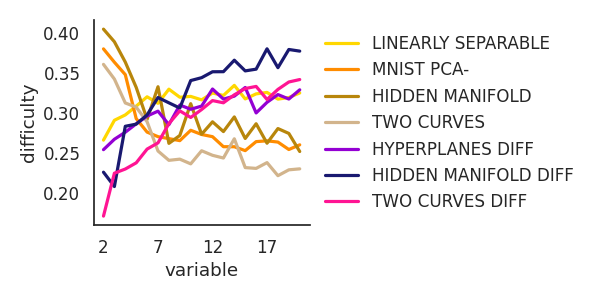}
        \caption{Average difficulty score with respect to the 22 difficult measures computed by the ECol package\footnote{https://github.com/lpfgarcia/ECoL} introduced in \citet{lorena2018data} for some of the datasets. The datasets ending in ``DIFF'' (blue, red and green curves) depend on a variable between $2$ and $20$ that we claim controls their difficulty, which is supported by the quantifier shown here. The other datasets vary the dimension, and -- with the curious exception from \textsc{linearly separable} -- decrease in difficulty when the input space gets larger. Note that the measures exhibit a huge variance, and the results from this or other data complexity measures should be interpreted with care.}
        \label{fig:profiling}
    \end{figure}

    As seen, some of the $10$ benchmarks consist of datasets that vary the input dimension where others vary parameters that supposedly control the complexity of the data. While the controversial debate about the best way of quantifying the complexity of data (for example, \citep{guan2022novel, smith2014instance, sotoca2005review}) lies outside of the scope of this paper, we give some support to the claim of an increasing complexity by reporting the average difficulty score of the measures proposed in \citet{lorena2018data} extending a seminal paper from 2002 \citep{ho2002complexity} in Figure~\ref{fig:profiling}. 

\section{Hyperparameter tuning}\label{sec:pipeline}

Hyperparameter optimisation is one of the most important steps in classical machine learning to allow models to reveal their true potential. Likewise, quantum machine learning models tend to show a wide variety in performance depending on hyperparameters such as the number of layers in the ansatz. Even seemingly inconspicious hyperparameters such as the learning rate of an optimiser can influence generalisation errors significantly \citep{Moussa2023}. As mentioned in Section~\ref{sec:rigour}, including a wide enough hyperparameter search grid can make baseline models match the performance of state-of-the-art methods. 

Hyperparameter optimisation is also one of the most cumbersome aspects of training (quantum) machine learning algorithms since it involves searching a large configuration space that is not amenable to gradient-based optimization\footnote{Although recently, to address such issues, techniques such as implicit differentiation have been adapted to perform gradient-based optimization of continuous-valued hyperparameters~\citep{pmlr-v108-lorraine20a, ahmed2022implicit}, but exploring these techniques exceed the scope of this study.}, while each point or hyperparameter configuration involves several training runs during procedures like cross-validation.

\begin{figure}
    \centering
    \includegraphics[width=.46\textwidth]{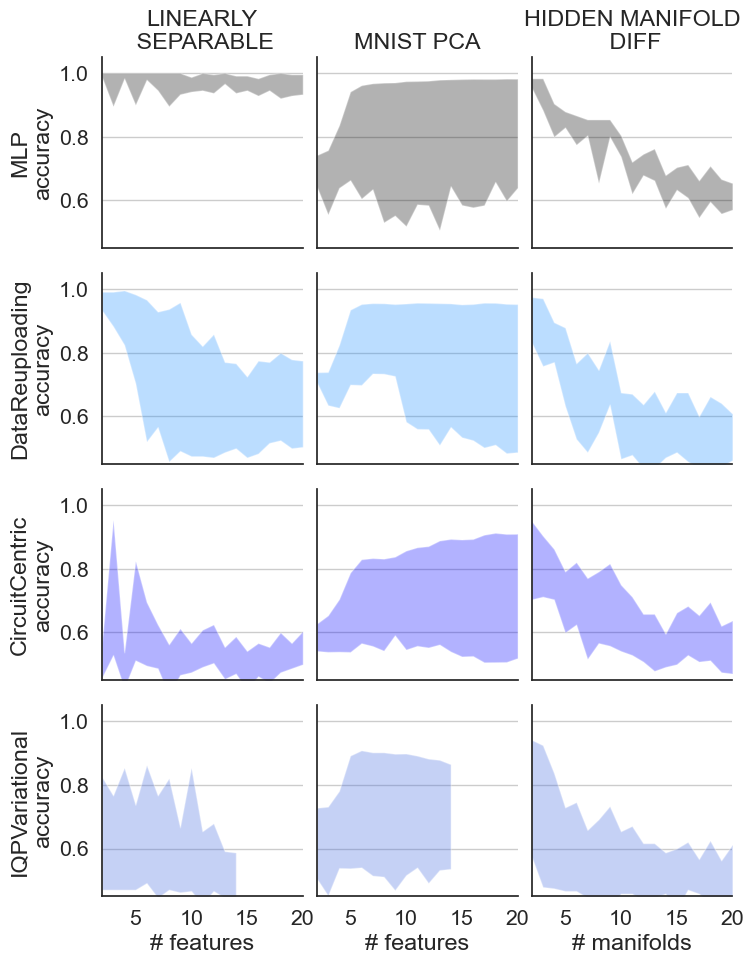}
    \caption{Ranges of test accuracies of models during grid search to find the best hyperparameters. There is a large variation in the results which becomes very important to consider as poor hyperparameter choices could lead to misleading conclusions about the power of (quantum) machine learning models.}
    \label{fig:hp-all-variations}
\end{figure}

We conduct an extensive hyperparameter search for all models and datasets in all our experiments, using a full grid search algorithm implemented by the \textit{scikit-learn} package~\citep{scikit-learn} with the default five-fold cross-validation, using the accuracy score\footnote{The accuracy can be an overly simplistic measure for some classification tasks and especially with unbalanced data \citep{provost1998case}. However, we utilise it here for its clarity in interpretation, and since our datasets are balanced.} to pick the best model. While there are more sophisticated techniques~\citep{optuna_2019, vizier}, a full grid search has the advantage of simplicity, and allows us to extract and study unbiased information about the hyperparameter landscapes. We remark that the number of hyperparameters varies significantly from model to model since our aim was to follow the proposals in the original papers. In some cases, we were forced to select a subset due to the exponential increase in grid size with the number of hyperparameters. In these cases, the most relevant hyperparameters were selected by analysing preliminary results first on smaller-scale data.  We describe the hyperparameter ranges for each model in Appendix~\ref{app:models} and summarise a few findings of potential interest here.

\begin{figure}
    \centering
    \includegraphics[width=0.40\textwidth]{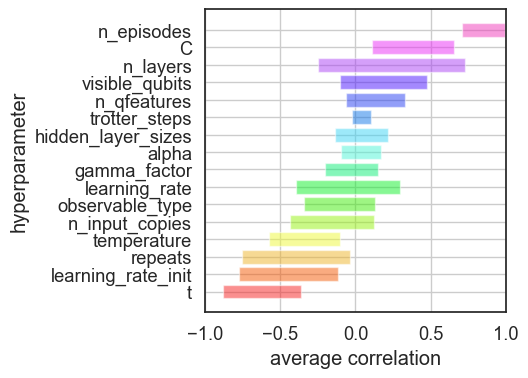}
    \caption{Correlation between values chosen for a particular hyperparameter and test accuracy during cross-validation. We show aggregated information across all classifiers, datasets and runs during grid search. Note that some hyperparameters only appear in a single model, while others -- such as the number of variational layers ({\emph{n\_layers}}) -- are averaged over several models.}
    \label{fig:hp-all-correlations}
\end{figure}

\begin{figure}
    \centering
    \includegraphics[width=.35\textwidth]{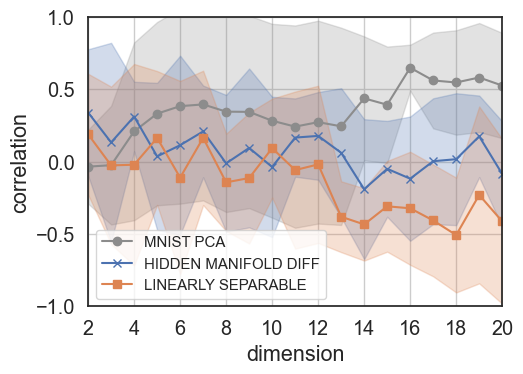}
    \caption{Correlation of the number of quantum layers and test set classification accuracy during cross-validation; averaged across classifiers that use this hyperparameter and all datasets in the three benchmarks reported. While intuitively, more layers should lead to better performance when the dimension of the inputs (\textsc{mnist pca}/\textsc{linearly separable}) or manifold (\textsc{hidden manifold diff}) grows, we see that the results can vary depending on the benchmark.}
    \label{fig:hp-corr-vs-dim}
\end{figure}

First, the performance of both classical and quantum models varies significantly across hyperparameter choices: Figure~\ref{fig:hp-all-variations} shows the ranges in the test accuracy during grid search for a select few classifiers and benchmarks, some of which lie between almost random guessing and near-perfect classification. This makes hyperparameter tuning crucial: a highly optimised architecture of one model type can easily outperform another model that has been given less resources to select the right configuration.

We also show the correlations (using the Pearson correlation coefficient) between the accuracy and hyperparameter values in Figure~\ref{fig:hp-all-correlations} to indicate the relative influence each hyperparameter has on the model performance. We find that aggregated over all datasets and models, some hyperparameters have a high correlation with the accuracy, e.g., increasing the number of episodes in the \texttt{QuantumKitchenSink} model seems to improve its performance, while decreasing the simulation time $t$ improves the \texttt{ProjectedQuantumKernel} model. However, the best hyperparameter choice can vary significantly with the dataset: Figure~\ref{fig:hp-corr-vs-dim} shows three different benchmarks where the correlation between the number of layers of a quantum model and the test accuracy shows very different trends. In case of the \textsc{mnist pca} benchmark, increasing the number of layers leads to higher accuracies, while for the \textsc{linearly separable} benchmark we observe the opposite effect. Both trends get stronger for higher input dimensions. At the same time, for the \textsc{hidden manifold diff} benchmark the correlation between accuracy and the number of layers is not significant. 

These simple insights from the hyperparameter study show that hyperparameter choice can be very non-intuitive, especially as models increase in size. The hyperparameter choices for a small datasets cannot be expected to be optimal for more complicated scenarios. In case of quantum models, hyperparameter exploration becomes computationally expensive even for moderate-sized models.

\section{Results}\label{sec:results}

We finally report our findings from the benchmark study. As a reminder, our goal is twofold: We were motivated to independently test the overwhelming evidence that quantum models perform better than classical ones emerging from benchmarks conducted in the literature. However, this only helps to judge where we currently are, not necessarily where the future potential of quantum models lies. A much more important question we are interested in is which ideas in current near-term quantum machine learning model design hold promise and which ones do not -- in other words, what research is worthwhile pursuing in order to use quantum computers effectively for learning. As we will see, the benchmark results give us a number of interesting clues that we will discuss in the next Section~\ref{sec:questions}.  

    \subsection{Out-of-the-box classical models outperform quantum models} 
        
\begin{figure}
    \centering
    \includegraphics[width=0.5\textwidth]{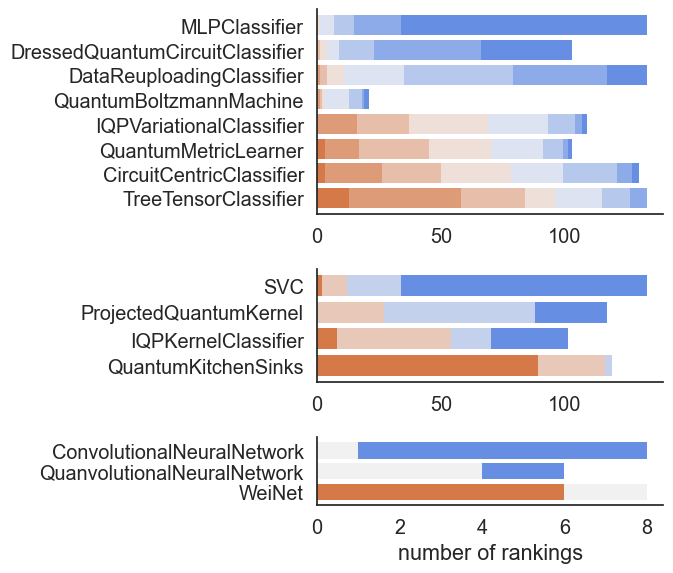} 
    \caption{Number of rankings (blue/first to red/last) across all datasets that a model was tested on, for the three model families. The models were sorted from top to bottom according to the average normalised rank. The three classical out-of-the-box models perform best. Note that the total number of benchmarks a model competed in varies due to runtime limitations, and since the convolutional architectures were only tested on \textsc{mnist cg} and \textsc{bars \& stripes}.}
    \label{fig:ranking-all}
\end{figure}

A very clear finding across our experiments was that the out-of-the-box classical models systematically outperform the quantum models. Figure~\ref{fig:ranking-all} shows the number of different rankings (first, second, and so forth) across all benchmark experiments we ran for the different models. The models within a family are sorted according to the expected normalised\footnote{We normalise the rank since not all models competed in all benchmarks, and it is easier to have a better rank when competing in experiments with fewer overall competitors.} rank: If a model came first in a benchmark with $10$ competitors (normalised ranks $0.1, 0.2,...,1$), and fourth in a benchmark with five competitors (normalised ranks $0.2, 0.4, 0.6, 0.8, 1$), its expected normalised rank is $(0.1+0.8)/2 = 0.45$. Note that this is one choice of many, and employing other reasonable ordering and aggregation mechanisms that we tested change the picture slightly, but not significantly: In all three model families, the prototypical classical models came first.

Breaking down the fully aggregated results, one finds that the rankings of models between benchmarks were surprisingly consistent (see Appendix~\ref{app:results} for a full report). For example, in four out of seven benchmarks used for QNN models, the \texttt{MLPClassifier} ranked first with the \texttt{DressedQuantumCircuitClassifier} coming second, whereas in the remaining three benchmarks the roles between these (conceptually very similar) models were reversed. The \texttt{DataReuploadingClassifier} and \texttt{QuantumBoltzmannMachine} (which was not run on the 10-d DIFF benchmarks) usually share places three and four. In five out of the seven benchmarks, the bottom ranks are taken by the \texttt{CircuitCentricClassifier} and \texttt{TreeTensorClassifier} as the worst-performing models -- interestingly the only two models based on amplitude encoding. 

For the kernel methods, the support vector machine (\texttt{SVC}) does best on 4 out of 7 benchmarks, showing a very similar behaviour to the other two SVM-based classifiers. Consistently worst-performing is \texttt{QuantumKitchenSinks}, a model that uses a quantum circuit mapping to computational basis samples as a random, non-trainable feature map.\footnote{It is important to mention how Figure~\ref{fig:hp-all-correlations} revealed that the number of episodes in the model, which controls the number of feature vectors created from quantum circuit samples, correlated positively with the test accuracy, and allowing for significantly more than $2000$ episodes may have boosted this model's performance.}

The two quantum convolutional neural networks were also outperformed by the vanilla \texttt{ConvolutionalNeuralNetwork} model. Surprisingly, \texttt{WeiNet} failed entirely to learn the \textsc{bars \& stripes} data; a task which we considered easy for a model of its kind. 

    \begin{figure*}
    \centering
    \includegraphics[width=0.5\textwidth]{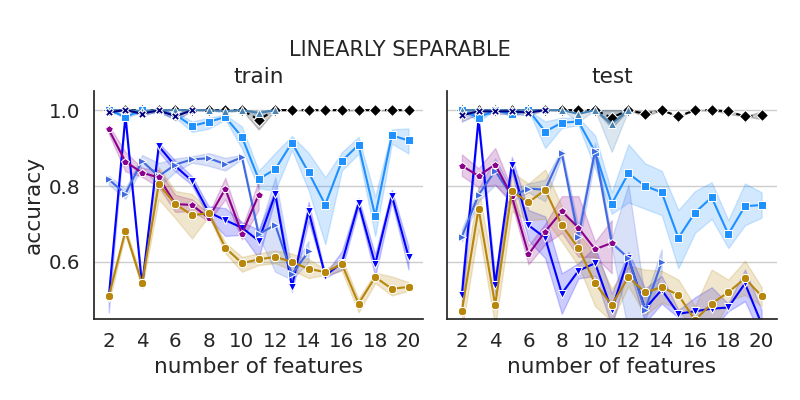}\includegraphics[width=0.5\textwidth]{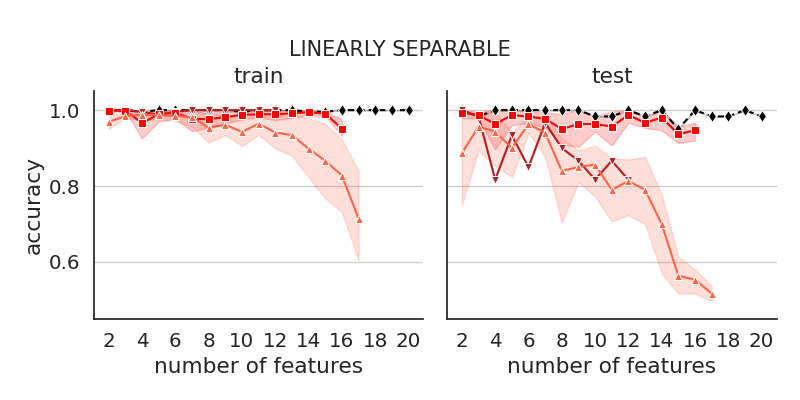}
    \includegraphics[width=0.5\textwidth]{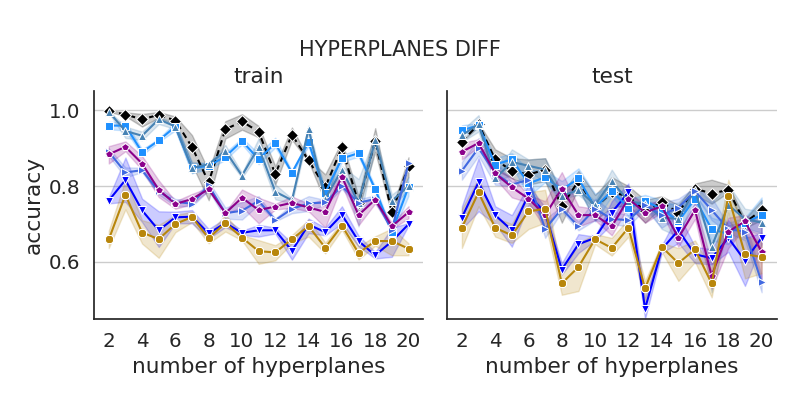}\includegraphics[width=0.5\textwidth]{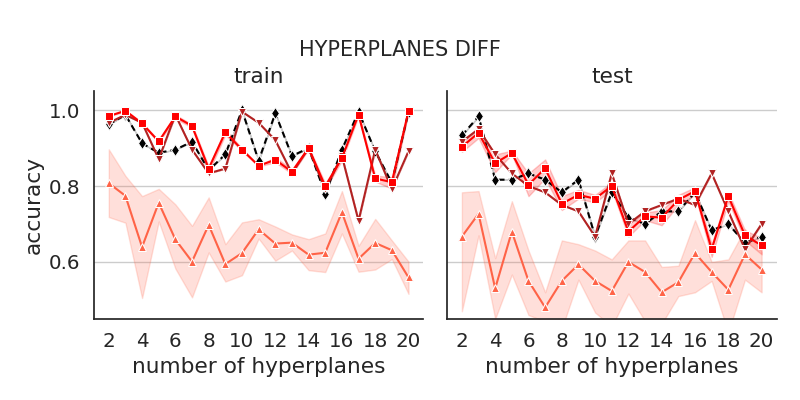}
    \includegraphics[width=0.5\textwidth]{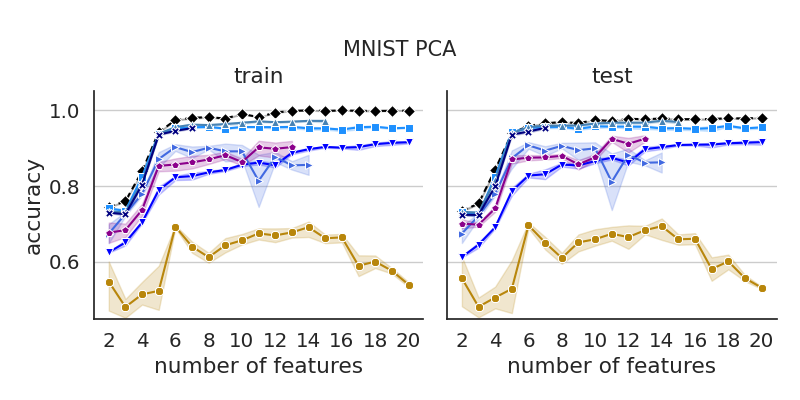}\includegraphics[width=0.5\textwidth]{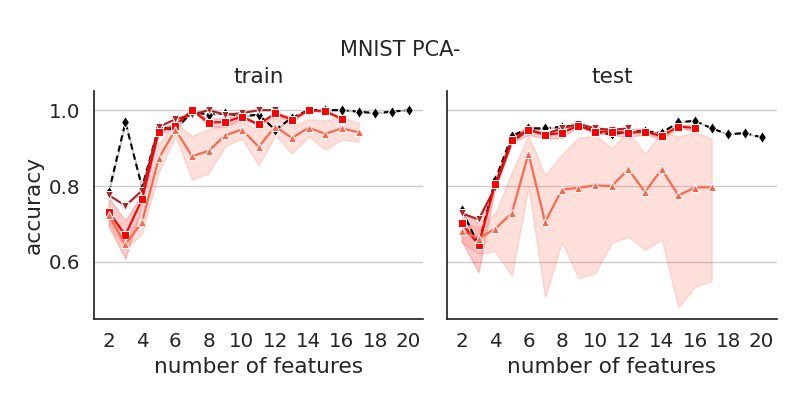}
    \includegraphics[width=0.5\textwidth]{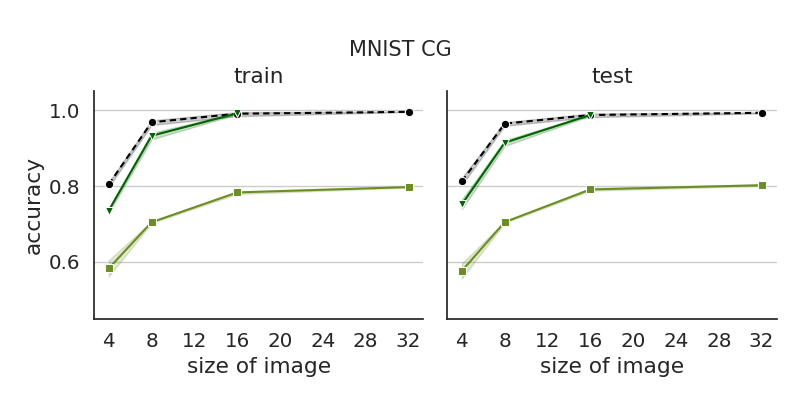}\includegraphics[width=0.5\textwidth]{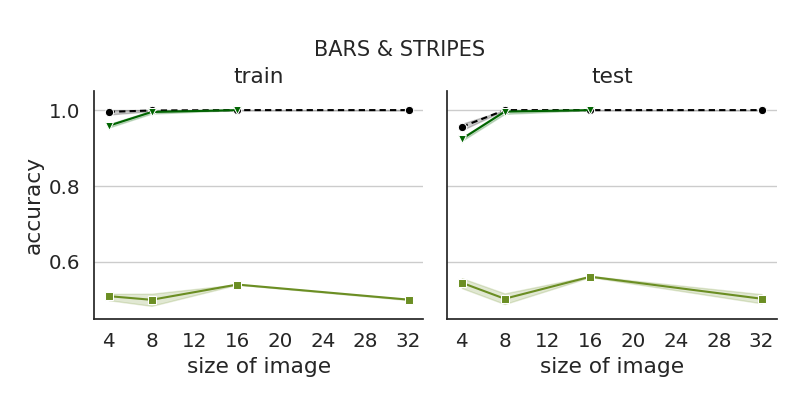}
    \includegraphics[width=0.3\textwidth]{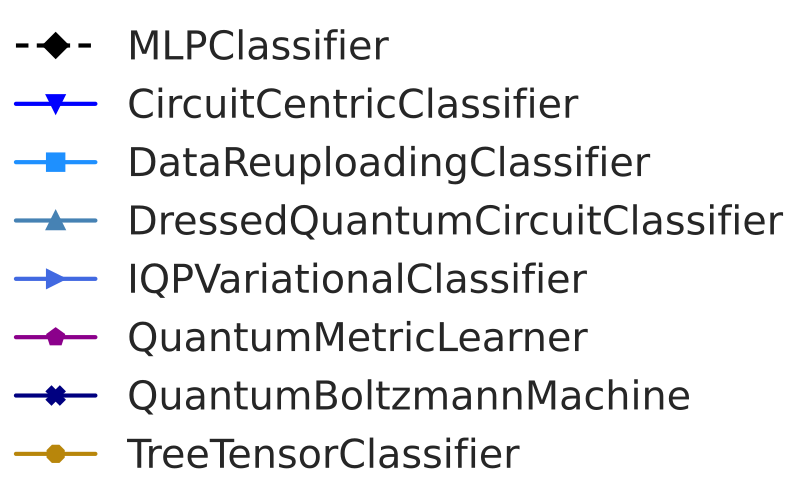}~~~~~~~~~~~~~\includegraphics[width=0.23\textwidth]{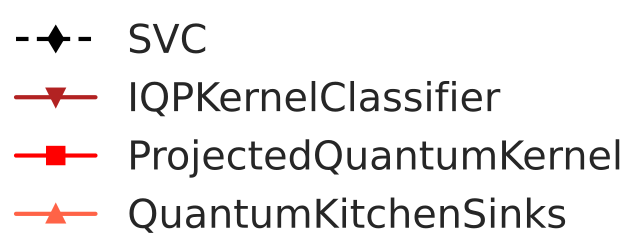}~~~~~~~~~~~~~\includegraphics[width=0.25\textwidth]{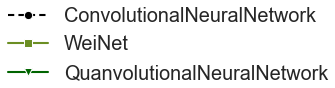}
    \caption{Selected detailed train and test accuracies for some of the benchmarks. As reflected in the aggregated results in Figure~\ref{fig:ranking-all}, the three classical baseline models usually outperform the quantum models across the quantity that is varied. However, there are nuances: while quantum models perform particularly poorly on the \textsc{linearly separable} benchmark, most of them follow the classical model performance closely on the \textsc{hyperplanes diff} benchmark. The accuracies of the QNN family on \textsc{mnist pca} mimic the trend of the classical neural network (\texttt{MLPClassifier}), but with an offset towards lower scores, while most quantum kernel methods perform as well as the SVM (\texttt{SVC}).}
    \label{fig:scores-datasets}
\end{figure*}

    \subsection{No systematic overfitting or scaling effects}

There are three interesting general observations of effects one might expect but which we did \textit{not} observe. These are exemplified with the selected detailed benchmark results shown in  Figure~\ref{fig:scores-datasets}. 
Firstly, although not all quantum models -- in particular the QNN methods -- use explicit regularisers, they do not show systematic overfitting of the training data. (Some exceptions can be seen in the \textsc{hidden manifold}, \textsc{hidden manifold diff} and \textsc{two curves diff} benchmarks shown in Appendix~\ref{app:results}, for which also the classical models struggle with overfitting.)

Secondly, we do not observe any improvement in performance of the quantum models relative to the classical baselines for increasing problem difficulties. For the difficulty-controlled benchmarks, \textsc{hyperplanes diff} (Figure~\ref{fig:scores-datasets}) and \textsc{hidden manifold diff} (Appendix~\ref{app:results}), the trends of the quantum models' test accuracies generally follow the trend of the classical model. For the difficulty-controlled benchmark \textsc{two curves diff} (Appendix~\ref{app:results}), the quantum models perform worse than the corresponding classical method as the difficulty is increased. Interestingly, for the hardest datasets from this benchmark, the classical baseline models achieve high ($>90\%$) test accuracy whereas all quantum models appear to struggle. This is somewhat surprising, since the embedded curve that defines the structure of the data is a Fourier series, and one may expect quantum models to have a natural bias for this kind of data \cite{schuld2021effect}. 

Thirdly, except from the \textsc{linearly separable} benchmark we do not observe a significant scaling effect with the size of the dimension; also here quantum models do not get significantly better or worse in performance compared to the classical baseline.

    \subsection{Quantum circuits without entanglement do well}

An important question for quantum machine learning benchmarks is how the performance of a model depends on properties that we consider to be ``quantum''. There are many different definitions of this notion (such as ``experiments that produce non-classical statistics''), and without being explicitly stated very often, the definition of ``not classically simulatable'' dominates the thinking in the quantum machine learning community. An experimental design to probe the question is therefore  to replace the quantum model architecture with a circuit that is classically tractable (i.e., it can be simulated at scale) and measure if the performance deteriorates. If not, we have evidence that other properties than ``quantumness'' are responsible for the performance we see -- at least in the small-scale datasets chosen here. 

To put ``quantumness'' to the test in our benchmarks we add the \texttt{SeparableVariationalClassifier} and \texttt{SeparableKernelClassifier} described in Section~\ref{sec:models} to our quiver of non-convolutional models. These are fully disentangled $n$-qubit models that can be divided into $n$ separate single-qubit circuits, and hence easily classically simulated at scale. We do not add a separable convolutional model since the \texttt{ConvolutionalNeuralNetwork} itself can be seen as a special case of one, since it is equivalent to replacing the quantum layer of the \texttt{QuanvolutionalNeuralNetwork} model by a circuit that implements the identity transformation. We already know that this model does not perform worse than the entanglement-using \texttt{QuantumConvolutionalNeuralNetwork} and \texttt{WeiNet}.

\begin{figure}
    \centering
    \includegraphics[width=0.24\textwidth]{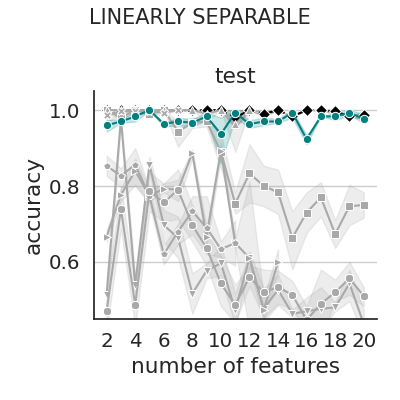}\includegraphics[width=0.24\textwidth]{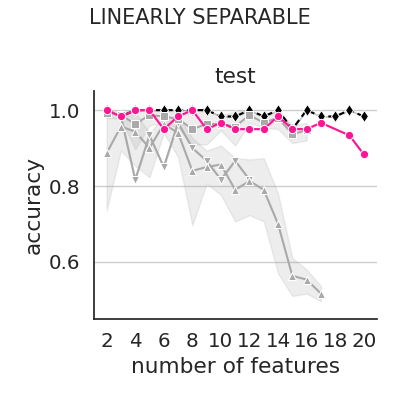}
    \includegraphics[width=0.24\textwidth]{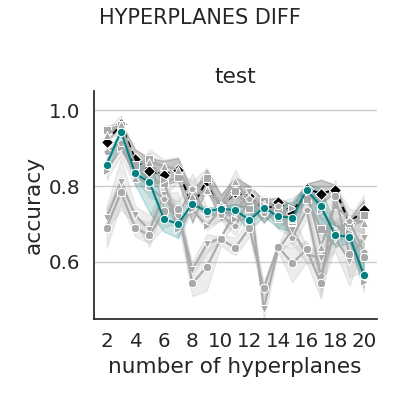}\includegraphics[width=0.24\textwidth]{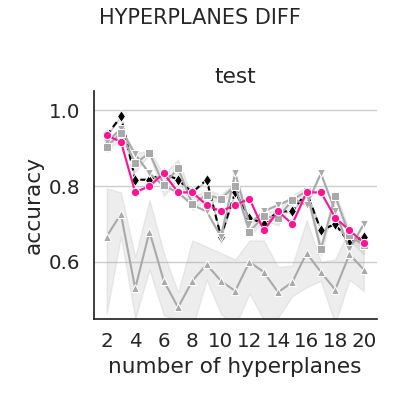}
    \includegraphics[width=0.24\textwidth]{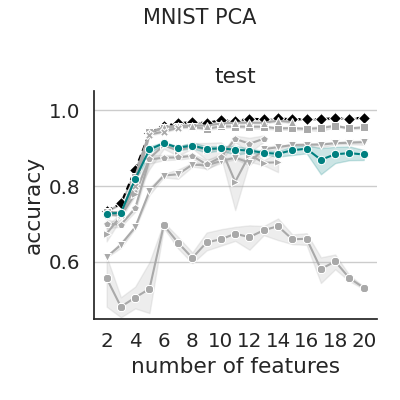}\includegraphics[width=0.24\textwidth]{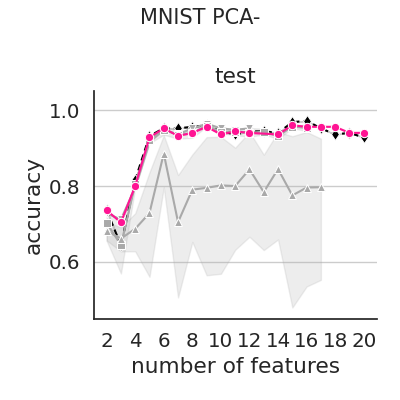}
    \includegraphics[width=0.25\textwidth]{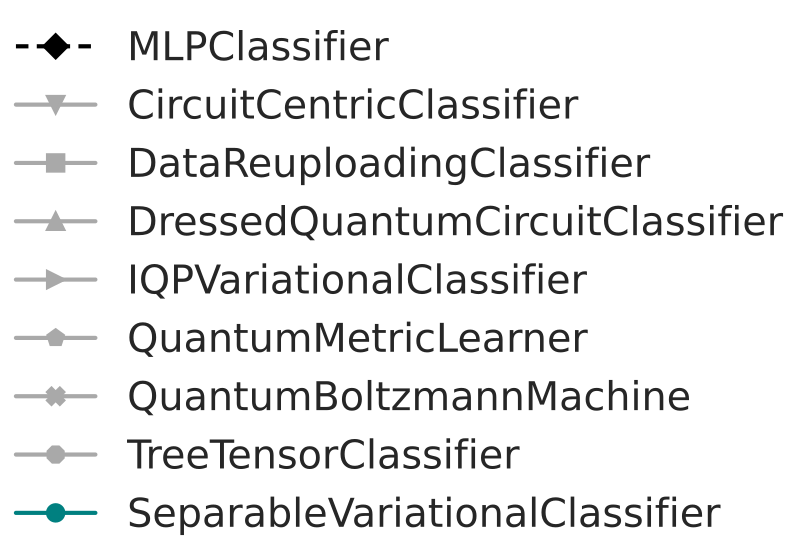}\includegraphics[width=0.2\textwidth]{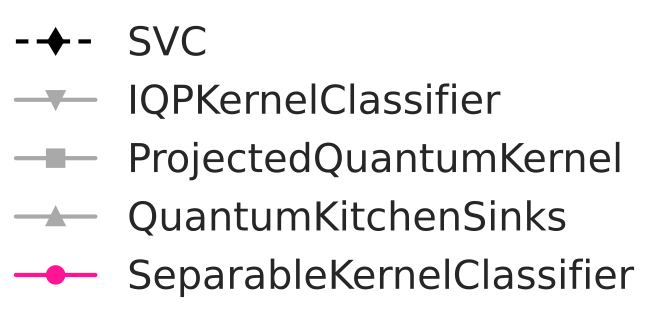}
    \caption{Replotting the test accuracies from Figure~\ref{fig:scores-datasets} while adding \texttt{SeparableVariationalClassifier} and \texttt{SeparableKernelClassifier}. These fully classically simulatable models perform similarly or better than most other quantum models.}
    \label{fig:scores-separable}
\end{figure}

Replotting the test accuracies from Figure~\ref{fig:scores-datasets} with the new models added we see in Figure~\ref{fig:scores-separable} that the non-entangled models do surprisingly well. This can be confirmed by including the separable models in the ranking results from Figure~\ref{fig:ranking-all} (see Appendix~\ref{app:results}). One finds that compared to the quantum kernel methods, the \texttt{SeparableKernelClassifier} takes second rank after the \texttt{SVC}. Among the QNNs, the \texttt{SeparableVariationalClassifier} is only consistently beaten by the \texttt{MLPClassifier}, \texttt{DressedQuantumCircuitClassifier} and \texttt{QuantumBoltzmannMachine}, as well as occasionally by \texttt{DataReuploadingClassifier}.  

\begin{figure}
    \centering
    \includegraphics[width=0.40\textwidth]{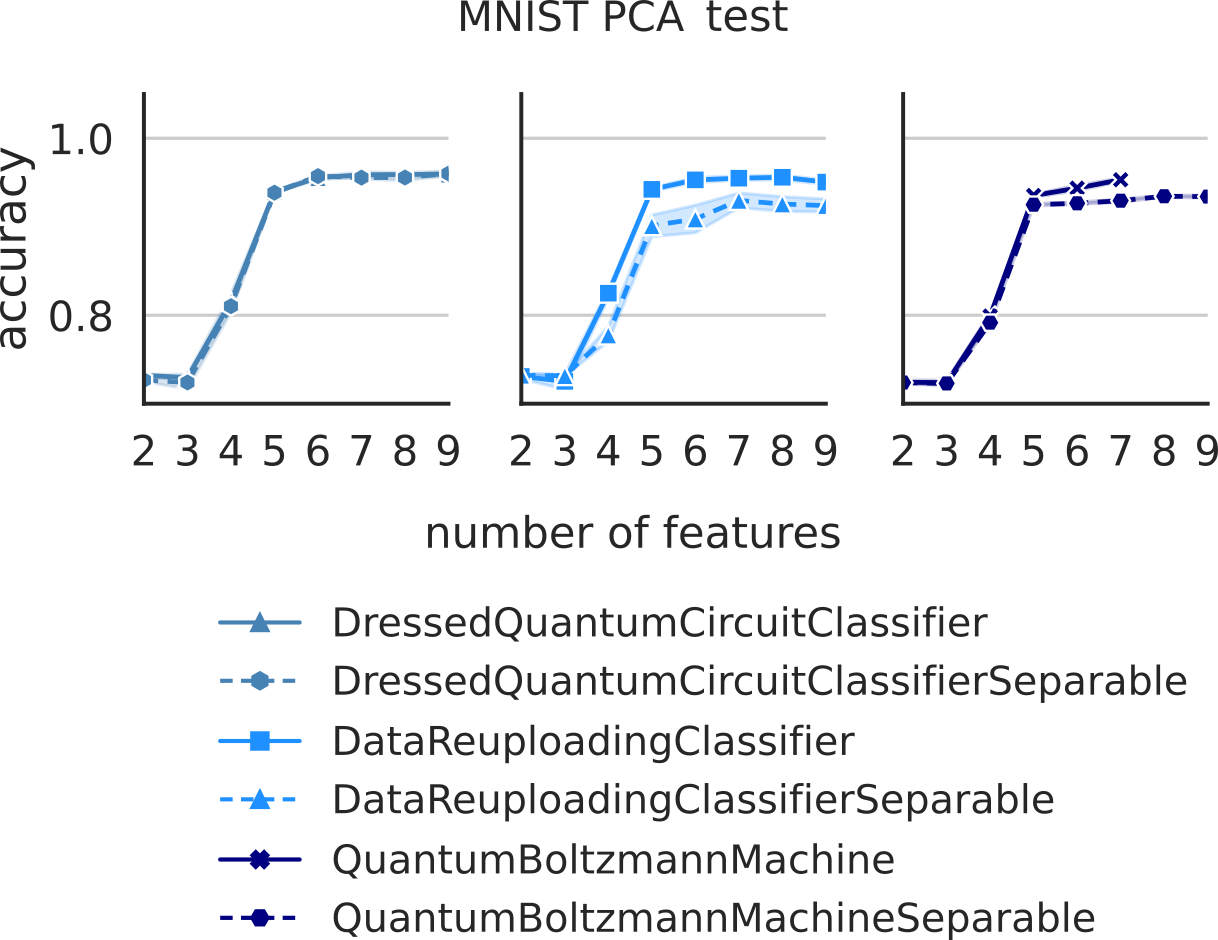}
    \caption{Comparison of the test accuracy of the three QNN models that performed better than the \texttt{SeparableVariationalClassifier}, shown on \textsc{mnist pca} up to 9 dimensions, with variations of the models that remove any entangling operations from the circuits. The \texttt{DressedQuantumCircuitClassifier} shows no drop in performance, while \texttt{DataReuploadingClassifier}, and to some extent \texttt{QuantumBoltzmannMachine}, do worse without entanglement. }
    \label{fig:drc-dqcc-separable}
\end{figure}

Are these three QNNs better than our disentangled QNN because of their entanglement, or is this due to other design choices? Figure~\ref{fig:drc-dqcc-separable} compares the original implementations of these models with variations that remove any entangling gates or measurements from the quantum circuits they use.\footnote{We chose to run this experiment on \textsc{mnist pca} dimensions as the quantum models showed a prototypical performance with respect to their overall rankings, and the variations between individual datasets of different input dimensions were small.} The results suggest that the entangling gates do not play a role in the top-performing \texttt{DressedQuantumCircuitClassifier}. However, removing entanglement \textit{does} decrease the test accuracy of \texttt{QuantumBoltzmannMachine} and \texttt{DataReuploadingClassifier}. Whether the ``quantumness'' of the entangling gates is the deciding factor, or whether the removal of certain gates could be mitigated by a better non-entangling design that enriches the expressivity of the models is an important subject for further studies. 

\section{Questions raised about quantum model design}\label{sec:questions}

Benchmarks cannot only give us evidence on which model is better than another, but open up paths to more qualitative questions, for example by systematically removing parts of a model, or by visualising simple cases. We want to give a few examples here.

\subsection{Do quantum components in hybrid models help?}

\begin{figure}
    \centering
    \subfloat[Transformation of a 2-dimensional moons dataset throughout the trained layers of the \texttt{DressedQuantumCircuitClassifier} (top row), compared to a model where we replaced the quantum circuit by another classical layer (bottom row).]{
    \includegraphics[width=0.9\columnwidth]{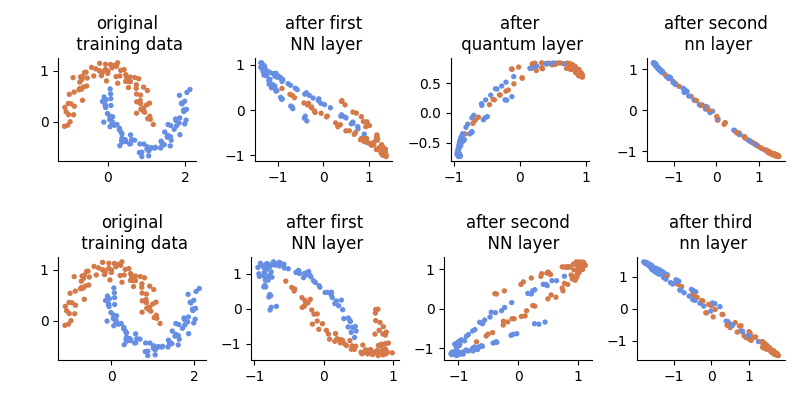}
    \label{fig:nn-layers}}\\
    \subfloat[The left-most images are two examples of the \textsc{MNIST-CG} data for a $16\times 16$ pixel grid. The three images to the right show three channels of the data after being feature mapped by the initial quantum layer of \texttt{QuanvolutionalNeuralNetwork}. The quantum feature map appears to produce different, noisier versions of the input image.]{
        \includegraphics[width=0.9\columnwidth]{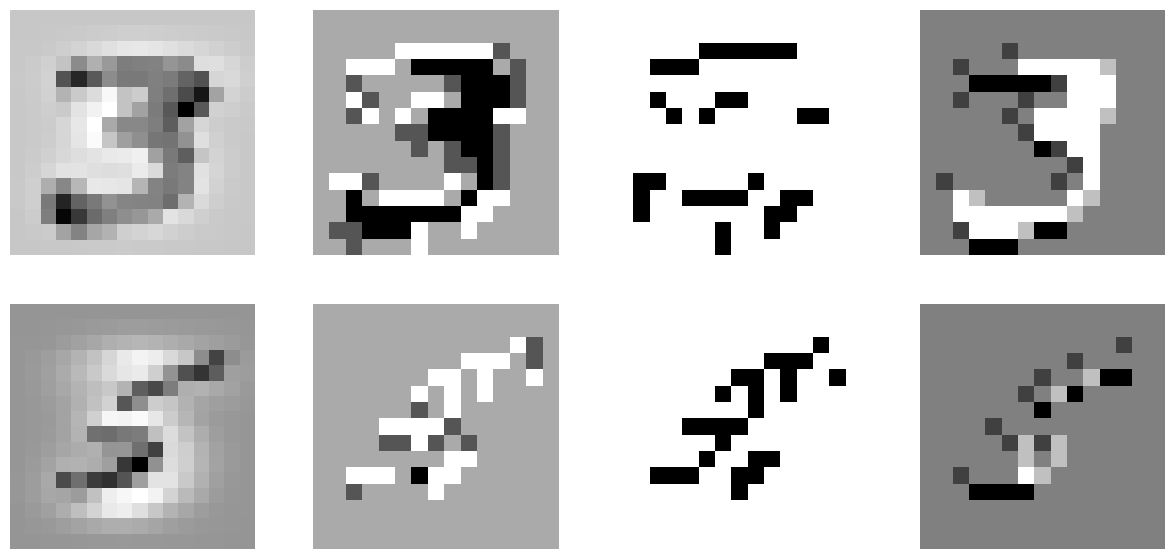}
    \label{fig:quanv-feature-map}
    }
    \caption{The effect of quantum layers in hybrid classical-quantum neural network models.}
\end{figure}

By far the best QNN model is the \texttt{DressedQuantumCircuitClassifier}, which replaces a layer of a neural network with a standard variational quantum model. The central question for such a hybrid model is whether or not the ``quantum layer'' plays a qualitatively different role to a possible classical layer. Figure~\ref{fig:nn-layers} shows the input transformations of the two neural network layers and the quantum layer for a very simple 2d dataset, and compares it with the same model in which we exchanged the quantum layer by a neural network of the same architecture as the first layer. In this small experiment, the qualitative effect of both kinds of models is similar, namely to reshape the data to a one-dimensional manifold that facilitates the subsequent linear classification. This is consistent with the fact that in most experiments, the \texttt{DressedQuantumCircuitClassifier}'s performance followed the classical neural network closely. 

The \texttt{QuanvolutionalNeuralNetwork} is a hybrid model of similar flavour since it adds a quantum circuit as the first layer to a convolutional neural network. In Figure~\ref{fig:quanv-feature-map} we show the result of this layer for a model with $\emph{n\_qchannels}=3$ for two input examples of $16 \times 16$ pixels. It is unclear if the first quantum layer is generally useful for learning from image data, since in most cases the map seems to simply create a noisy version of the original image. Given that the model performs worse than the \texttt{ConvolutionalNeuralNetwork} (at least on the small datasets we were able to get results for), it seems that this feature map actually degrades the data so that it is subsequently more difficult for the classical convolution to learn from. 

For the wide range of studies into hybrid quantum-classical neural network architectures, an important question is therefore whether the quantum layer introduces a genuinely different kind of transformation that helps the layer-wise feature extraction, or if it is simply `not doing much harm'.

    \subsection{What makes data reuploading work?}

Besides the hybrid neural network architectures (and the computationally costly \texttt{QuantumBoltzmannMachine} for which we unfortunately only have limited data available) the \texttt{DataReuploadingClassifier} performs relatively well compared to other quantum neural networks of a similar design. What features of the model explain these results? 

While the term ``data reuploading'' is often used rather generally to describe the layer-wise repetition of an encoding, there are a few other distinctive features in the original model we implemented here. For example, the inputs are individually rescaled by classical trainable parameters before feeding them into quantum gates, which in the picture of quantum models as a Fourier series \citep{schuld2021effect} re-scales the frequency spectrum that the Fourier series is built from. Furthermore, there is no separation between the embedding and variational part of the circuit; instead the embedding layer is trainable (see Eq.~\ref{reupload_gate} in App.~\ref{app:models}), leading to a sort of ``trainable data encoding'' (a feature that the \texttt{QuantumMetricLearner} also exhibits). Furthermore, the cost function differs from standard approaches as it measures the fidelity to certain states associated with a class label, and contains more trainable parameters. Which of these features is important for the success of the model - or is it a combination of them? 

\begin{figure}
    \centering
    \includegraphics[width=0.25\textwidth]{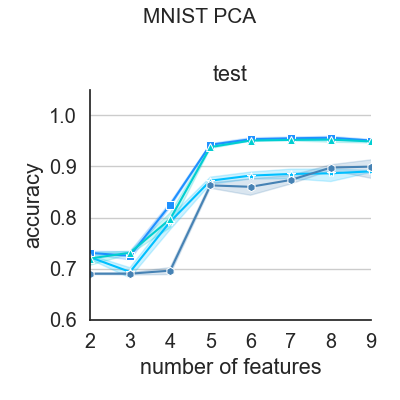}
    \includegraphics[width=0.4\textwidth]{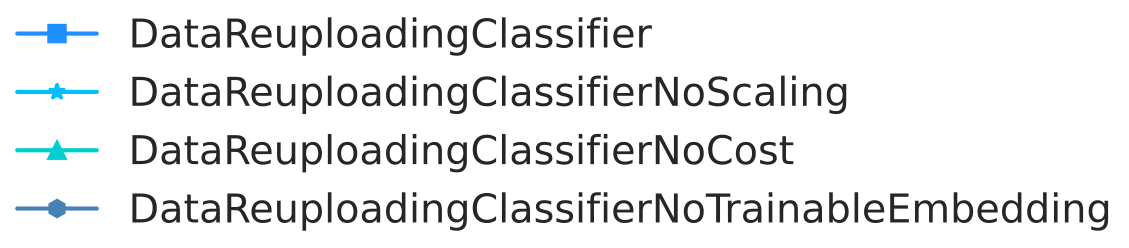}
    \caption{Comparison of the test accuracy of the Data Reuploading Classifier on \textsc{mnist pca} up to 9 dimensions with three modifications of the original implementation: The \texttt{NoCost} variation replaces the special cost function with a standard cross entropy cost function, the \texttt{NoScaling} version removes the trainable parameters multiplied to the inputs, and the \texttt{NoTrainableEmbedding} variation first applies all data encoding gates and then all variational gates. While the former has only a small influence on the performance, the latter two seem to both change the accuracy scores of higher dimensions significantly. }
    \label{fig:drc-ablation}
\end{figure}

As a first probe into this question, Figure~\ref{fig:drc-ablation} shows the test accuracy of the \texttt{DataReuploadingClassifier} when we remove the three properties -- the trainable rescaling, the mixing of variational and embedding circuits, and the peculiar cost function -- individually from the model in a small ablation study. We use the \textsc{mnist pca} benchmark up to $9$ dimensions once more. 

The results suggest that both the trainable re-scaling and embedding are crucial for performance, while the special cost function is not. This is particularly interesting, since follow-up papers often only consider the trainable embedding as a design feature -- however, the interplay of these two features may be important.

\subsection{What distance measures do quantum kernels define?}
    
We observe that the quantum kernel methods (except \texttt{QuantumKitchenSinks}, which makes very special design decisions compared to the other two) have a surprisingly similar performance to the support vector machine with a Gaussian kernel. A kernel defines a distance measure on the input space. The distance measure is used to weigh the influence of a data point on the class label of another. What distance measure do the quantum kernels define, and are they similar to the Gaussian kernel? 

\begin{figure}
    \centering
    \includegraphics[width=0.5\textwidth]{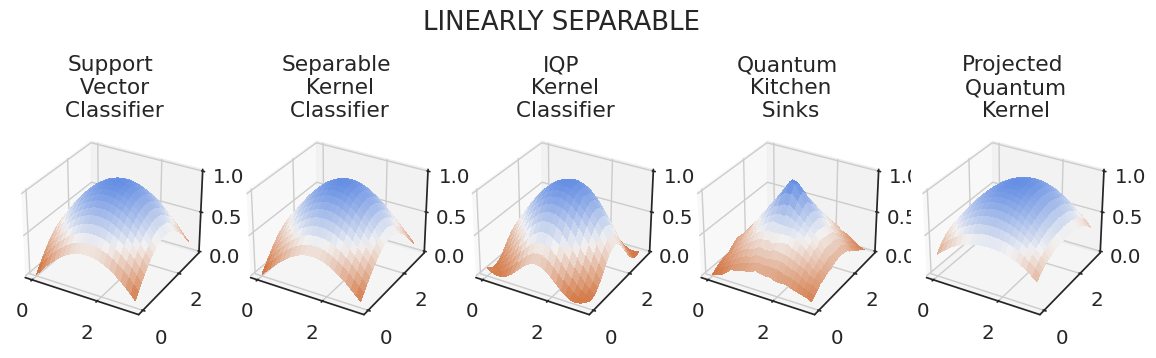}
    \includegraphics[width=0.5\textwidth]{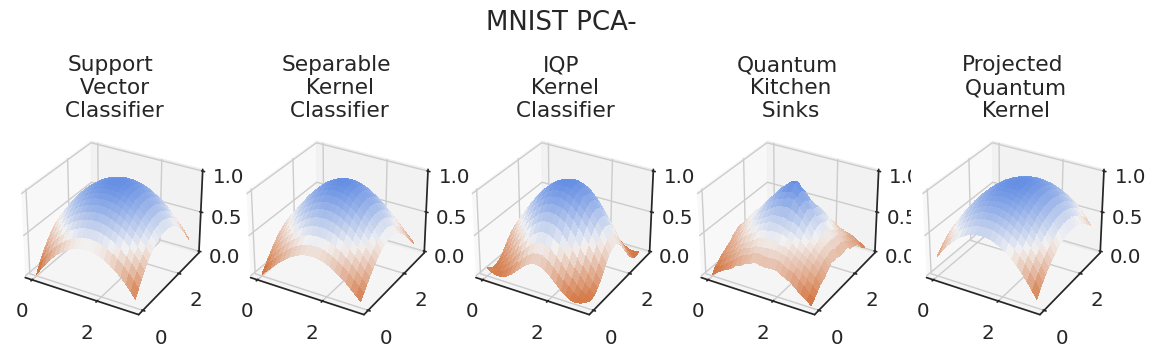}
    \includegraphics[width=0.5\textwidth]{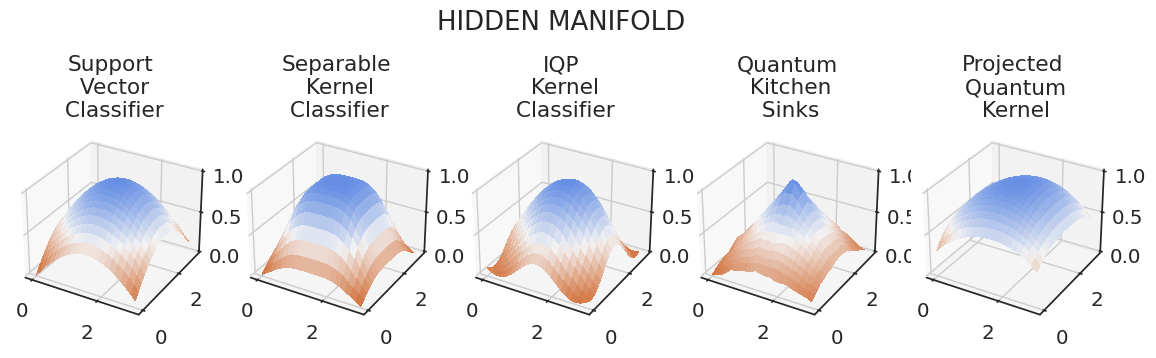}    \includegraphics[width=0.5\textwidth]{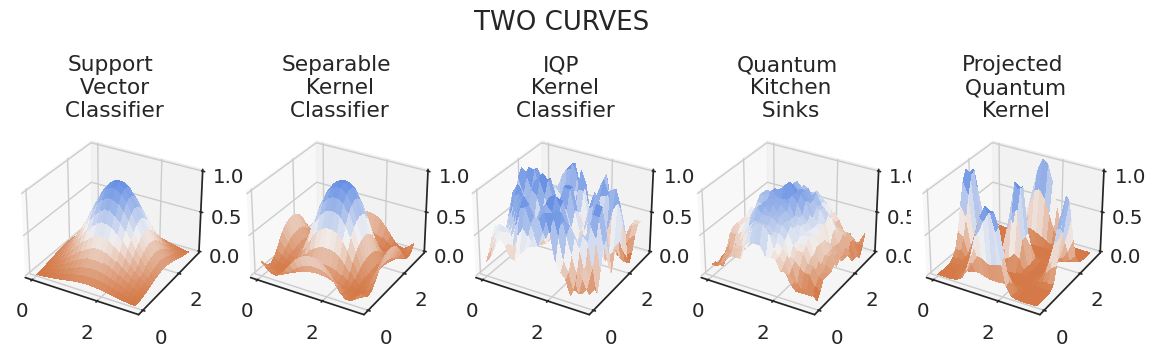}
    \caption{Kernels used by models for different 2-dimensional datasets, selecting the best hyperparameters found during grid search. The plots are generated by fixing one datapoint at $(\pi/2, \pi/2)$ in the x-y-plane and varying the other one, while plotting the kernel value for the two datapoints on the z-axis. The kernel value shows how much weight the classifier gives a potential datapoint in the x-y-plane when predicting the class of the fixed point. While there are some variations, most quantum kernel methods have a similar radial shape to the Gaussian kernel. }
    \label{fig:kernels-2d}
\end{figure}

Figure~\ref{fig:kernels-2d} shows the shape of the kernels used by models trained on 2-dimensional versions of our benchmarks. We include the \texttt{SeparableKernelClassifier} for interest, and define the kernel 
of \texttt{QuantumKitchenSinks} as the inner product of the feature vectors created by the quantum circuit. With a few occasional exceptions -- notably on \textsc{two curves}, a dataset that seems to require very narrow bandwidths in kernels and encourages quantum kernels to extend into their periodic regions -- the kernel shapes do indeed resemble the \texttt{SVC}'s Gaussian kernel.  

Does that mean that quantum kernels are just approximations to a method that has been around for decades? While 2-dimensional visualisations of the kernel function can help us gain some understanding, we need to look into higher dimensions. Here geometric visualisations become tricky to interpret, and it can be useful to compare the actual Gram matrices $G$ with entries $G_{ij} = \kappa(\mathbf{x}_i, \mathbf{x}_i)$ for pairs of training points $\mathbf{x}_i, \mathbf{x}_j$, which are used when fitting the models\footnote{We found that the test set Gram matrices do not lead to different results.}. A popular measure is the kernel alignment \citep{cristianini2001kernel} that computes the product of corresponding entries of two matrices. However, this measure makes it hard to distinguish regions in which one Gram matrix has finite values and another near-zero ones from regions where both have near-zero values. To achieve a more insightful comparison we rescale the Gram matrices to have entries in $[0, 1]$ and use the distance 
$$d(G|G') = \frac{\sum_{ij} (G_{ij} - G'_{ij})^2}{|G|}, $$
where $|G|$ refers to the number of entries in $G$ (which is the same as $|G'|$). This can be seen as a ``difference measure'' where $0$ signals identical Gram matrices, and $1$ maximally different ones. 

\begin{figure}
    \centering
    \includegraphics[width=0.48\textwidth]{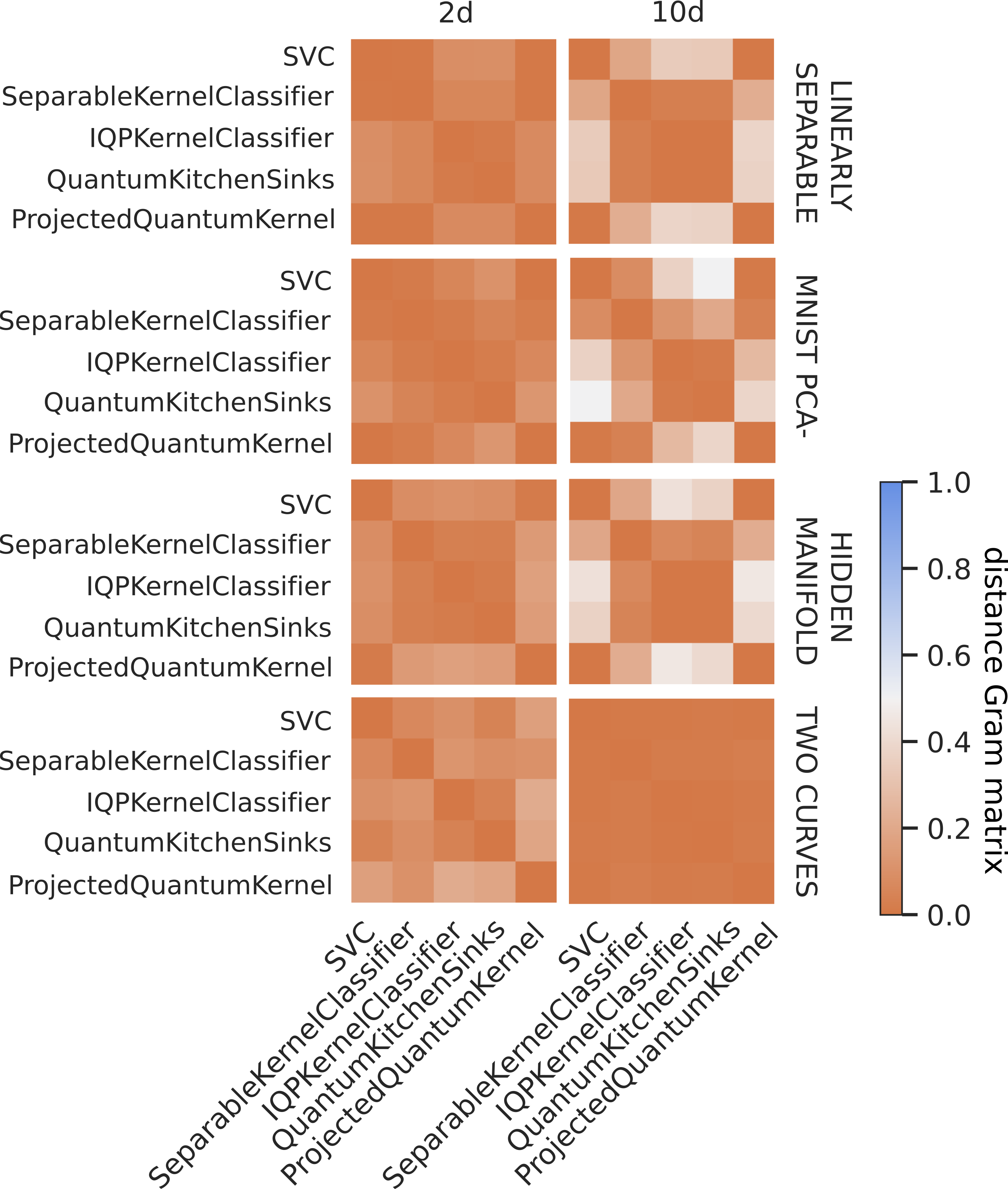} 
    \caption{Average squared difference $d(G|G')$ between the entries of the Gram matrices of different kernels. The Gram matrices are constructed from the training data using the best hyperparameters found during grid search for the particular dataset. Here we show the four benchmarks that contained datasets in both $2$ and $10$ dimensions. In higher dimensions, the Gram matrices of quantum models tend to look similarly, with the exception of the \texttt{ProjectedQuantumKernel} model, that in turn tends to resemble the \texttt{SVC}. }
    \label{fig:gram-difference}
\end{figure}

The results, of which some representative examples in 2d versus 10d are shown in Figure~\ref{fig:gram-difference}, give a slightly different picture that can only faintly be seen in the 2d kernel plots. In higher dimensions, only the \texttt{ProjectedQuantumKernel} resembles the \texttt{SVC} model, while  the other three quantum kernels resemble \textit{each other}.\footnote{An exception is the \textsc{two curves} benchmark, where all Gram matrices are similar.} Does this mean that the projected quantum circuit is not so much responsible for learning, but rather the subsequent Gaussian kernel applied to the features computed by the quantum circuit? The other three quantum kernels, in turn, produce very similar Gram matrices in high dimensions. Do most ``non-projected'' quantum kernel designs share this behaviour? 

Overall, attempting to understand the distance measure induced by a quantum kernel in high dimensions, rather than only focusing on its classical intractability, is an important open challenge in which benchmarks can help.

\subsection{Why are polynomial features not working?}

Another consistent finding on the lower end of the spectrum is that the two QNN models that encode data via amplitude encoding, the \texttt{CircuitCentricClassifier} and \texttt{TreeTensorClassifers}, perform poorly. A ready explanation for the \texttt{TreeTensorClassifier} is that neither the amplitude encoding nor the variational unitary evolution of the model can change the distance between (pre-processed) input data points. Moreover, due to the goal of avoiding vanishing gradients, the model uses a very limited number of parameters that scales only logarithmically with the number of qubits; for example, for a problem with 16 features, one has only 7 variational parameters to train. This severely limits the expressivity of the model. 

However, these arguments do not hold for the \texttt{CircuitCentricClassifier}, which uses an expressive variational circuit and several copies of the amplitude encoded state. This creates an overall state that is a tensor product $\mathbf{x} \times \mathbf{x} \times ...$ of the (pre-processed) input. Since we allow for up to three copies in hyperparameter search, this model has the ability to build up to $3$rd order polynomial features like $\mathbf{x}_1^3$ or $\mathbf{x}_1 \mathbf{x}_6^2$ through the embedding. It is therefore interesting to ask whether the lack in performance of the \texttt{CircuitCentricClassifier} is due to polynomial features not being useful in the datasets we consider here, or if a degree of $3$ is too low. Looking at the hyperparameter optimisation (compare also the low correlation of \emph{n\_input\_copies} with the test accuracy in Figure~\ref{fig:hp-all-correlations}), we find that the optimal number of copies is often $1$ or $2$ instead of the maximum possible $3$, which means that the model does not systematically prefer more complex features. 

\begin{figure}
\centering
\includegraphics[width=0.8\columnwidth]{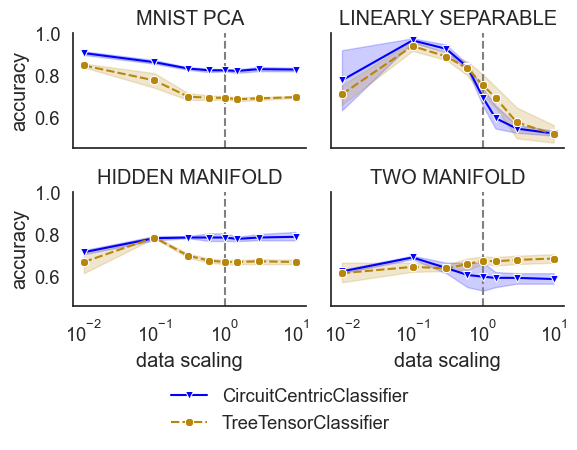}
\caption{The average test accuracy for selected datasets of input dimension six when multiplying the input features by different scaling factors. The models were trained 5 times on each dataset, and the best hyperparameters were taken from our regular hyperparameter search. The default scaling we used in the previous results is shown by the dashed line, and is not always the optimal value, which could sometimes explain the poor performance of models based on amplitude encoding. }
    \label{fig:scaling}
\end{figure}

Note that another explanation is that we missed an important hyperparameter during grid search. Figure~\ref{fig:scaling} plots the average accuracy of relevant models over all datasets that have an input dimension of 6, when scaling the input data by different amounts. Scaling the data to smaller values than we used as default can have a beneficial effect, even when selecting the best hyperparameters found using this default scaling. 

Overall, an important question is whether there are datasets and hyperparameter configurations for which the ability to construct high-order polynomial features using quantum circuits would be interesting -- or is amplitude embedding just not a good design choice?

    \subsection{Why do quantum models struggle with the linearly separable benchmark?}

\begin{figure}
    \centering
    \includegraphics[width=0.4\textwidth]{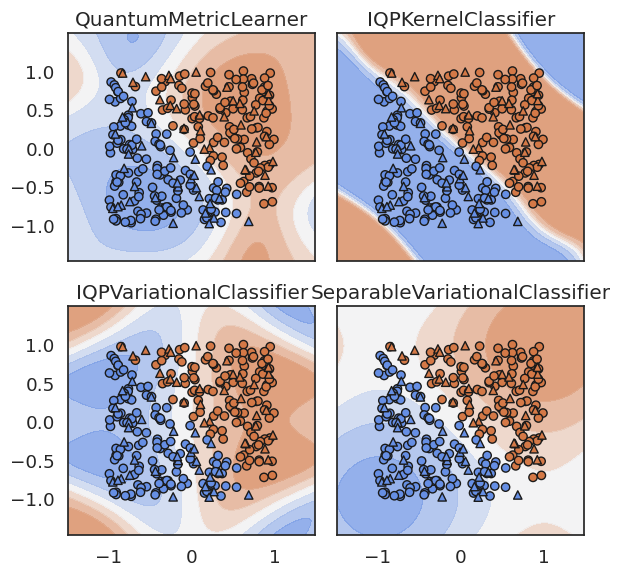}
    \caption{Decision boundary of selected models trained on the 2d \textsc{linearly separable} dataset. Training points are shown in round and test points in triangular shape. All four models use angle embedding, therefore forming decision boundaries based on periodic functions. This may introduce an inductive bias that is not suited for data requiring linear decision boundaries in higher dimensions. }
    \label{fig:decision-boundaries-linsep}
\end{figure}

As discussed above, the behaviour of both QNNs and quantum kernel methods on the  \textsc{linearly separable} benchmark is a notable outlier, as the performance of nearly all quantum models is poor and gets worse with the dimensionality of the data while the classical models achieve near-perfect classification.\footnote{It is interesting to note that in a similar manner, \texttt{WeiNet} -- the only non-hybrid quantum convolutional neural network we tested -- performs badly on \textsc{bars \& stripes} which we considered to be a very simple task.} But is this really a property of linearly separable datasets? It is more likely that many of the quantum models in our selection, in particular those using angle embedding, have an inductive bias against linear decision boundaries like the one implanted into the \textsc{linearly separable} data generation procedure. (Instead, they may do very well on linearly separable data of Gaussian blobs.) For example, Figure~\ref{fig:decision-boundaries-linsep} shows the decision boundaries for selected models on the 2-dimensional dataset of the \textsc{linearly separable} benchmark using the best hyperparameters found during grid search -- all models try to fit the data with periodic regions. While this may work for low dimensions, the constraints could be incompatible with linear decision boundaries in higher dimensions. An interesting theoretical study would be to analyse which kinds of data this behaviour is or is not suited for, and what the resource requirements, such as more embedding layers, are necessary to overcome the problem.

    In contrast, most quantum models performed almost similar to the classical baselines on the \textsc{hyperplanes diff} benchmark, suggesting that perhaps none of the models contained a bias that was particularly aligned with this data. As a reminder, the data generation procedure was intended to test the ability of models to detect a ``global'' labeling rule, for which the positioning of several hyperplanes in a low-dimensional manifold is relevant.

\section{Conclusion}\label{sec:conclusion}

One of the most important lessons one learns when undertaking a study of this kind is that benchmarking is a subtle art that is filled with difficult decisions and potential pitfalls. As such, the benchmarking of new quantum machine learning proposals should be considered an extreme challenge, rather than as a task that can be safely given to lesser experienced researchers or relegated to the afterthought of a study. It is hard to coax robust and meaningful results from systems as complex as machine learning models trained on data, and non-robust claims can have a profound impact on where the community searches for good ideas. The single most effective remedy is scientific rigour in the methodological design of studies, including extensive reporting on the choices made and their potential bias on the results.

Perhaps the most important question regarding the future of benchmarking of quantum models is what kind of data to choose. More studies that focus on questions of structure in data are crucial for the design of meaningful benchmarks: What mathematical properties do real-world applications of relevance have? How can we isolate, downscale and artificially reproduce them? How can we connect them to the mathematical properties of quantum models? This is a task shared with classical machine learning research, but further exacerbated by the fact that the areas in which quantum computers can unlock new capabilities of learning are not yet identified. Using the right data and finding quantum models with an advantage hence becomes a ``chicken and egg problem'' that is best tackled from two sides; however in the current literature, the focus on model design dominates by far. 

Aside from these conceptual challenges, benchmarking quantum machine learning models also poses a formidable challenge to current quantum software. On the one hand, this is due to the immense resource requirements of hyperparameter optimisation. On the other hand, quantum machine learning models are usually elaborate pipelines of hybrid quantum-classical systems, each of which requires different logics for performance tools like compilation, parallelisation, caching or GPU use. Furthermore, results on small datasets cannot  be used to reason about larger datasets, as we know from deep learning that big data leads to surprisingly different behaviour. There is hence a need to study how results scale to larger datasets, which typically pushes the number of qubits to the limits of what is possible today.

Finally, instead of considering rankings only, benchmarks can help us to gain qualitative insights into which parts of a model design are crucial and which ones replaceable. Since the question of quantum advantage is undercurrent to almost all studies in quantum machine learning, a particularly important experiment that should become a standard in benchmarking is to remove ``quantumness'' from a model in a non-invasive manner and test if the results hold. Of course, there are other ways than removing entanglement to make models classically tractable or ``non-quantum'', such as limiting gates to the Clifford family, replacing unitary transformations by stochastic ones (see Appendix in \citep{abbas2023quantum}) or using Matrix Product State simulators with low bond dimension. Comparing to such circuit designs will provide invaluable information into the promise of ideas around variational quantum circuits.

\begin{acknowledgments}
Our computations were performed on the Cedar supercomputer at the SciNet HPC Consortium. SciNet is funded by Innovation, Science and Economic Development Canada; the Digital Research Alliance of Canada; the Ontario Research Fund: Research Excellence; and the University of Toronto.
\end{acknowledgments}

\bibliography{references}

\appendix

\section{CO\textsubscript{2} Emission Table}\label{sec:appconduct}

The following calculations are made using energy consumption data for the Cedar supercomputing cluster at the SciNet HPC Consortium, on which the vast majority of our simulations were run. 

\begin{table}[h]
\label{tab:cotwo}
\begin{tabular}[b]{l c}
\toprule
\textbf{Numerical simulations} & \\
\midrule
Total CPU usage [core years] &$\approx 27$ \\
Cluster energy consumption (per core) [W] &$\approx 11$\\
Total Energy Consumption Simulations [kWh] &$\approx 2600$ \\
Average Emissions Of CO$_2$ In Canada [$\mathrm{kg/kWh}$]& $\approx 0.13$\\
Total CO$_2$ Emissions For numerical simulations [kg] & $\approx 340$ \\
\midrule
\textbf{Transport} & \\
\midrule
Total CO$_2$ Emission For Transport [$\mathrm{kg}$] & 0\\
\midrule
Total CO$_2$ Emission [kg] & $\approx 340$ \\
Were The Emissions Offset? & No \\
\bottomrule
\end{tabular}
\end{table}

\section{Glossary of quantum machine learning concepts}\label{app:glossary}

\label{glos:ampemb}\emph{Amplitude embedding}---An input data vector $\vec{x}$ is said to be amplitude embedded into a pure quantum state $\ket{\psi(\vec{x})}$ if the quantum state takes the form 
\begin{align}
    \ket{\psi(\vec{x})} = \vec{x}\oplus\vec{c}/ \mathcal{N}
\end{align}
where $\vec{c}$ is a vector with constant entries and $\mathcal{N}=\sqrt{\vec{x}^\dagger\cdot \vec{x}+\vec{c}^{\dagger}\cdot \vec{c}}$ is the state normalization. \\

\label{glos:angleemb}\emph{Angle embedding}---An input data vector $\vec{x}=(x_j)$ is said to be angle embedded into a pure quantum state $\ket{\psi(\vec{x})}$ if the quantum state takes the form 
\begin{align}
    \ket{\psi(\vec{x})}=\prod_j \exp(-i G_j x_j)\ket{\psi_0}
\end{align}
where $\ket{\psi_0}$ is some initial quantum state and $G_j$ are Hermitian operators. If the operators $G_j$ act non-trivially on single qubits only, we call it a \emph{product angle embedding}. This process of angle embedding is sometimes repeated a number of times, which is often called \emph{data reuploading}. \\

\label{glos:crossentropy}\emph{Binary cross entropy loss}---Given class probabilities $P(\pm1\vert \vec{\theta},\vec{x}_i)$ and a label $y_i=\pm 1$, the cross entropy loss is 
\begin{align}
    \ell(\vec{\theta},\vec{x}_i,\vec{y_i}) = - \log P(y_i \vert \vec{\theta},\vec{x}_i).
\end{align}
Minimising the cross entropy loss on a dataset is equivalent to maximising the log likelihood of the data, and is the preferred loss in binary classification tasks. \\ 

\label{glos:gibbsstate}\emph{Gibbs state}---A Gibbs state is a quantum density matrix that is diagonal in the energy eigenbasis given by a Hamiltonian $H$, and whose probability distribution over energy eigenstates forms a Gibbs distribution. Mathematically we have 
\begin{align}
    \rho = \exp\big( -\frac{H}{k_b T}\big) /Z
\end{align}
where $T>0$ is a temperature, $k_b$ Boltzmann's constant and $Z=\text{tr }\exp\bigl(-\frac{H}{k_b T}\bigr)$. \\

\label{glos:iqp}\emph{Instantaneous Quantum Polynomial circuit}---A circuit that consists of input state preparation $\ket{0}$, quantum gates of the form  $\exp(-i G_X \theta)$ where $G_X$ is a product of $\sigma_X$ operators on a subset of qubits,
and measurement of a diagonal observable.  \\

\label{glos:linearloss}\emph{Linear loss}---Given class probabilities $P(\pm1\vert \vec{\theta},\vec{x}_i)$ and a label $y_i=\pm 1$, the linear loss is 
\begin{align}
    \ell(\vec{\theta},\vec{x}_i,\vec{y_i}) =  -P(y_i \vert \vec{\theta},\vec{x}_i).
\end{align}
Minimising the linear loss over a dataset is therefore equivalent to maximising the sum of the probabilities to classify each input correctly. \\

 
 \label{glos:maxmargin}\emph{Maximum margin (linear) classifier}---A linear classifier that separates the two classes such that the minimum distance of any training point to the hyperplane that defines the decision boundary is maxised. \\

 \label{glos:rbfkernel}\emph{Gaussian Kernel}---A kernel of the form 
 \begin{align}
     k(\vec{x},\vec{x}') = \exp(-\gamma \vert\vert \vec{x}-\vec{x}'\vert\vert)
 \end{align}
 where $\gamma$ is a free hyperparameter. \\

\label{glos:squareloss}\emph{Square loss}---Given a model whose output is $f(\vec{\theta},\vec{x})$, the square loss is 
\begin{align}
    \ell(\vec{\theta},\vec{x}_i,\vec{y_i}) = (f(\vec{\theta},\vec{x}_i)-y_i)^2.
\end{align}\\

\label{glos:vanishinggrads}\emph{Vanishing gradients}---A phenomenon that commonly occurs for deep or highly expressive variational quantum circuits whereby the expected magnitude of a typical gradient component decreases exponentially with the number of qubits when initialising all parameters uniformly at random \cite{bps}.

\section{Detailed description of the models\label{app:models}}
In this appendix we describe each model used in the study in further detail. Unless otherwise specified all variational models are  trained via  gradient-descent  using the Adam optimizer~\citep{kingma2014adam} implemented in Optax~\citep{deepmind2020jax} with default parameters except for the learning rate which we vary during hyperparameter search.

    \subsection{CircuitCentricClassifier \citep{circuitcentric}}
    
    An input vector $\mathbf{x}$ of dimension $d$ gets amplitude embedded into a suitable number of qubits (including padding and normalisation, see Glossary above). Note that this preprocessing strategy induces a qualitatively different behaviour when $d$ is a power of $2$, since without padding, normalisation looses all information about the length of the inputs.
    
    A hyperparameter allows for the pre-processed input to be encoded into multiple registers, which creates copies of amplitude encoding states. This effectively creates tensor products of the pre-processed inputs. A variational circuit that uses arbitrary single qubit rotations followed by cascades of controlled arbitrary single qubit rotations is followed by a Z measurement of the first qubit.
    
    The following is an example of the quantum circuit used in the \texttt{CircuitCentricClassifier} for two copies of an input $\mathbf{x} \in \mathbb{R}^4$ embedded into state $\ket{\psi_{\mathbf{x}}}$ and PennyLane's \texttt{StronglyEntanglingLayers} template implementing the variational ansatz:
    \begin{center}
    \includegraphics[width=0.25\textwidth]{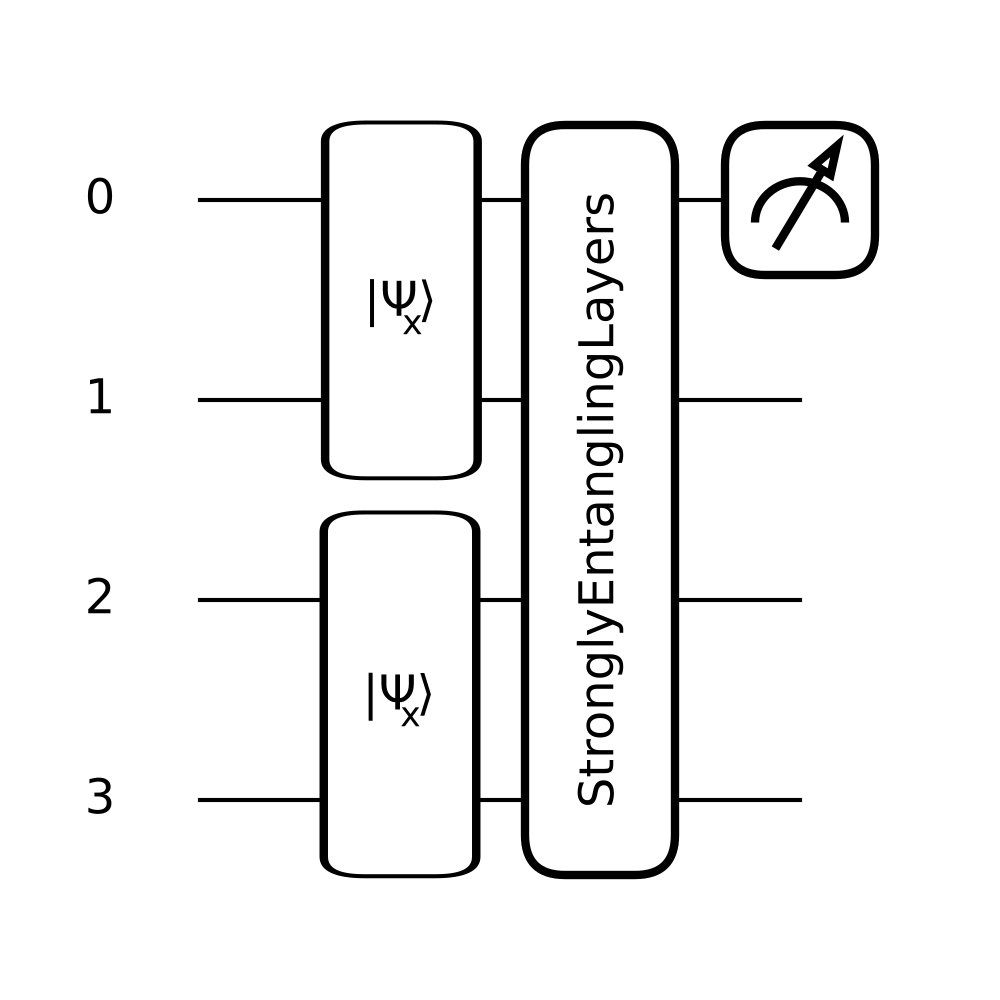}
    \end{center}

    The expected value of the measurement is added to a trainable classical bias. The parameters of the variational circuit, as well as the bias, are optimised using the square loss. \\
     
    \begin{tabular}{ll}
        \hline
        hyperparameter  & values \\
        \hline
        {\emph{learning\_rate}} & [0.001, 0.01, 0.1] \\
        {\emph{n\_layers}} (variational layers) & [1, 5, 10] \\
        {\emph{n\_input\_copies}} &  $[1, 2, 3]$ 
    \end{tabular}

    \subsection{DataReuploadingClassifier \citep{datareuploading}}
This model uses successive, trainable angle embeddings of data. Each qubit embedding gate takes a vector $\vec{x}$ of three features, two trainable three-dimensional real vectors $\vec{w}$ and $\vec{\theta}$, and encodes them as
\begin{align}\label{reupload_gate}
 U(\vec{x}\circ\vec{\omega}+\vec{\theta})
\end{align}
where 
\begin{align}
    U(\vec{\phi}) = e^{iZ\phi_1/2}e^{iY\phi_2/2}e^{iZ\phi_3/2}
\end{align}
parameterizes a general SU(2) rotation on a single qubit, and $\circ$ denotes element-wise multiplication of vectors. 

To encode data input $\vec{x}\in \mathbb{R}^d$, we therefore split $\vec{x}$ into $\lceil\frac{d}{3}\rceil$  vectors of size 3, and feed each vector into a distinct qubit embedding gate (padding input vectors with zero if necessary). The number of qubits is therefore set by the dimension of the input data features. This is followed by a sequence of CZ gates in a ladder structure (See Figure~5 of \citep{datareuploading}), and the process is repeated for a number of layers, which increases the expressivity of the model. 

The following is an example of the quantum circuit used in the \texttt{DataReuploadingClassifier} for the input $\mathbf{x} = (0.1, 0.2, 0.3, 0.4)^T$ embedded in $3$ trainable layers, and with the scaling parameters $\vec{\omega}$ all set to $1$.
 \begin{center}
 \includegraphics[width=0.45\textwidth]{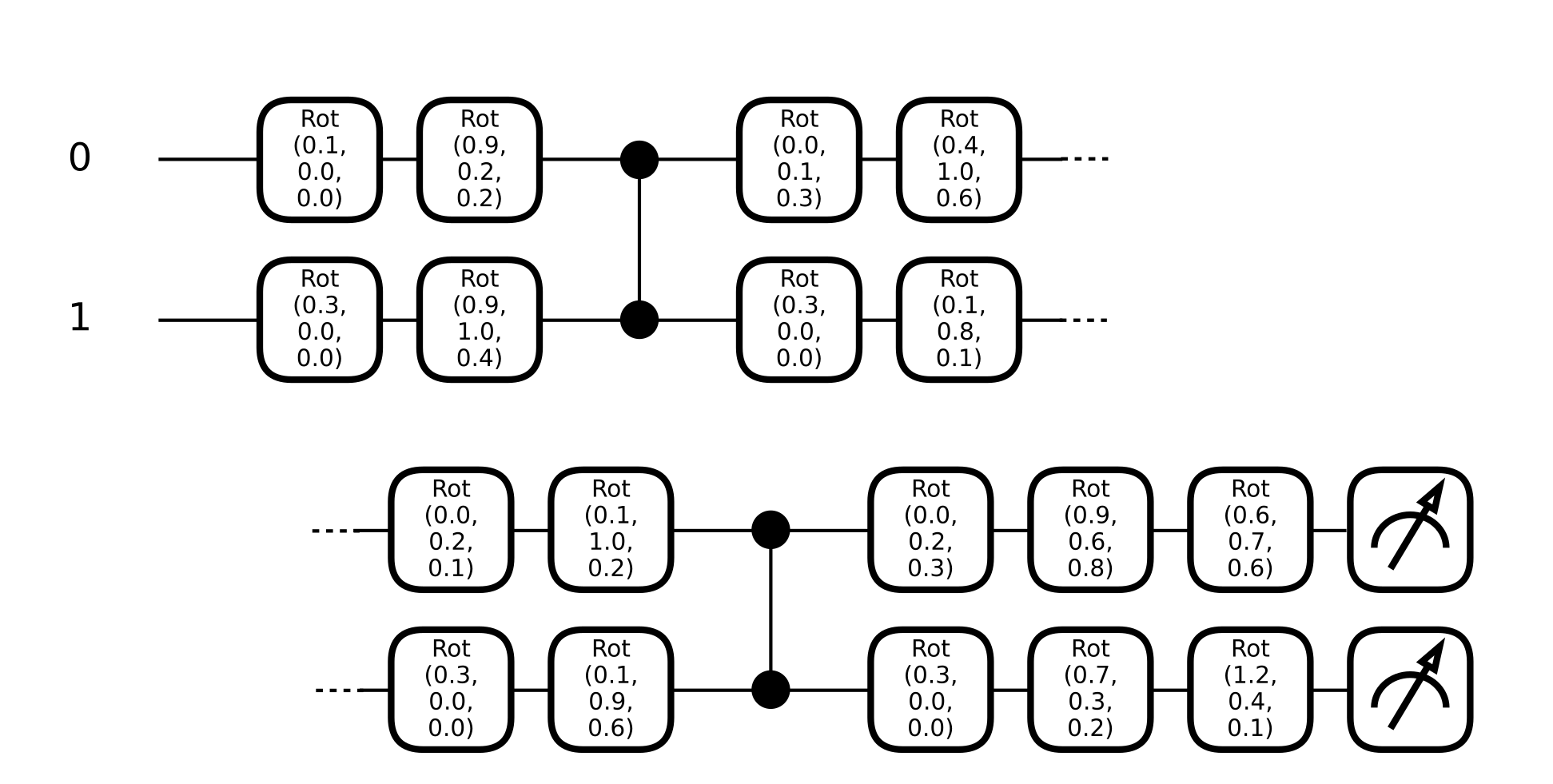}
 \end{center}

Training is based on the fidelities of the output qubits to one of two class states: here either $\ket{0}$ or $\ket{1}$. Defining $F_j^{0}(\vec{x})$, $F_j^{1}(\vec{x})$ as the fidelity of the $j^{th}$ output qubit to the state $\ket{0}$, $\ket{1}$, the loss $\ell$ for a single data point is given by
\begin{align}
    \ell(\vec{\theta},\vec{\omega},\vec{\alpha},\vec{x}_i) = \sum_{j=1}^{n_{\text{max}}}(\alpha_j^0F^0_j-(1-y_i))^2 + (\alpha_j^1F^1_j-y_i)^2
\end{align}
where the $\alpha_j^0$, $\alpha_j^1$ are trainable parameters, and $n_{\text{max}}$ determines the number of qubits to use for training and prediction. 

For prediction, we use the average fidelity to either $\ket{0}$ or $\ket{1}$ over the same qubits:
\begin{align}
    y_{\text{pred}} = \text{argmax}(\langle F_j^0 \rangle, \langle F_j^1 \rangle), 
\end{align}
where $\langle F_j^0 \rangle=\frac{1}{n_{\text{max}}}\sum_{j=1}^{n_{\text{max}}}F_j^0$. This is not specified in \citep{datareuploading}, but is a natural generalisation of the $n_{\text{max}}=1$ case they focus on.  The choice of the hyperparameter {\emph{observable\_type}} determines the number $n_{\text{max}}$ of qubits used to evaluate the weighted cost function. \\

\begin{tabular}{ll}
    \hline
    hyperparameter  & values \\
    \hline
    {\emph{learning\_rate}} & [0.001, 0.01, 0.1] \\
    {\emph{n\_layers}} (reuploading layers) & [1, 5, 10, 15] \\
    {\emph{observable\_type}} & [single, half, full]
\end{tabular}

    \subsection{DressedQuantumCircuitClassifier \citep{dressed}}

This model maps an input data point $\vec{x}$ two a 2-dimensional vector via 
\begin{align}\nonumber
    f(\vec{\theta}, \vec{W}_{\text{in}}, \vec{W}_{\text{out}},\vec{x}) = f_{\text{out}}(\vec{W}_{\text{out}}, f_Q(\vec{\theta},f_{\text{in}}(\vec{W}_{\text{in}}, \vec{x}))).
\end{align}
The functions $f_{\text{in}}(\vec{W}_{\text{in}}, \cdot), f_{\text{out}}(\vec{W}_{\text{out}}, \cdot)$ are single layer fully connected feed-forward neural networks with weights $\vec{W}_{\text{in}}\in \mathbb{R}^{d\times n}$, $\vec{W}_{\text{out}}\in\mathbb{R}^{n\times 2}$, where $f_{\text{in}}$ has a $\tanh$ activation scaled by $\pi/2$, and $f_{\text{out}}$ has no activation. The function $f_{Q}$ corresponds to a parameterised quantum circuit where input features are angle-encoded into individual qubits, followed by layers of single-qubit $Y$ rotations and CNOT gates applied in a ring pattern. The output of the circuit is an $n$-dimensional vector whose elements are the expectation values of single-qubit $Z$ measurements on each qubit. 

The following is an example of the quantum circuit used in the \texttt{DressedQuantumCircuitClassifier} for the input $\mathbf{x} = (0.1, 0.2, 0.3, 0.4)^T$ and 3 variational layers:
\begin{center}
\includegraphics[width=0.49\textwidth]{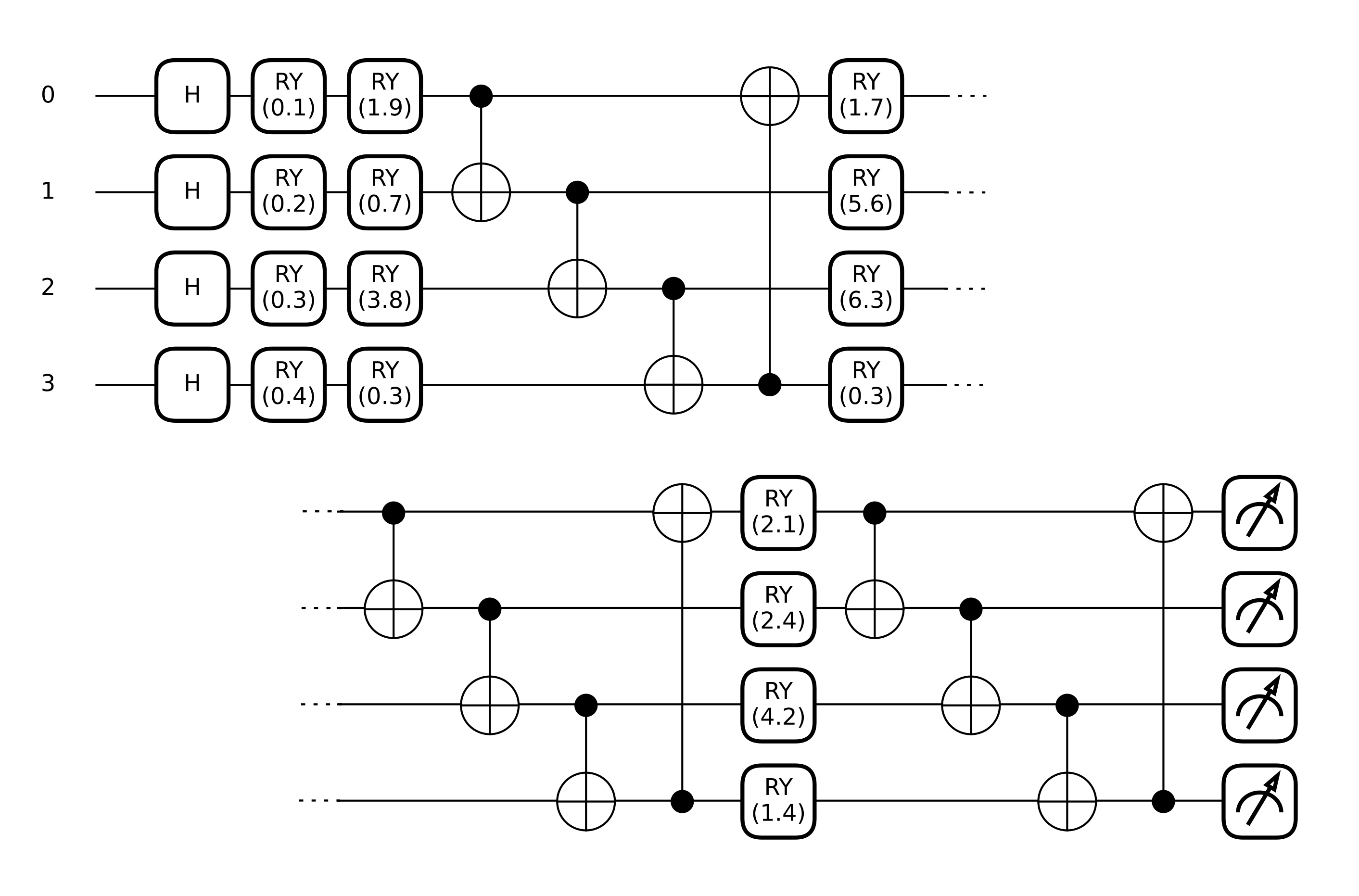}
\end{center}

The output vector is sent through a softmax layer that is used for prediction and gives the class probabilities, and the cross entropy loss is used to train $\vec{W}_{\text{in}}, \vec{W}_{\text{out}}$ and $\vec{\theta}$ simultaneously. \\

\begin{tabular}{ll}
    \hline
    hyperparameter  & values \\
    \hline
    {\emph{learning\_rate}} & [0.001, 0.01, 0.1] \\
    {\emph{n\_layers}}  & [1, 5, 10, 15] \\
\end{tabular}


    \subsection{IQPVariationalClassifier \citep{iqp}}
This model uses angle encoding $V(\vec{x})$ inspired from IQP circuits, which is implemented by PennyLane's \texttt{IQPEmbedding} class. This is followed by a trainable parameterised circuit $U(\vec{\theta})$, implemented by PennyLane's \texttt{StronglyEntanglingLayers} class. 

Prediction is given by measurement of $Z_1 Z_2$ on the first two qubits:
\begin{align}
    y_{\text{pred}} = \text{sign}f(\vec{\theta},\vec{x}),
\end{align}
where 
\begin{align}
f(\vec{\theta},\vec{x}) = \bra{0}V^\dagger(\vec{x})U^\dagger(\vec{\theta})Z_1 Z_2 U(\vec{\theta})V(\vec{x})\ket{0}.
\end{align}
The loss is equal to the linear loss:
\begin{align}
    \ell(\vec{\theta},\vec{x}) = (1-y\cdot f(\vec{\theta},\vec{x}))/2.
\end{align} \\

The following is an example of the quantum circuit used in the \texttt{IQPVariationalClassifier} for the input $\mathbf{x} = (0.1, 0.2, 0.3, 0.4)^T$:

\begin{center}
\includegraphics[width=0.25\textwidth]{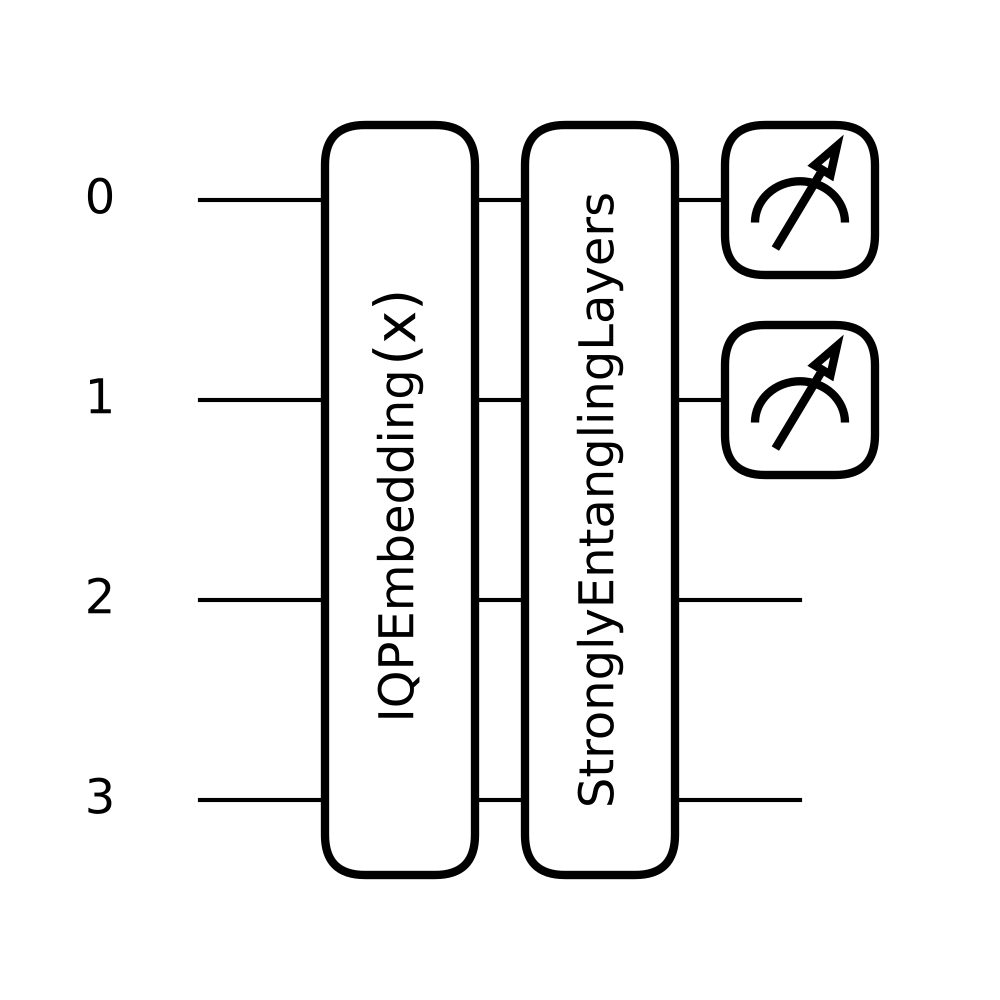}
\end{center}

\begin{tabular}{ll}
    \hline
    hyperparameter  & values \\
    \hline
    {\emph{learning\_rate}} & [0.001, 0.01, 0.1] \\
    {\emph{n\_layers}} (variational layers) & [1, 5, 10, 15] \\
    {\emph{repeats}}(embedding layers) & [1, 5, 10]
\end{tabular}

    \subsection{QuantumBoltzmannMachine \citep{qboltzmann}}
This model encodes data into a Gibbs state of a $n$-qubit Hamiltonian. We use a Hamiltonian that is a natural generalization of the one studied in the two-qubit example in \citep{qboltzmann}:
\begin{multline}
    H(\vec{\theta},\vec{x}) = \sum_j Z_j \vec{\theta}^Z_j\cdot \vec{x} + \sum_j X_j \vec{\theta}^X_{i}\cdot \vec{x} + \sum_{j>k} Z_jZ_k \vec{\theta}_{jk}\cdot \vec{x}
\end{multline}
where $\vec{\theta}^Z_j$, $\vec{\theta}^X_j$, $\vec{\theta}_{jk}$ are vectors of trainable parameters that we collect into $\vec{\theta}$. We take $n=d$ so that the number of qubits scales with the number of features. 

Since Gibbs state preparation is hard, the paper gives a recipe to parameterize a trial state for the Gibbs state and perform variational imaginary time evolution to approximate the desired state. Since this is quite computationally involved, we assume (as they do in \citep{qboltzmann}) that we have access to the perfect Gibbs state. It is therefore unclear whether the full algorithm can be expected to perform as well as our implementation. 

For prediction, a diagonal $\pm1$ valued observable $O$ is measured on a subset of qubits of size $n_{\text{vis}}$ (called the visible qubits, controlled by hyperparameter \emph{visible\_qubits}). The sign of the expectation value determines the label:
\begin{align}
    y_{\text{pred}} = \text{sign}\left( \text{tr}\left[ \rho(\vec{\theta},\vec{x})O  \right]\right),
\end{align}
where 
\begin{align}
    \rho (\vec{\theta}) = \frac{\exp \left(-\frac{H(\vec{\theta},\vec{x})}{T}\right)}{\text{tr}\left[ \exp \left(-\frac{H(\vec{\theta},\vec{x} )}{T}\right)\right]}
\end{align}
is a Gibbs state, and we choose $O$ to be
\begin{align}
    O = \frac{1}{n_{\text{vis}}}\sum_{j=1}^{n_{\text{vis}}}Z_j
\end{align}
(note \citep{qboltzmann} does not recommend a general form for the observable).

Training is done with a binary cross entropy loss. To define the probability of $y=1$, we use:
\begin{align}
    P(y=1\vert \vec{\theta},\vec{x})= \frac{1 +\langle O \rangle}{2}
\end{align}
which is in $[0,1]$ and agrees with the example in \citep{qboltzmann} for $n_{\text{vis}}=1$.

We note that since we are forced to work with mixed states, this implies a larger memory cost of simulation. As a result we were not able to test this model for as large qubit number as others. It is also the only model we implemented without the use of a PennyLane circuit, but rather by constructing the density matrix directly. \\

\begin{tabular}{ll}
    \hline
    hyperparameter  & values \\
    \hline
    {\emph{learning\_rate}} & [0.001, 0.01, 0.1] \\
    {\emph{temperature}} (T) & [1, 10, 100] \\
    {\emph{visible\_qubits}} & $[\text{single}, \text{half}, \text{all}]$
\end{tabular}

    \subsection{QuantumBoltzmannMachineSeparable}
    This model is a version of \texttt{QuantumBoltzmannMachine} that does not use entanglement. Once again we take $n=d$ so that the number of qubits scales with the number of features. The Hamiltonian is 
    \begin{align}
    H(\vec{\theta},\vec{x}) = \sum_j Z_j \vec{\theta}^Z_j\cdot \vec{x} + \sum_j X_j \vec{\theta}^X_{i}\cdot \vec{x},
    \end{align}
    whose corresponding Gibbs state is a product mixed state by virtue of the Hamiltonian being product. The model is equivalent to \texttt{QuantumBoltzmannMachine} otherwise and uses the same hyperparameter grid.

    \subsection{QuantumMetricLearner \citep{metriclearner}}

    The \texttt{QuantumMetricLearner} works quite differently from other quantum neural networks. It uses a trainable, layer-wise embedding inspired by the QAOA algorithm implemented by PennyLane's \texttt{QAOAEmbedding} template, which employs one more qubit than there are features. The ansatz for one layer encodes input features into X rotations, followed by parametrised ZZ and Y rotations. The additional qubit with a constant angle is used as a ``latent feature''.\\
    
    Training of the embedding is performed by measuring the overlap between a pair of embedded data points from the same [different] classes and minimising [maximising] their fidelity.

    More precisely, if $\ket{\phi_{\theta}(\mathbf{x})}$ is the quantum state embedding an input vector $\mathbf{x}$, and $AA$, $BB$ [$AB$] are sets of randomly sampled training input pairs $(\mathbf{x}, \mathbf{x}')$ from training data in the same class $A$ or $B$ [different classes], the cost function is defined as 

    \begin{multline}
        c(A, B) = 1 - \sum_{(\mathbf{a}, \mathbf{b}) \in AB} |\langle\phi_{\boldsymbol{\theta}}(\mathbf{a})|\phi_{\boldsymbol{\theta}}(\mathbf{b})\rangle|^2 \\ 
        +0.5  \sum_{(\mathbf{a}, \mathbf{a}') \in AA} |\langle\phi_{\boldsymbol{\theta}}(\mathbf{a})|\phi_{\boldsymbol{\theta}}(\mathbf{a}')\rangle|^2  \\ 
        + 0.5 \sum_{(\mathbf{b}, \mathbf{b}') \in BB} |\langle\phi_{\boldsymbol{\theta}}(\mathbf{b})|\phi_{\boldsymbol{\theta}}(\mathbf{b}')\rangle|^2. 
    \end{multline}

    Note that in the original paper, all possible pairs of datapoints within a random batch are compared, however to have a better control of how many circuits are run we sample a batch of random pairs instead.

    Once trained, the embedding is directly used for prediction: A new input is embedded and the resulting state compared to a random batch of embedded training data points $A'$ and $B'$ from each class. The class that it is closest to on average is assigned. This rule corresponds to the ``fidelity classifier'' from the paper:
    
    \begin{equation}
        f(\boldsymbol{\theta}, \mathbf{x}) =  \sum_{\mathbf{a} \in A'} |\langle\phi_{\boldsymbol{\theta}}(\mathbf{a})|\phi_{\boldsymbol{\theta}}(\mathbf{x})\rangle|^2 -  \sum_{\mathbf{b} \in B'}|\langle\phi_{\boldsymbol{\theta}}(\mathbf{b})|\phi_{\boldsymbol{\theta}}(\mathbf{x})\rangle|^2.
    \end{equation}

    The final prediction is made by taking the sign,
    \begin{align}
    y_{\text{pred}} = \text{sign} \;f(\boldsymbol{\theta}, \mathbf{x}).
    \end{align}

    The following is the quantum circuit used to evaluate overlaps in the \texttt{QuantumMetricLearner} for two inputs $\mathbf{x}, \mathbf{x'} \in \mathbb{R}^4$:
    \begin{center}
    \includegraphics[width=0.2\textwidth]{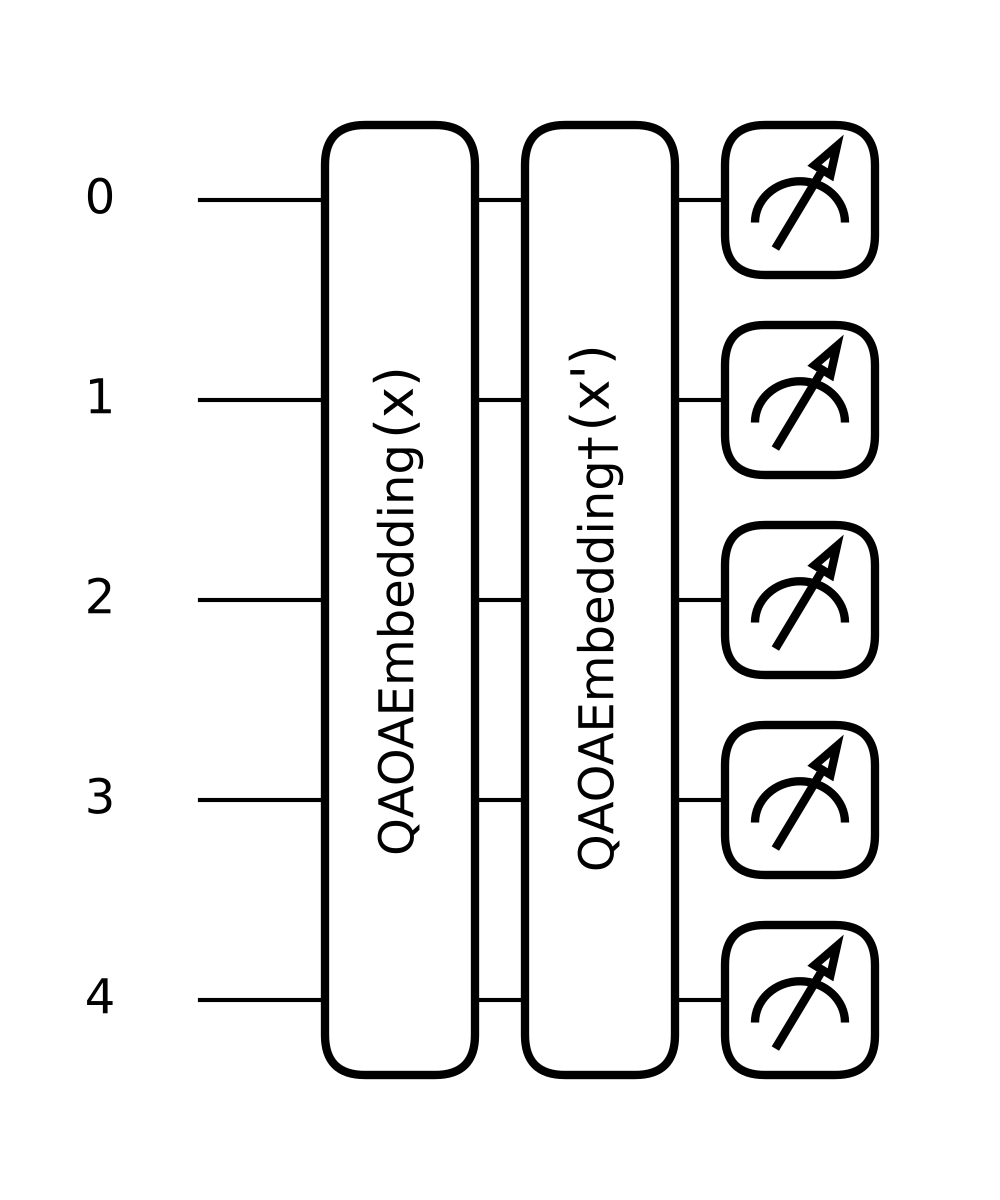}
    \end{center}

    The number of circuits run for each training step and prediction scales linearly with the size of the batch of example pairs used -- in this paper we fixed $32$ for both. Larger batch sizes allow a more reliable estimate of the cost and predicted label. \\
     
    \begin{tabular}{ll}
        \hline
        hyperparameter  & values \\
        \hline
        {\emph{learning\_rate}} & [0.001, 0.01, 0.1] \\
        {\emph{n\_layers}} (embedding layers) & [1, 3, 4] \\
    \end{tabular}

    \subsection{TreeTensorClassifier \citep{treetensor}}
This model was designed to avoid the phenomenon of barren plateaus. We implement the `tree tensor' structure shown in Figure~1 of \citep{treetensor}. The variational circuit in this model has a tree-like structure and therefore requires a number of qubits that is a power of 2. In \citep{treetensor} one first optimizes a variational circuit that finds a state that approximates an amplitude-encoded data state. The reason for this is to make the algorithm more efficient; here, to avoid an additional variational optimisation, we assume direct access to the exact amplitude encoded state $V(\vec{x})\ket{0}$ (padding with constant values $1/2^n$ with $n$ the number of qubits when necessary). 

This state is then fed into the variational circuit $U(\vec{\theta})$ consisting of trainable single-qubit $Y$ rotations and CNOTs. The tree structure means that there are few parameters; for a circuit with $n$ qubits one has only  $2n-1$ parameters. 

Prediction is given by measurement of $Z$ on the first qubit,
\begin{align}
    y_{\text{pred}} = \text{sign}f(\vec{\theta},\vec{x}),
\end{align}
where 
\begin{align}
f(\vec{\theta},\vec{x}) = \bra{0}V^\dagger(\vec{x})U^\dagger(\vec{\theta})Z_1 U(\vec{\theta})V(\vec{x})\ket{0},
\end{align}
and training is via the square loss:
\begin{align}
    \ell(\vec{\theta},\vec{x},y) = (f(\vec{\theta},\vec{x})-y)^2. 
\end{align}
The following is an example of the quantum circuit used in the \texttt{TreeTensorClassifier} for a $4$-dimensional input encoded into state $\ket{\psi_{\mathbf{x}}}$:

\begin{center}
\includegraphics[width=0.3\textwidth]{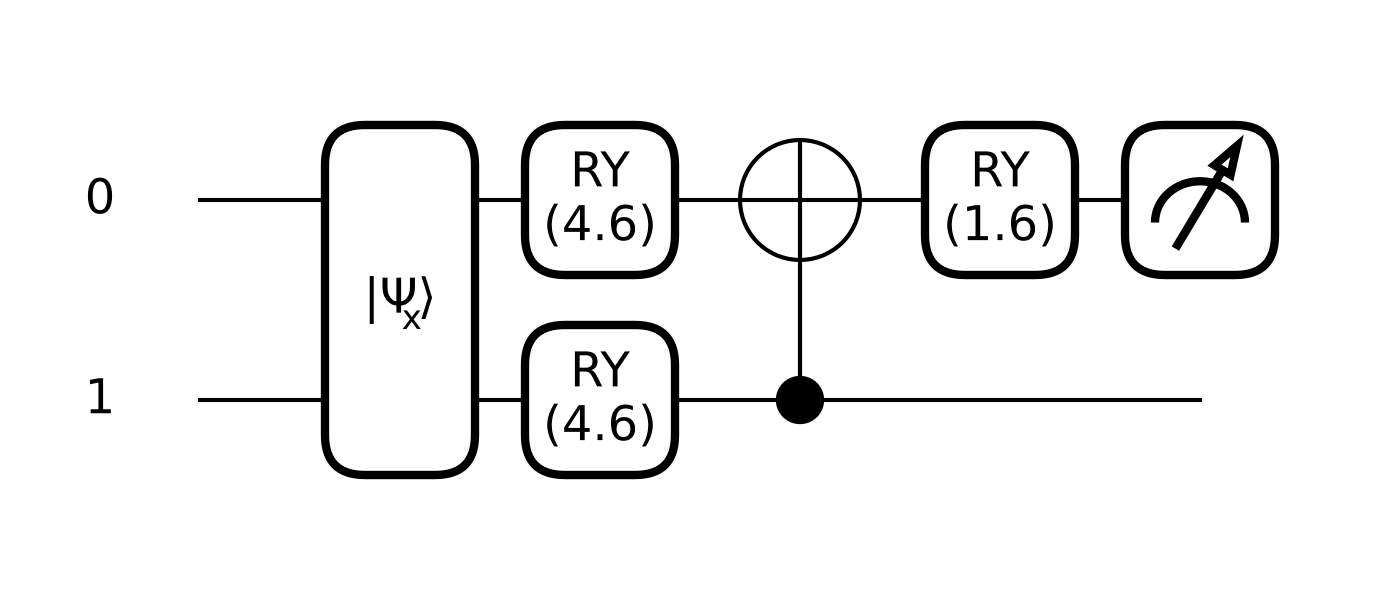}
\end{center}

This model has few suggested hyperparameters, so we vary only the learning rate. \\

\begin{tabular}{ll}
    \hline
    hyperparameter  & values \\
    \hline
    {\emph{learning\_rate}} & [0.001, 0.01, 0.1]
\end{tabular}

    \subsection{IQPKernelClassifier \citep{iqp}}
This model is a kernel equivalent of the IQP variational model. The embedding $V(\vec{x})\ket{0}$ is the same IQP-inspired embedding given by PennyLane's \texttt{IQPEmbedding} template. This defines a kernel 
\begin{align}
    k(\vec{x},\vec{x}') = \text{tr}[\rho(\vec{x})\rho(\vec{x}')] = \vert\bra{0}V^{\dagger}(\vec{x})V(\vec{x}')\ket{0}\vert^2,
\end{align}
which we evaluate by applying the unitary $V^{\dagger}(\vec{x})V(\vec{x}')$ to an input state $\ket{0}$ and calculating the probability to measure $\ket{0}$. 

The following is an example of the quantum circuit used in the \texttt{IQPKernelClassifier} for two inputs $\mathbf{x}, \mathbf{x} \in \mathbb{R}^4$:

\begin{center}
\includegraphics[width=0.25\textwidth]{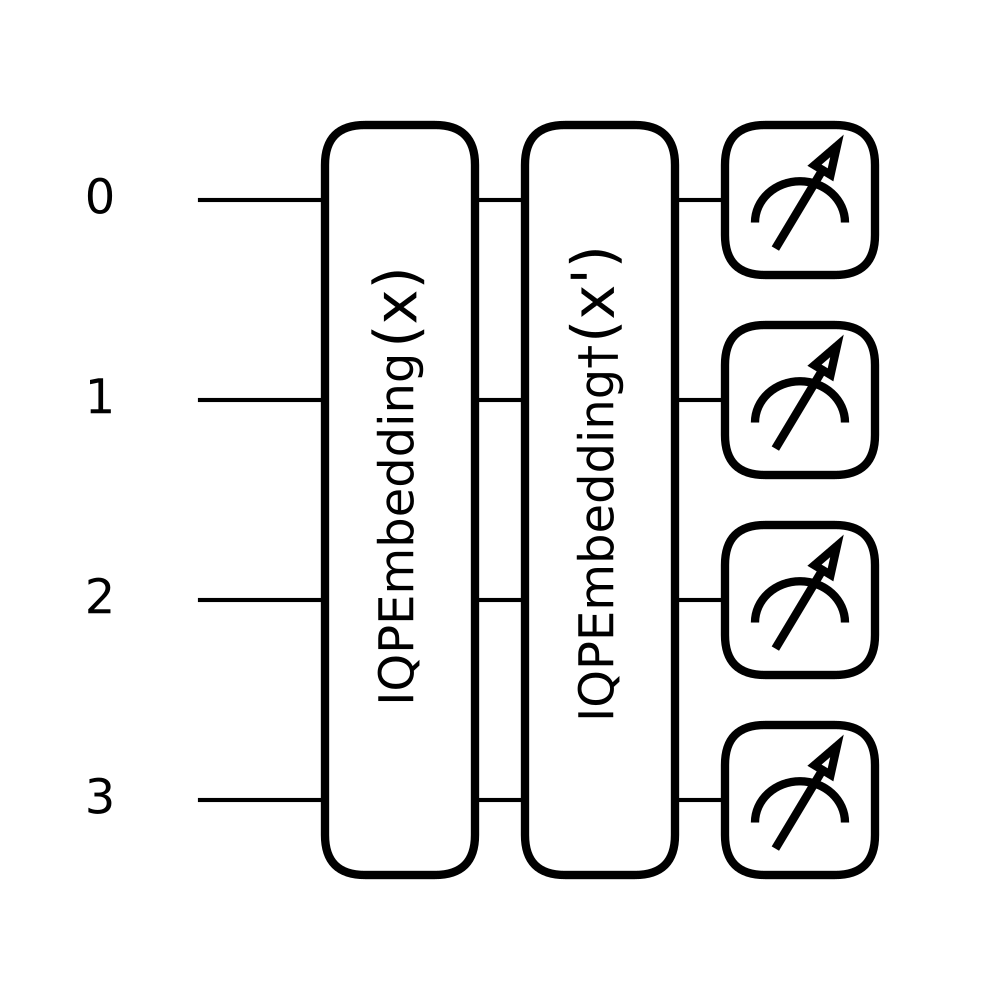}
\end{center}

The kernel matrix $K$ is fed to scikit-learn's SVC class, which trains a support vector machine classifier. Prediction is given by

\begin{align}
    y_{\text{pred}} = \text{sign}\;\sum_{i}\alpha_i k(\vec{x}_i,\vec{x}),
\end{align}

with $\alpha_i$ the weights of the support vector machine. \\

The hyperparameter search values are as follows: \\

\begin{tabular}{ll}
    \hline
    hyperparameter  & values \\
    \hline
    {\emph{repeats}} (embedding layers) & [1, 5, 10] \\
    {\emph{C}} (SVC regularization) & [0.1,1,10,100]
\end{tabular}

    \subsection{ProjectedQuantumKernel \citep{projectedquantumkernel}}
This model is a kernel method that uses a Hamiltonian-inspired data embedding that resembles a Trotter evolution of a 1D-Heisenberg model with random couplings. This consists of applying random single-qubit rotations to an input state $\ket{0}$ of $n=d+1$ qubits, followed by $L$ layers of two-qubit rotations with generators $XX$, $YY$, $ZZ$ to pairs of adjacent qubits, with angles given by the elements of $\vec{x}$:
\begin{equation}
     \prod_{j=1}^n \exp\left(-\mathrm{i} \frac{t}{L} x_{j} \left(X_j X_{j+1} + Y_j Y_{j+1} + Z_j Z_{j+1}\right)\right),
\end{equation}
where $t$ is a hyperparameter of the model. Writing the embedded states as $\rho(\vec{x}_i)$, the kernel function $k(\vec{x}_i,\vec{x}_j)$ is
\begin{align}
    \exp \left(-\gamma  \sum_{k=1}^n \sum_{P \in \{X, Y, Z\}} \left(\text{tr}[P \rho_k(\vec{x}_i)] - \text{tr}[ P \rho_k(\vec{x}_j)]\right)^2\right)
\end{align}
where $\rho_k$ is the reduced state of $k^{th}$ qubit of $\rho$. This is simply the RBF kernel with bandwidth $\gamma$ applied to feature mapped vectors $\vec{\phi}(\vec{x})$ with elements $ (\text{tr}[P \rho_k(\vec{x})])$.

The following is an example of the quantum circuit used to compute feature vectors from measurements in \texttt{ProjectedQuantumKernel} for the input $\mathbf{x} = (0.1, 0.2, 0.3, 0.4)^T$:

\begin{center}
\includegraphics[width=0.49\textwidth]{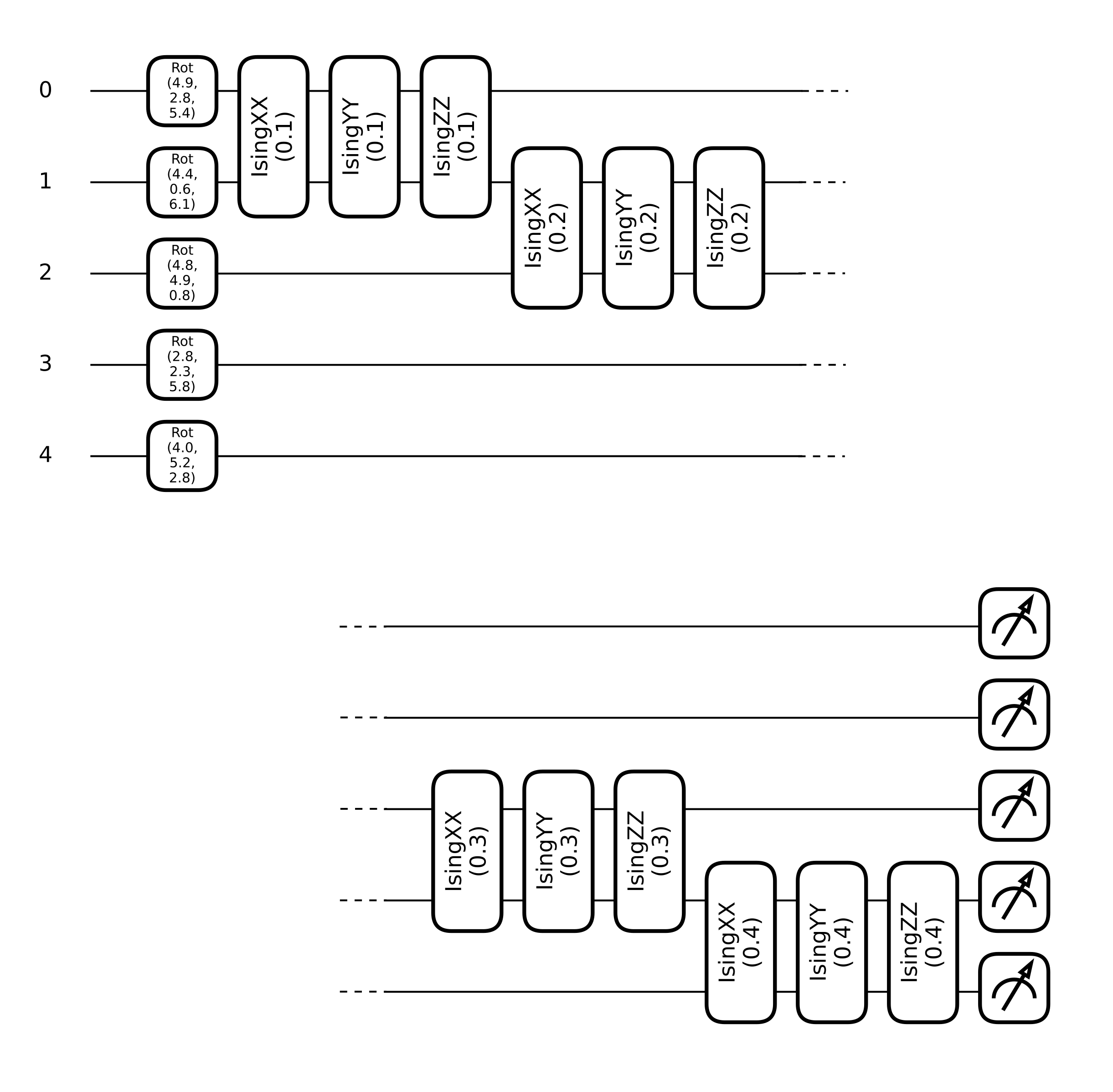}
\end{center}

According to \citep{projectedquantumkernel}, the default $\gamma$ value is set as 
\begin{align}
    \gamma_0 = \frac{1}{\text{Var}(\vec{\phi})d}
\end{align}
where $\text{Var}(\vec{\phi})$ is the variance of the elements of all the vectors $\vec{\phi}(\vec{x}_i)$. We include another hyperparameter `gamma\_factor' that scales this default value.  

A support vector machine is then trained using scikit-learn's SVC class, and prediction is given by
\begin{align}
    y_{\text{pred}} = \text{sign}\;\sum_{i}\alpha_i k(\vec{x}_i,\vec{x}). 
\end{align} \\

\begin{tabular}{ll}
    \hline
    hyperparameter  & values \\
    \hline
    {\emph{trotter\_steps}}(embedding layers) $L$ & [1, 3, 5] \\
    {\emph{C}} (SVC regularization) & [0.1, 1, 10, 100] \\
    $\emph{t}$ (time) & [0.01, 0.1, 1.0] \\
    {\emph{gamma\_factor}}  & [0.1,1,10]
\end{tabular}

    \subsection{QuantumKitchenSinks \citep{kitchensinks}}
This model uses a quantum circuit to define a feature map given by the concatenation of its output bit-strings. These features are then used to train a linear classifier. The feature map procedure works as follows:
\begin{itemize}
    \item Linearly transform the input feature vector $\vec{x}$ as $\vec{x}'_k=W_k\vec{x} + \vec{b}_k$ using randomly sampled $W_k$, $\vec{b}_k$ for $k=1,\cdots,k_{\text{max}}$. Here $W_k$, $\vec{b}_k$ are such that $\vec{x}'$ has dimension $n$ which may be different from $d$. 
    \item Feed each $\vec{x}'_k$ into a circuit (described below) that returns a single measurement sample $\vec{z}_k\in\{0,1\}^n$. The concatenated vector $\vec{z}_1 \oplus \cdots \oplus \vec{z}_{k_{\text{max}}}$ is the feature mapped vector of size $n\cdot k_{\text{max}}$. 
\end{itemize}
The circuit used in the second step above consists of angle encoding the feature mapped vectors with $X$ rotations on individual qubits, and applying two layers of CNOT gates: the first between adjacent qubits $(j,j+1)$; the second between qubits $(j,j+2)$ a distance 2 apart. This choice is a natural generalisation of the example present in \citep{kitchensinks}, which does not present a specific circuit structure for circuits beyond 4 qubits.

The following is an example of the quantum circuit used to compute feature vectors \texttt{QuantumKitchenSinks} for the input $\mathbf{x} = (0.1, 0.2, 0.3, 0.4)^T$ (here without first applying the classical linear transformation):

\begin{center}
\includegraphics[width=0.30\textwidth]{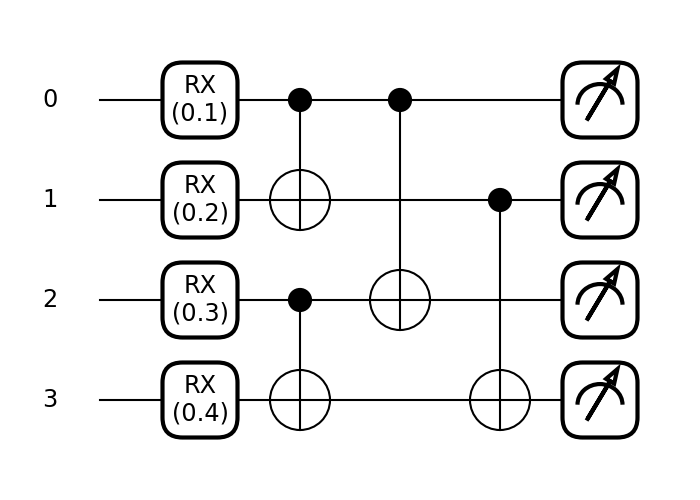}
\end{center}

The feature mapped vectors are then passed to a linear classifier, which we implement using scikit-learn's \texttt{LogisticRegression} class (note logistic regression can be used for linear classification via the cross entropy loss). 

For prediction, a new feature-mapped vector $\vec{z}$ is created via the same process, and 
\begin{align}
    y_{\text{pred}} = \text{sign} \;\vec{w}\cdot\vec{z}
\end{align}
where $\vec{w}$ is the linear classifier found during training.\\

The hyperparameter search values are as follows: \\

\begin{tabular}{ll}
    \hline
    hyperparameter  & values \\
    \hline
    {\emph{n\_qfeatures}} (circuit size) & [$d$, $\lfloor d/2 \rfloor$] \\
    {\emph{n\_episodes}} (number of circuits $k_{\text{max}}$) & [10, 100, 500, 2000] 
\end{tabular}

    \subsection{QuanvolutionalNeuralNetwork \citep{quanvolutional}}
This model consists of a fixed quantum feature map followed by a trainable convolutional neural network. The data is first scaled to lie in the range $[-1,1]$ and a binary threshold function (with threshold zero) is applied and the data scaled by $\pi$ so that it takes values in $\{0, \pi\}$.  One then applies a convolutional layer to the data, where the convolutional filters are given by a $n_q^2$ qubit quantum circuits that take as input $n_q\times n_q$ sized windows of the input data, in an analogous manner to a classical convolutional filter. The quantum circuits consists of random gates that we implement via PennyLane's \texttt{RandomLayers} class, and the output of the circuits is the number of ones in the bitstring that has the highest probability to be sampled from the circuit. As with convolutional neural networks, we allow for more than one channel in this layer, controlled by the hyperparameter \emph{n\_qchannels}. 

The following is an example of the quantum circuit used in the \texttt{QuanvolutionalNeuralNetwork} for a $2 \times 2$-dimensional input window, where the thresholded and rescaled pixel values form a pre-processed input vector $(0, 0, \pi, \pi)^T$:

\begin{center}
\includegraphics[width=0.25\textwidth]{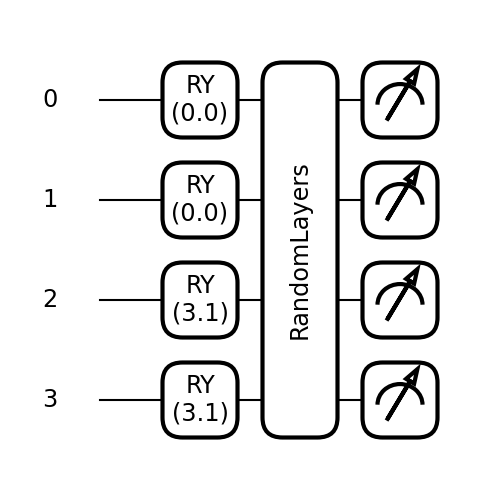}
\end{center}

The resulting data is then fed into a convolutional neural network with the same specifications as for the \texttt{ConvolutionalNeuralNetwork}. \\

\begin{tabular}{ll}
    \hline
    hyperparameter  & values \\
    \hline
    {\emph{learning\_rate}} & [0.0001, 0.001, 0.01] \\
    {\emph{n\_qchannels}}  & [1, 5, 10 ]\\
    {\emph{qkernel\_shape}} ($n_q$) & [2, 3] \\
    {\emph{kernel\_shape}} (CNN filter size) & [2, 3, 5]
\end{tabular}

    \subsection{WeiNet \citep{weinet}}
This model (that we call WeiNet following the first author of the paper) implements a convolutional layer as a unitary operation that acts on input data that is amplitude encoded into a quantum state. The model has two registers: the ancilliary register and the work register. The ancilliary register is used to parameterise a 4 qubit state which in turn controls a number of unitaries that act on the work register, where the data is encoded via amplitude encoding. Note that in figure 2 of \citep{weinet}, the Hadamard gates on the ancilla register have no effect since we trace this register out. The effect of this register is then to simply perform a classical mixture of the unitaries $Q_i$ defined therein on the work register. For simplicity (and to save qubit numbers), we parameterise this distribution via $16$ real trainable parameters. 

Two of the qubits are then traced out, which is equivalent to a type of pooling. All single and double correlators $\langle Z \rangle$ and $\langle ZZ \rangle$ are measured, and a linear model on these values is used for classification.

The following is an example of one of the quantum circuits used in the \texttt{WeiNet} model for a $4 \times 4$-dimensional input window encoded into state $\ket{\psi_{\mathbf{x}}}$ acted on by ``filter unitary'' $U$:

\begin{center}
\includegraphics[width=0.25\textwidth]{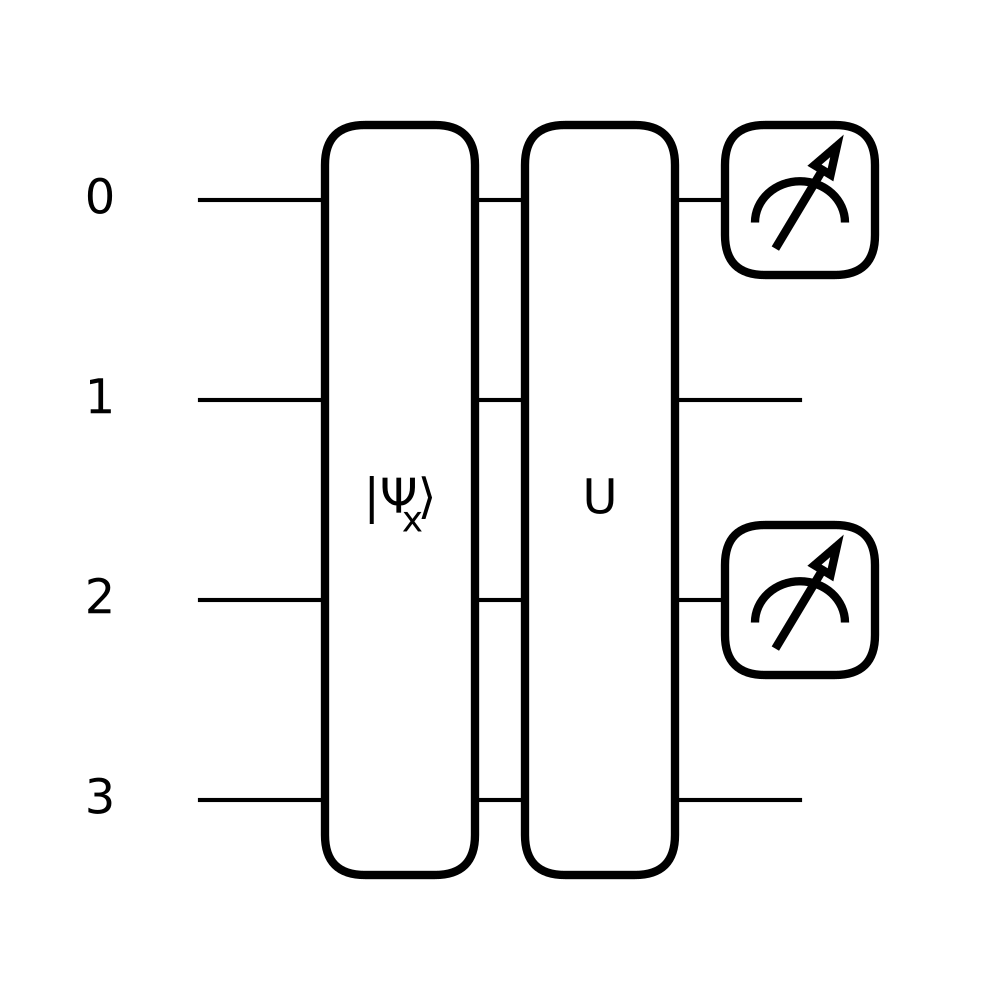}
\end{center}

The paper does not specify the training loss; like the other convolutional models we use the binary cross entropy with a sigmoid (logistic) activation. The corresponding probabilities from the sigmoid activation are used for prediction. \\

\begin{tabular}{ll}
    \hline
    hyperparameter  & values \\
    \hline
    {\emph{learning\_rate}} & [0.0001, 0.001, 0.01] \\
    {\emph{filter\_type}} & [edge\_detect, smooth, sharpen ]
\end{tabular}

    \subsection{MLPClassifier}
This is a multiplayer perception model implemented via scikit-learn's \texttt{MLPClassifier} class. An input feature vector $\vec{x}$ is transformed using a sequence of linear transformations $W_l$ and element wise-activation functions $a$ as 
\begin{align}
    f(\vec{x}) = a(W_L\cdots a(W_2(a(W_1\vec{x})))\cdots).
\end{align}
`We use the rectified linear unit activation $a(x)=\max(0,x)$. The trainable parameters of the model are the weights of the matrices $W_k$. We vary the number of layers $L$ and the shapes of the weight matrices $W_k$ via the model parameter \emph{hidden\_layer\_size}, which we set to be one of $[(100,), (10, 10, 10, 10), (50, 10, 5)]$. Here, each element of the tuple corresponds to a different layer and the values give the output dimensions of the corresponding matrices. The last layer is not included here since it is always of dimension 1. Training is done via gradient descent with the binary cross entropy loss using the adam update and the default class fit method. We vary the initial learning rate and regularisation strength \emph{alpha}. All other parameters are set to the class defaults, except the maximum number of iterations, \emph{max\_iter}, that we set to 3000. \\

\begin{tabular}{ll}
    \hline
    hyperparameter  & values \\
    \hline
    \emph{learning\_rate} & [0.001, 0.01, 0.1] \\
    \emph{hidden\_layer\_sizes} & [(100,), (10,10,10,10), (50,10,5) ]\\
    {\emph{alpha}} (regularisation) & [0.01, 0.001, 0.0001] \\
\end{tabular}

    \subsection{SVC}
This model is a support vector machine classifier implemented via scikit-learn's SVC class. We use the radial basis function kernel: 
\begin{align}
    k(\vec{x},\vec{x}') = \exp(-\emph{gamma}\vert\vert x-x' \vert\vert^2).
\end{align}
During hyperparameter search, we vary the bandwidth parameter $\emph{gamma}$ and the regularisation strength $\emph{C}$. \\

\begin{tabular}{ll}
    \hline
    hyperparameter  & values \\
    \hline
    \emph{C} (SVC regularization) & [ 0.1, 1, 10, 100] \\
    \emph{gamma} & [0.001, 0.01, 0.1, 1]
\end{tabular}

    \subsection{ConvolutionalNeuralNetwork}
This model is a vanilla implementation of a convolutional neural network (CNN), written in flax. The structure of the network is as follows 

\begin{itemize}
    \item a 2D convolutional layer with 32 output channels
    \item a max pool layer
    \item a 2D convolutional layer with 64 output channels
    \item a max pool layer
    \item a two layer fully connected feedforward neural network with 128 hidden neurons and one output neuron
\end{itemize}

The probability of class 1 is given by 
\begin{align}
    P(+1\vert \vec{w},\vec{x}) = \sigma(f(\vec{w}),\vec{x})
\end{align}
where $\vec{w}$ are the weights of the model, $f(\vec{w})$ is the value of the final neuron, and $\sigma$ is the logistic function. These probabilities are fed to binary cross entropy loss for training. \\

\begin{tabular}{ll}
    \hline
    hyperparameter  & values \\
    \hline
    \emph{learning\_rate} & [0.0001, 0.001, 0.01 ] \\
    {\emph{kernel\_shape}} & [2, 3, 5] 
\end{tabular}

    \subsection{SeparableVariationalClassifier}
    
This is a simple quantum neural network model that does not use entanglement. The data encoding $V($ consists of $L$ layers, where in each layer arbitrary trainable single-qubit rotations are performed followed by a product angle embedding of the data via Pauli Y rotation gates. The encoding is proceeded by another layer of trainable single-qubit rotations, and prediction is given by measurement of $ O = \frac{1}{n}(Z_1 +\cdots +Z_n)$, i.e.\
\begin{align}
    y_{\text{pred}} = \text{sign}f(\vec{\theta},\vec{x})
\end{align}
where 
\begin{align}
f(\vec{\theta},\vec{x}) =\langle \frac{1}{n}(Z_1 +\cdots +Z_n) \rangle,
\end{align}
where $\vec{\theta}$ represents all trainable parameters in the $L$ layers. 

The following is an example of the single-qubit quantum circuit used in the \texttt{SeparableVariationalClassifier} model to process the first value of the input vector $\mathbf{x} = (0.1, 0.2, 0.3, 0.4)^T$ (whereas the remaining features are processed by similar circuits):

\begin{center}
\includegraphics[width=0.3\textwidth]{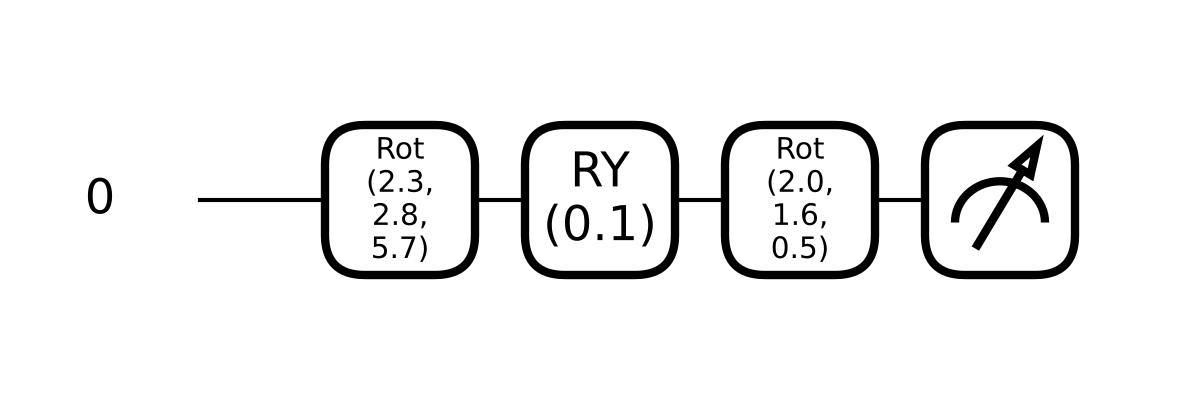}
\end{center}

To train the model, we define class probabilities via the logistic function
\begin{align}
    P(+1\vert \vec{\theta},\vec{x}) = \sigma(6\langle O \rangle)
\end{align}
where we multiply the observable value by $6$ since the sigmoid function varies significantly over the range $[-6,6]$. These probabilities are then used in a binary cross entropy loss function. \\

\begin{tabular}{ll}
    \hline
    hyperparameter  & values \\
    \hline
    {\emph{learning\_rate}} & [0.001, 0.01, 0.1] \\
    {\emph{encoding\_layers}} ($L$) & [1,3,5,10] 
\end{tabular}

    \subsection{SeparableKernelClassifier}
This model is the kernel equivalent of the above. The data encoding consists $L$ layers, where in each layer an $X$ rotation with angle $\pi/4$ is applied to each qubit followed by a product of $Y$ rotations that encode each element of $\vec{x}$ into individual qubits (so $n=d)$. 

The kernel is given by \eqref{qkernelfunction}:
\begin{align}
    k(\vec{x},\vec{x}') = \text{tr}[\rho(\vec{x})\rho(\vec{x}')],
\end{align}
and the model is trained using scikit-learn's SVC class. 

The following is an example of the single-qubit quantum circuit used in the \texttt{SeparableKernelClassifier} model to process the first value of the inputs $\mathbf{x}, \mathbf{x'} = (0.1, 0.2, 0.3, 0.4)^T$, using 2 layers:

\begin{center}
\includegraphics[width=0.49\textwidth]{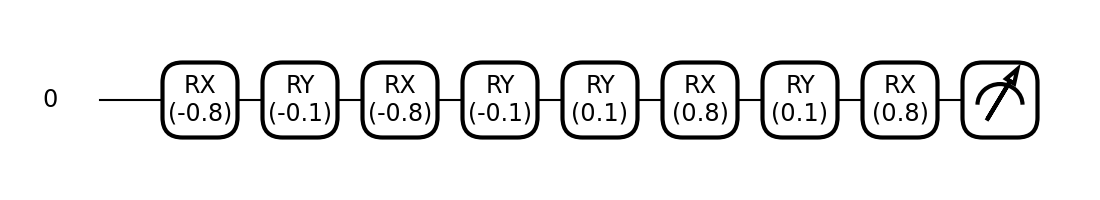}
\end{center}

\begin{tabular}{ll}
    \hline
    hyperparameter  & values \\
    \hline
    \emph{encoding layers} ($L$) & [1,3,5,10]  \\
    SVM regularization & [0.1, 1, 10, 100] 
\end{tabular}


\section{Convergence criteria of variational models}\label{app:convergence}
Here we describe the criterion used to decide convergence during training of variational models. This was used for all quantum neural network and quantum convolutional neural network models, as well as the vanilla convolutional neural network model. Convergence is decided as follows:

\begin{itemize}
    \item During training, the previous 400 values of the loss are stored, and at each step after step 400, the sample mean $\mu_1$ of the loss values for the first 200 and the sample mean $\mu_2$ of the loss values of last 200 steps is calculated, as well as the standard deviation $\sigma_2$ of the last 200 loss values. 
    \item If the model has converged, the statistics of the loss values in the two intervals should be approximately the same, so that the statistics of the sample means $\mu_1$ and $\mu_2$ are the same. Under an assumption of normality, the standard deviation of $\mu_2$ is $\sigma_2/\sqrt{200}$, and so for a converged model one expects
    \begin{align}
    \vert \mu_1 - \mu_2 \vert < \frac{\sigma_2}{\sqrt{200}}
    \end{align}
    with reasonable probability, whereas a model that is still training and whose loss decreases significantly in 200 steps is unlikely to satisfy the criterion.
    \item We adopted a slightly smaller interval than this to decrease the likelyhood of accidental early stopping. Namely, we used the criterion
    \begin{align}
        \vert \mu_1 - \mu_2 \vert < \frac{\sigma_2}{2\sqrt{200}}
    \end{align}
    to decide convergence. From visual inspection of loss plots, we found that this always gave a reasonable stopping criterion that corresponded to an approximately flat loss curve over the last 200 steps. 
\end{itemize}

\section{Details on the datasets used in this study}\label{app:data}

    We used 6 types of data generation procedures, five of which are purely synthetic, while one (MNIST) uses post-processing to reduce the dimension of the feature vectors. The data generation procedures provide different variables that can be tuned to create individual datasets. We refer to datasets for which one such variable is tuned as ``benchmarks'' and denote them by capital spelling.
    
    This section describes the procedures in more detail, and gives the settings for generating the benchmarks. The code can be found in the github repository.

    Note that unless otherwise stated, data was split into training and test sets using a ratio of test/train=$0.2$.
    
    \subsection{Linearly separable}\label{app:data-linearly-separable}

    The linearly separable data generation procedure creates data by a perceptron with fixed weights that labels input vectors sampled from a hypercube. The data is hence guaranteed to be linearly separable, and the dataset can be considered as the easiest or ``fruit-fly'' task in classification.\\
    
    \textit{Procedure}:
    \begin{enumerate}
        \item Sample input vectors $\mathbf{x} \in \mathbb{R}^d$ uniformly from a $d$-dimensional hypercube spanning the interval $[-1, 1]$ in each dimension. 
        \item Pick a weight vector for some $\mathbf{w} \in \mathbb{R}^d$ to define the linear decision boundary at $\mathbf{w}\mathbf{x} = 0$. Only retain the first $N$ input vectors that do not lie within a margin of size $0.02d$ around the decision boundary, or $\ \lVert \mathbf{w}\mathbf{x}\rVert > 0.02d$.
        \item Generate continuous-valued labels  $y' = \mathbf{wx}$.
        \item Binarise the labels via
        $$ y = \begin{cases} \phantom{-}1 \text{ if } y'-y_{\text{med}} > 0\\ -1 \text{ else},
        \end{cases}
        $$
        where $y_{\text{med}}$ is the median of all continuous labels. This standarisation procedure ensures that the classes are balanced.
    \end{enumerate}

    \textit{Settings for the \textsc{linearly separable} benchmark:} 
    \begin{itemize}
        \item Number of features $d \in \{2,...,20\}$
        \item Number of samples $N=300$
        \item Perceptron weights $\mathbf{w} = (1,...,1)^T$
    \end{itemize}

    \subsection{Bars and stripes}\label{app:data-bas}
    The bars and stripes generation procedure is intended to be a simple task for the three convolutional models. It creates gray-scale images of either vertical bars or horizontal stripe on a 2D pixel grid. \\

    \emph{Procedure}:
    \begin{enumerate}
        \item Sample $N$ labels $y_i=-1,1$ uniformly at random.
        \item For each $y_i$ create a pixel grid of shape $d\times d$ that will store the data $\vec{x}_i$. If $y_i=-1$, for each column, sample a random variable taking values $\pm 1$ with equal probability, and fill the column of $\vec{x}_i$ with this value. If $y_i=1$, do the same for the rows. 
        \item For each image $\vec{x}_i$ add independent Gaussian noise with standard deviation $\sigma$ and mean $0$. 
    \end{enumerate}

    \textit{Settings for the \textsc{bars \& stripes} benchmark:}    
    \begin{itemize}
        \item Image width $d=4, 8, 16, 32.$
        \item noise standard deviation $\sigma = 0.5$
        \item Number of data samples $N=1000$
    \end{itemize}
    
    \subsection{Downscaled MNIST}\label{app:data-mnist}
    
    The MNIST datasets are based on the famous handwritten digits data \citep{lecun1998mnist} using digits $3$ and $5$ which are amongst the hardest to distinguish. The ratio between test and training set for this procedure is test/train=$0.17$. The original data was processed by different methods for dimensionality reduction.\\
    
    The \textsc{mnist pca}, \textsc{mnist pca}$^-$ benchmarks use principal component analysis (PCA) to reduce dimensions.\\
    
    \textit{Procedure}:
    \begin{enumerate}
        \item Flatten and standarise the inputs images, which is important for PCA to work well. The standarisation parameters are derived from the training set, and then used to standarise the test set.\footnote{This best practice takes into account that in applications one does not necessarily have access to the test set at training time.}
        \item Compute the $d$ largest principal components of the pre-processed training set inputs via Principal Component Analysis. 
        \item Project the training and test set inputs onto those components to gain new input vectors of dimension $d$. This is the \textsc{mnist pca} dataset.
        \item For the \textsc{mnist pca}$^-$ dataset, we sampled a subset of $250$ data points for each of the training and test set from \textsc{mnist pca}.
    \end{enumerate}

    \textit{Settings for the \textsc{mnist pca} benchmark:}
    \begin{itemize}
        \item Number of features $d \in \{2,...,20\}$
    \end{itemize}
    
    \textit{Settings for the \textsc{mnist pca}$^-$ benchmark:}
    \begin{itemize}
        \item Number of features $d \in \{2,...,20\}$
        \item Number of samples $N =250$ for training and test set each. 
    \end{itemize}
    
    The \textsc{mnist cg} benchmark coarse-grains the pixels of the original images to try and preserve the correlation structure used by convolutional neural networks. As before, we use the digits 3 and 5 only. \\
    
    \textit{Procedure}:
    \begin{enumerate}
        \item Resize the original 28x28 pixel data to a pixel grid of size $H\times H$, using bilinear interpolation. 
        \item Flatten and standardize the images.
    \end{enumerate}

    \textit{Settings for the \textsc{mnist cg} benchmark:}
    \begin{itemize}
    \item Pixel grid height/width $H~\in~\{4, 8, 16, 32\}$. 
    \end{itemize}

    \subsection{Hidden manifold model}\label{app:data-hidden-manifold}

    This data generation procedure is based on \citet{goldt2020modeling}, who classify data sampled from a $m$-dimensional manifold by a neural network, and then embed the data into a $d$-dimensional space. The properties of this data generation process allow the authors to compute analytical generalisation error dynamics using tools from statistical physics. The structure intends to mimic datasets used in image recognition (such as MNIST), which have been shown to effectively use low-dimensional manifolds.\\    

    \textit{Procedure}:

    \begin{enumerate}
        \item Randomly sample $N$ feature vectors $\mathbf{c}^m \in \mathbb{R}^m$ with entries from a standard normal distribution. These vectors lie on the ``hidden manifold''.
        \item Create an embedding matrix $F \in \mathbf{R}^{d \times m}$. 
        \item Embed the feature vectors via
        $$
        \mathbf{x} = \mathbf{\phi}(\mathbf{Fc}/\sqrt{m})
        $$
        where $\mathbf{\phi}_i(\mathbf{x}) =  \tanh(x_i-b_i)$
        \item Generate continuous-valued labels using a neural network applied to the vectors on the manifold
        $$y' =  \mathbf{v}^T \mathbf{\varphi}(\mathbf{Wc}/\sqrt{m})$$
        using component-wise $tanh$ functions as the activation $\varphi$, and the entries in $W \in \mathbb{R}^{m, m}, \mathbf{v} \in \mathbb{R}^{m}$ sampled from a standard distribution. 
        \item In order to get balanced classes, we rescale the data by subtracting the median $y_{\text{med}}$ of all labels and then apply a thresholding function
        $$  
        y = \begin{cases} \phantom{-}1 \text{ if } y'-y_{\text{med}} > 0\\ -1 \text{ else}     
        \end{cases}
        $$
    \end{enumerate}

    \textit{Settings for the \textsc{hidden manifold} benchmark:}    
    \begin{itemize}
        \item Number of features $d \in \{2,...,20\}$
        \item Number of samples $N=300$
        \item Manifold dimension $m=6$ 
        \item Entries of feature matrix $F$ sampled from a standard distribution
    \end{itemize}
    \textit{Settings for the \textsc{hidden manifold diff} benchmark:}    
    \begin{itemize}
        \item Number of features $d=10$
        \item Number of samples $N=300$
        \item Manifold dimension $m\in \{2,...,20\}$ 
        \item Entries of feature matrix $F$ sampled from a standard distribution
    \end{itemize}

    \subsection{Two curves} \label{app:data-two-curves}
    
    This data generation procedure is inspired by \citet{buchanan2021deep}, who consider data sampled from two curves -- one for each class -- embedded into a $d$-dimensional space to prove that the maximum curvature and minimum distance of these curves determine the resources required by a neural network to generalise well. We can hence understand curvature and distance as two variables that influence the difficulty of the data. \\
    
    To control the curvature we use a one-dimensional Fourier series of a maximum degree $D$ in each dimension. To control the average distance of the curves we use the same embedding, but shift one curve by some constant. \\

    \textit{Procedure}:

    \begin{enumerate}
    \item Sample $N$ values $t \in \mathbb{R}$ uniformly at random 
    from the interval $[0, 1]$. (This value defines the position of a data point on the curve we embed.)
    \item To create the inputs for class $1$, embed half of the $t$-values into a $d$-dimensional space via a Fourier series defined in every dimension,
    $$x_i = \sum_{n = 0}^D \alpha^i_n \cos(nt) + \beta^i_n \sin(nt)  + \epsilon, $$
    where $D$ is the maximum degree of the Fourier series and $\{\alpha^i_n\}, \{\beta^i_n\}$ are real-valued Fourier coefficients that we sample uniformly from the interval $[0, 1]$. The noise factor $\epsilon$ determines the variance of a random ``spread'' added to the curves around their trajectory in the high-dimensional space.
    \item To create the inputs for class $-1$, embed the other half of the $t$-values using the same procedure and Fourier coefficients, but adding an offset of $\Delta$ to each dimension.
    \end{enumerate}
    
    \textit{Settings for the \textsc{two curves} benchmark:}    
    \begin{itemize}
        \item Number of features $d \in \{2,...,20\}$
        \item Number of samples $N=300$
        \item Noise factor $\epsilon = 0.01$
        \item Maximum degree $D=5$
        \item Curve offset $\Delta= 0.1$
    \end{itemize}
    \textit{Settings for the \textsc{two curves diff} benchmark:}    
    \begin{itemize}
        \item Number of features $d =10$
        \item Number of samples $N=300$
        \item Noise factor $\epsilon = 0.01$
        \item Maximum degree $D \in \{2,...,20\}$
        \item Curve offset $\Delta= \frac{1}{2D}$
    \end{itemize}

    \subsection{Hyperplanes and parity}\label{app:data-hyperplanes}
    
    We created an artificial dataset that classifies low-dimensional feature vectors by whether they lie on the ``positive'' side of an even or odd number of a set of $k$ hyperplanes. The feature vectors are then embedded into a higher-dimensional space via a linear transform. The result is a division of the space into regions of different classes that are delineated by hyperplane intersections. The parity operation makes sure that a model implicitly has to learn all hyperplane positions to guess the right label. The difficulty of the classification problem is expected to increase with the number of hyperplanes.\\

    \textit{Procedure}:

    \begin{enumerate}
        \item Sample $N$ feature vectors $\mathbf{c} \in \mathbb{R}^m$ from a standard normal distribution.
        \item Embed each feature vector into $\mathbb{R}^d$ by multiplying the feature vectors with a matrix $\mathbf{M} \in \mathbb{R}^{d \times m}$,
        $$ \mathbf{x} = \mathbf{M}\mathbf{c}.$$
        \item Compute $k$ predictions for the $m$-dimensional feature vectors via
        $$ p^{(j)} = \begin{cases} 1 \text{ if } \mathbf{w}^{(j)} \mathbf{c} + b^{(j)} > 0\\ -1 \text{ else}     
        \end{cases} j=1,...,k
        $$
        using uniformly sampled weight vectors $\{\mathbf{w}^{(j)} \in \mathbb{R}^m\}$ and biases $\{b^{(j)}\}$.
        \item The final label is defined as the parity of these predictions, or whether the number of 1-predictions is even:
        $$
         y = \begin{cases} \phantom{-}1 \text{ if } \sum\limits_{j=1}^k \frac{p^{(j)} + 1}{2}  \text{ even} \\ -1 \text{ else}     
        \end{cases}
        $$
        \item To ensure balanced classes we initially sample a larger number of datapoints from which we subsample the desired number for each class.
        \item Standarise the inputs. 
    \end{enumerate}

    \textit{Settings for the \textsc{hyperplanes diff} benchmark:}    
    \begin{itemize}
        \item Dimension $d=10$
        \item Number of hyperplanes $k \in \{2,...,20\}$
        \item Number of data samples $N=1000$
        \item Dimension of hyperplane and initial feature vectors $m=3$
        \item Entries of $\mathbf{M}$ are uniformly sampled from $[0, 1]$
    \end{itemize}

    \section{Collection of detailed results}\label{app:results}

    We add the ranking and accuracy plots of all benchmarks here for readers who are interested in the detailed results. 

    \begin{figure*}
        \centering
        \includegraphics[width=0.4\textwidth]{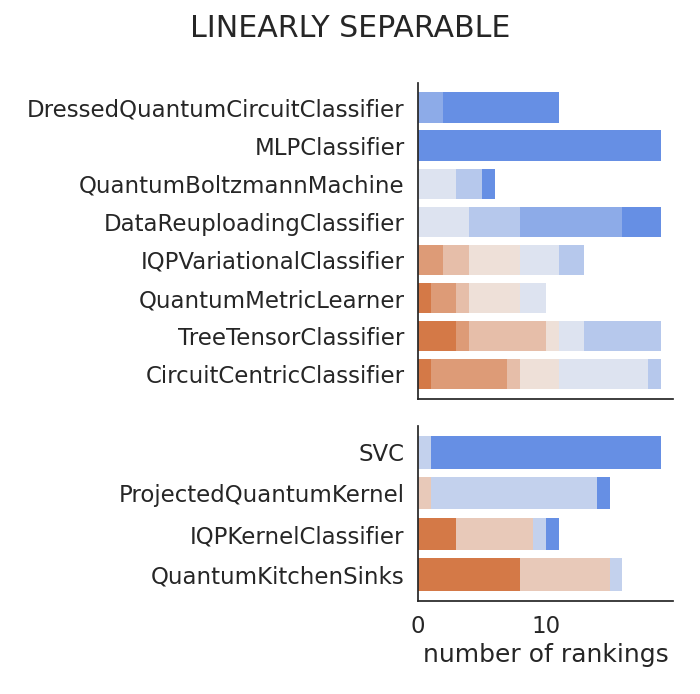}
        \includegraphics[width=0.4\textwidth]{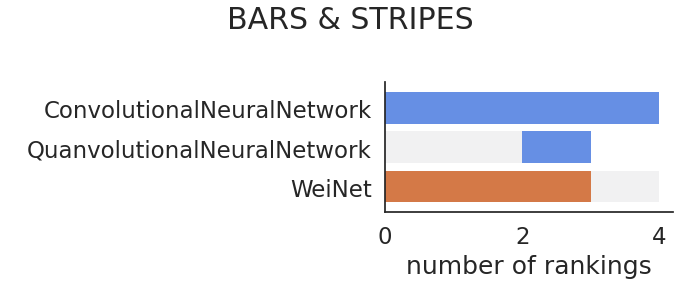}
        \includegraphics[width=0.4\textwidth]{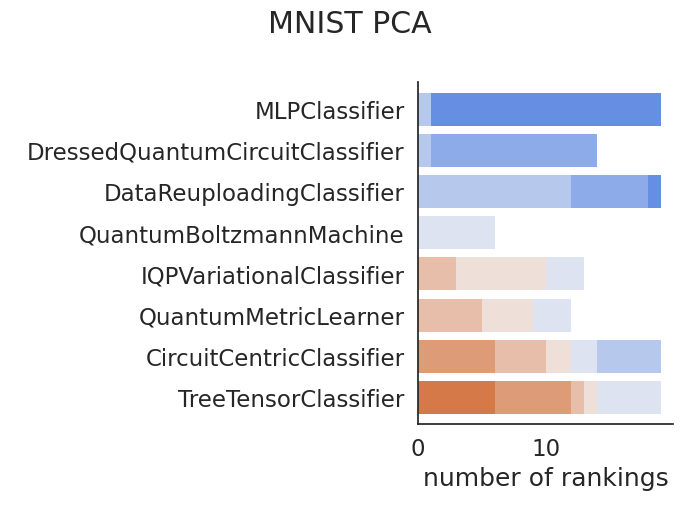}
        \includegraphics[width=0.4\textwidth]{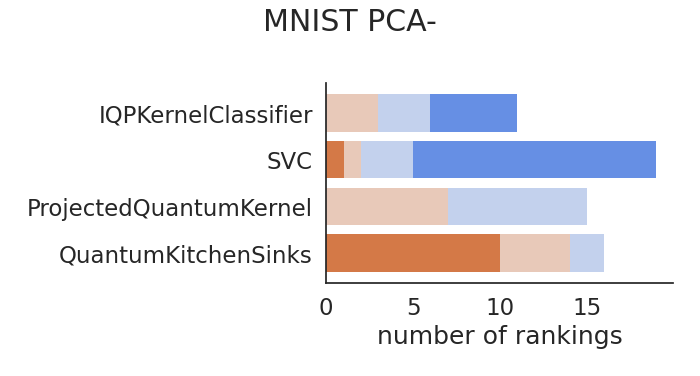}
        \includegraphics[width=0.4\textwidth]{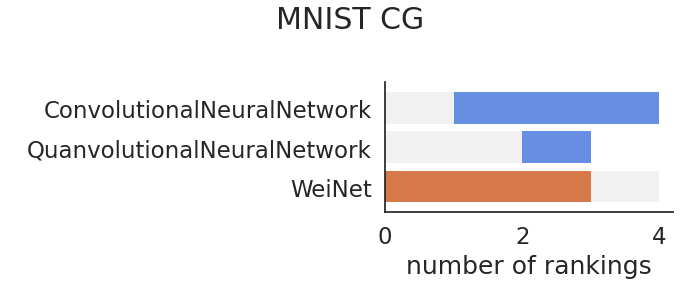}
        \caption{Ranking plots like shown in Figure~\ref{fig:ranking-all} for selected benchmarks.}
        \label{fig:rankings-benchmarks1}
    \end{figure*}

    \begin{figure*}
        \centering
        \includegraphics[width=0.4\textwidth]{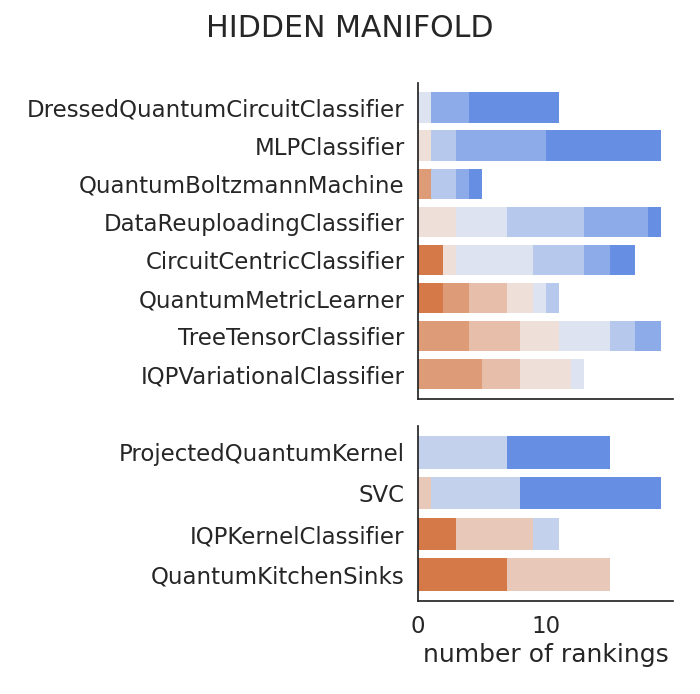}
        \includegraphics[width=0.4\textwidth]{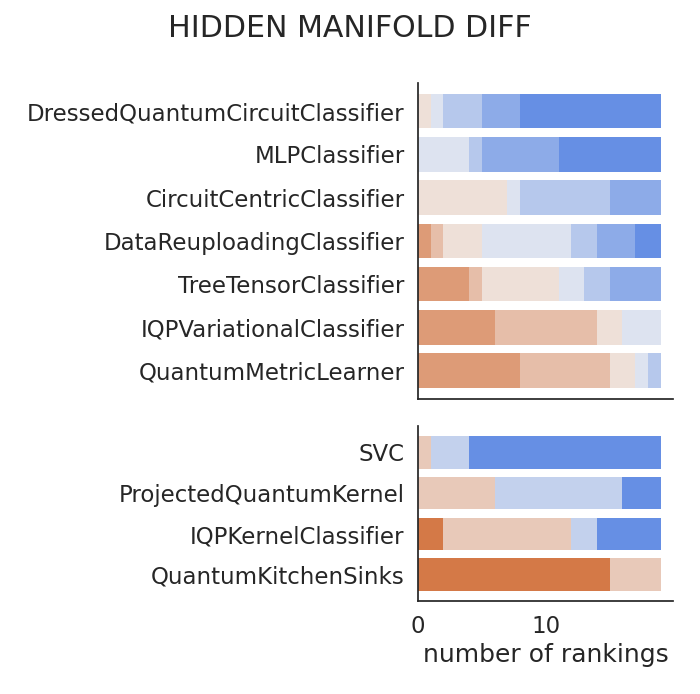}
        \includegraphics[width=0.4\textwidth]{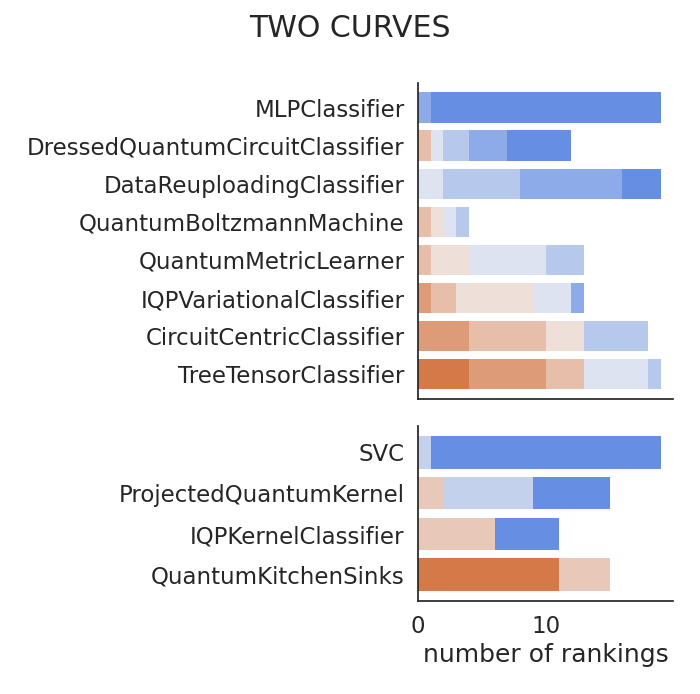}
        \includegraphics[width=0.4\textwidth]{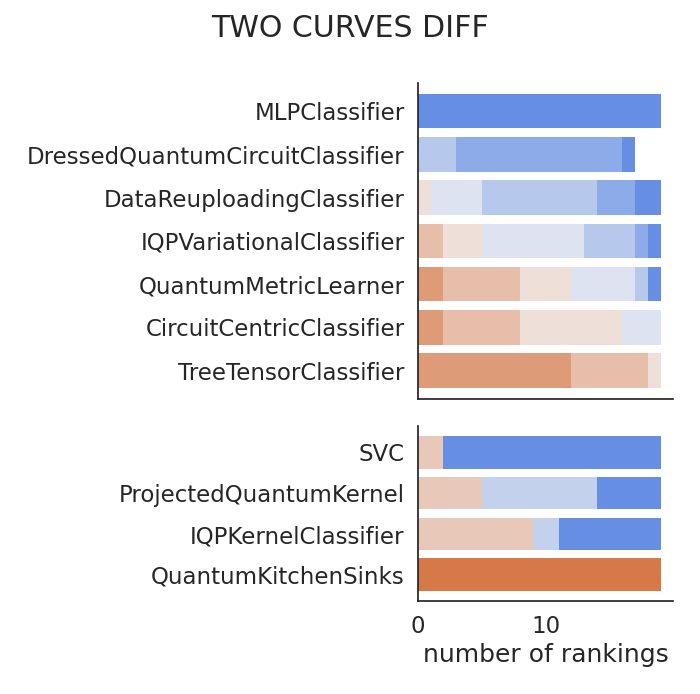}
        \includegraphics[width=0.4\textwidth]{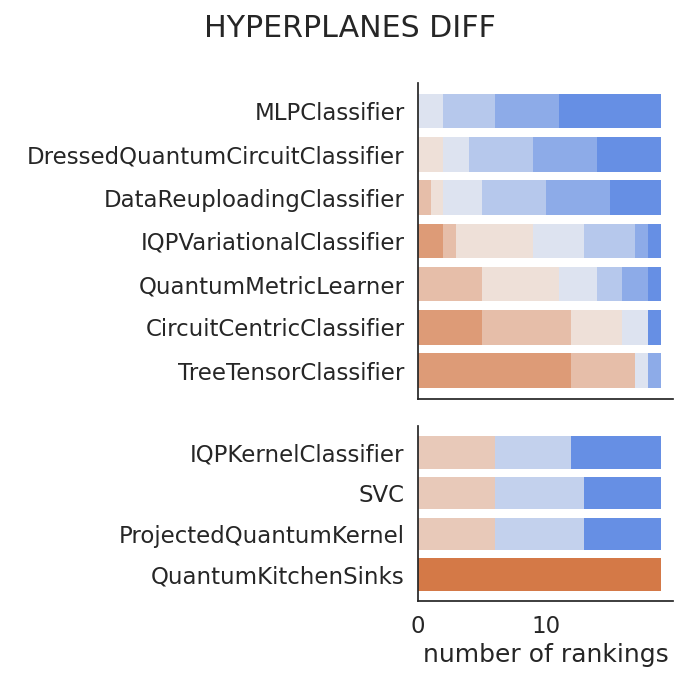}
        \caption{Ranking plots like shown in Figure~\ref{fig:ranking-all} for selected benchmarks (continued).}
        \label{fig:rankings-benchmarks1}
    \end{figure*}

    \begin{figure*}
        \centering
        \includegraphics[width=0.49\textwidth]{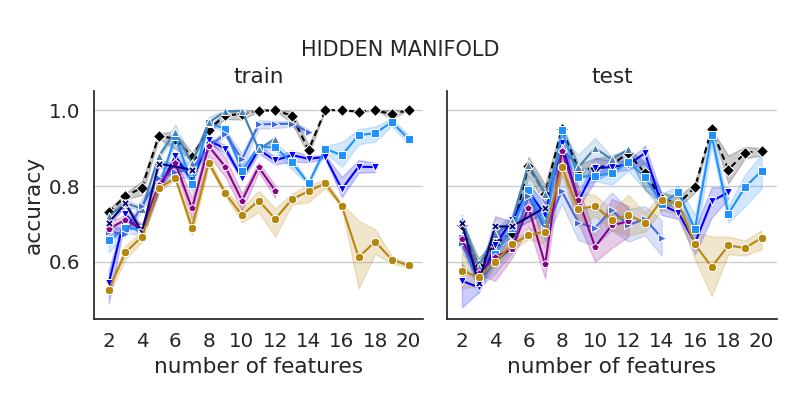}\includegraphics[width=0.49\textwidth]{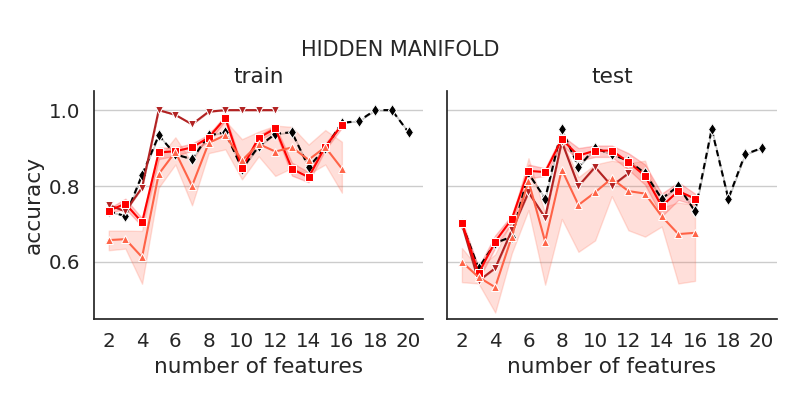}
        \includegraphics[width=0.49\textwidth]{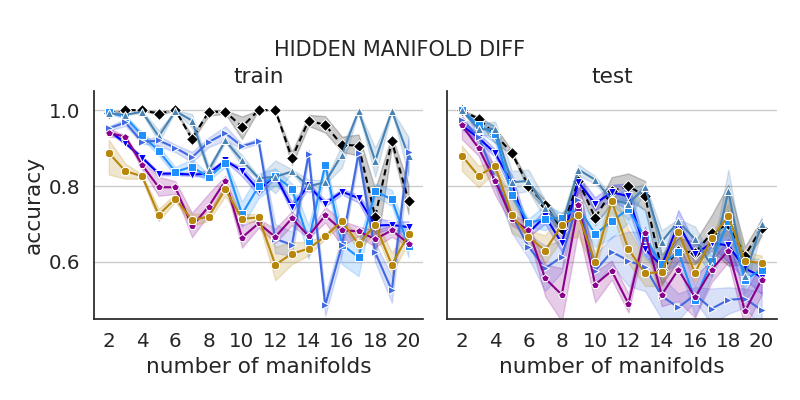}\includegraphics[width=0.49\textwidth]{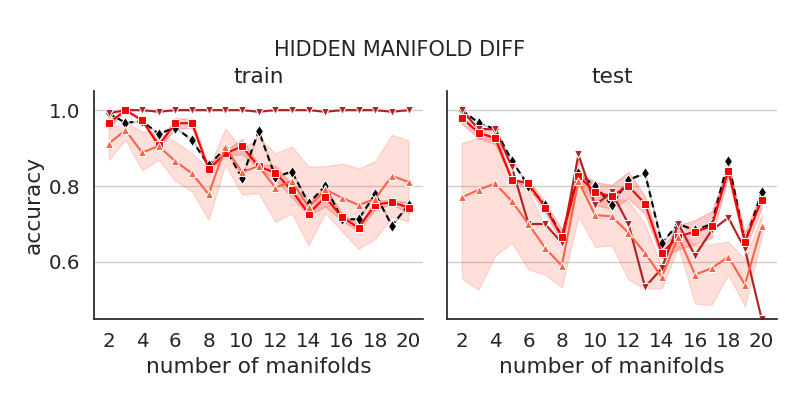}
        \includegraphics[width=0.49\textwidth]{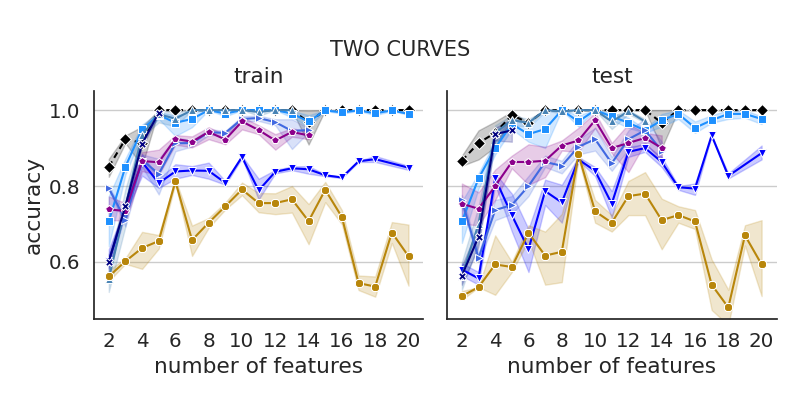}\includegraphics[width=0.49\textwidth]{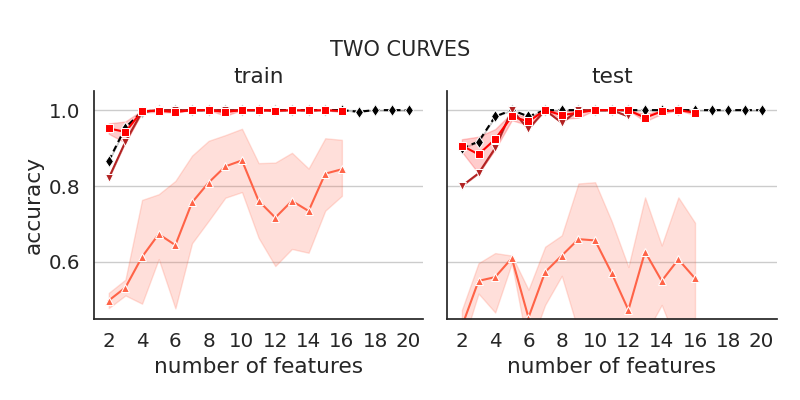}
        \includegraphics[width=0.49\textwidth]{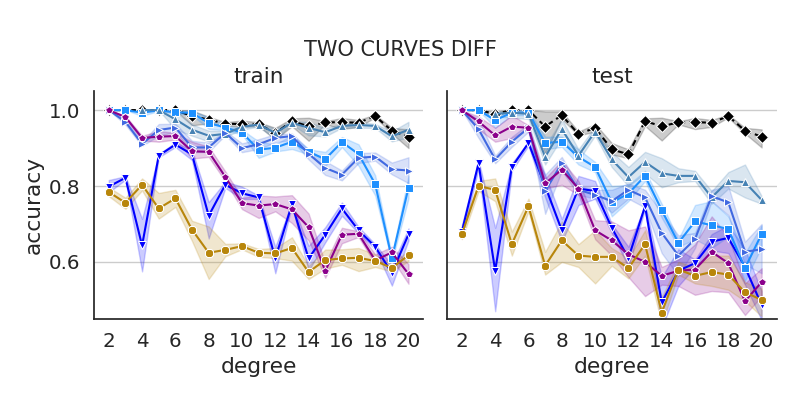}\includegraphics[width=0.49\textwidth]{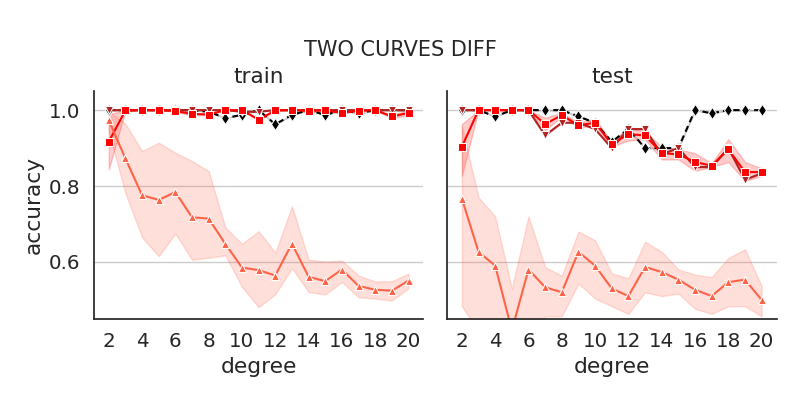}        
        \includegraphics[width=0.3\textwidth]{figures/score-qnn-legend.png}~~~~~~~~~~~~~~~~~~~~~~~~~~~~\includegraphics[width=0.23\textwidth]{figures/score-kernel-legend.png}
        \caption{Detailed training and test accuracies for the benchmarks not shown in Figure~\ref{fig:scores-datasets}.}
        \label{fig:scores-datasets-rest}
    \end{figure*}

    \begin{figure}
        \centering
        \includegraphics[width=0.24\textwidth]{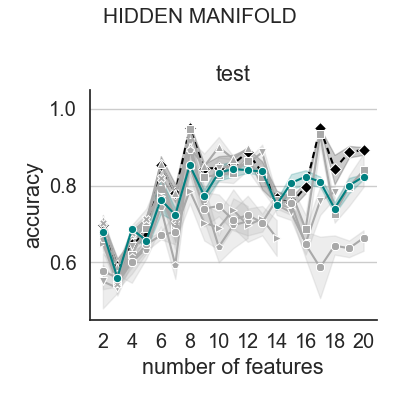}\includegraphics[width=0.24\textwidth]{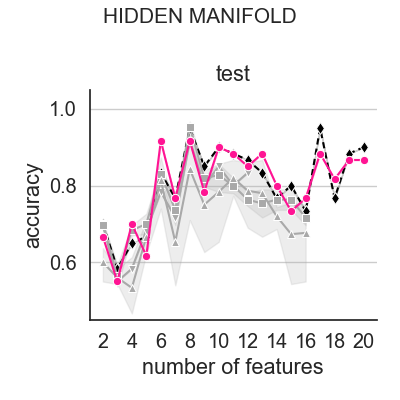}
        \includegraphics[width=0.24\textwidth]{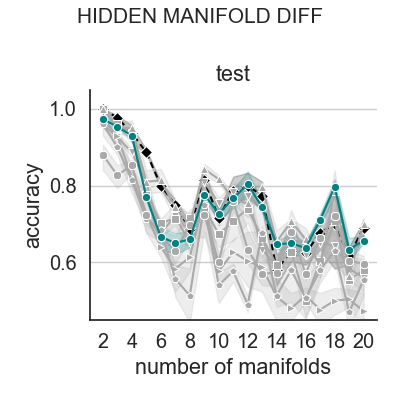}\includegraphics[width=0.24\textwidth]{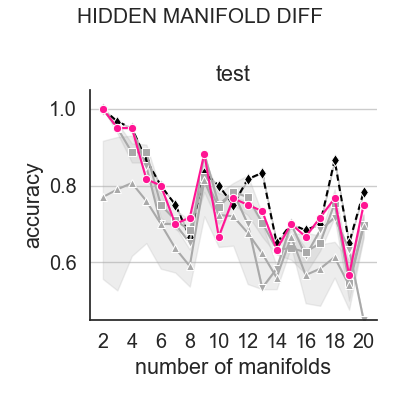}
        \includegraphics[width=0.24\textwidth]{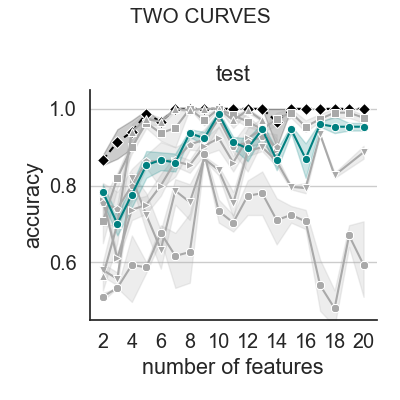}\includegraphics[width=0.24\textwidth]{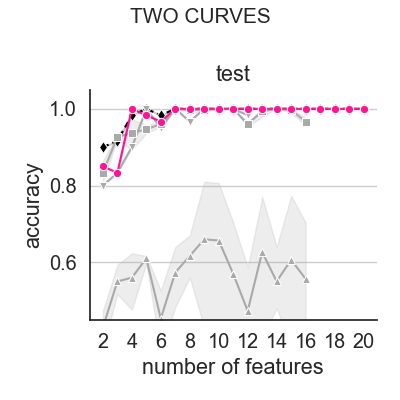}
        \includegraphics[width=0.24\textwidth]{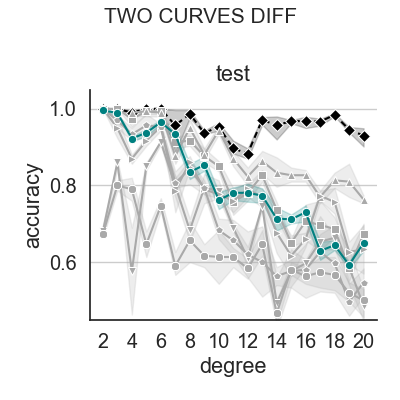}\includegraphics[width=0.24\textwidth]{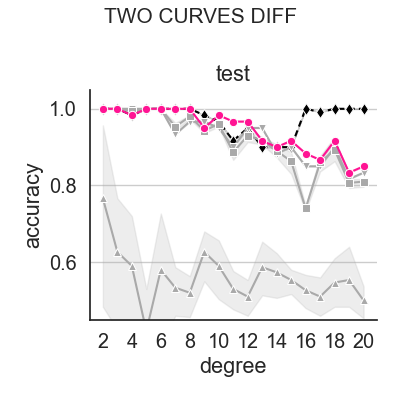}
        \includegraphics[width=0.25\textwidth]{figures/score-separable-qnn-legend.png}\includegraphics[width=0.2\textwidth]{figures/score-separable-kernel-legend.png}
        \caption{Results for separable models for the benchmarks not shown in Figure~\ref{fig:scores-separable}.}
        \label{fig:scores-separable-rest}
    \end{figure}

    \begin{figure}
        \centering
        \includegraphics[width=0.49\textwidth]{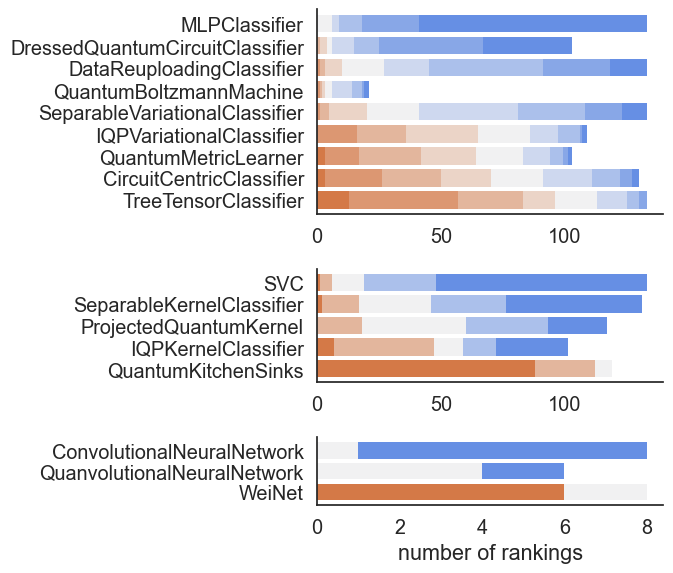}
        \caption{Total ranking results as reported in Figure~\ref{fig:ranking-all}, but with the two separable models included.}
        \label{fig:ranking-all-sep}
    \end{figure}

\end{document}